\documentclass[english,aps,manuscript]{revtex4}
\usepackage[T1]{fontenc}
\usepackage[latin9]{inputenc}
\usepackage{color}
\usepackage{amsmath}
\usepackage{amssymb}
\usepackage{graphicx}
\usepackage{esint}

\makeatletter
\@ifundefined{textcolor}{}
{%
 \definecolor{BLACK}{gray}{0}
 \definecolor{WHITE}{gray}{1}
 \definecolor{RED}{rgb}{1,0,0}
 \definecolor{GREEN}{rgb}{0,1,0}
 \definecolor{BLUE}{rgb}{0,0,1}
 \definecolor{CYAN}{cmyk}{1,0,0,0}
 \definecolor{MAGENTA}{cmyk}{0,1,0,0}
 \definecolor{YELLOW}{cmyk}{0,0,1,0}
 }

\@ifundefined{definecolor}
 {\usepackage{color}}{}

\makeatother

\usepackage{babel}
\begin{document}

\title{Efficient finite-time measurements under thermal regimes}

\author{Carlos Alexandre Brasil}

\affiliation{Instituto de Física \textquotedbl{}Gleb Wataghin\textquotedbl{},
Universidade Estadual de Campinas, P.O. Box 6165, 13083-970 Campinas,
SP, Brazil}

\author{Leonardo Andreta de Castro}

\author{Reginaldo de Jesus Napolitano}

\affiliation{Instituto de Física de São Carlos, Universidade de São Paulo, P.O.
Box 369, 13560-970 São Carlos, SP, Brazil}
\begin{abstract}
Contrary to conventional quantum mechanics, which treats measurement
as instantaneous, here we explore a model for finite-time measurement.
The main two-level system interacts with the measurement apparatus
in a Markovian way described by the Lindblad equation, and with an
environment, which does not include the measuring apparatus. To analyse
the environmental effects on the final density operator, we use the
Redfield approach, allowing us to consider a non-Markovian noise.
In the present hybrid theory, to trace out the environmental degrees
of freedom, we use a previously-developed analytic method based on
superoperator algebra and Nakajima-Zwanzig superoperators. Here, we
analyse two types of system-environment interaction, phase and amplitude
damping, which allows us to conclude that, in general, a finite-time
quantum measurement performed during a certain period is more efficient
than an instantaneous measurement performed at the end of it, because
the rate of change of the populations is attenuated by the system-measurement
apparatus interaction. 
\end{abstract}


\maketitle

\section{Introduction}

The measurement problem in quantum mechanics is responsible for most
of the controversy surrounding this theory \cite{key-5-1,key-6}.
While in classical physics the observation of an object does not presuppose
any changes of its properties, in quantum mechanics this interaction,
far from insignificant, causes profound changes on the object under
observation. Besides changing the state of the system, the measurement
of a given property can influence the value of another (the case of
\emph{conjugate observables} such as position and momentum). Another
aggravating aspect is the fact that it is not possible to predict
with certainty the result of a specific measurement: quantum mechanics
rules allow us to know only the several possible results and their
respective probabilities - it is an \emph{ensemble theory}, as found
by Born \cite{key-7,key-8} and Dirac \cite{key-9,key-10}.

To describe a measurement, it is necessary to expand the wave-function
in terms of the eigenstates of the observable to be measured: the
possible results of the measurement will be the respective eigenvalues
of the observable, and the corresponding probabilities will be the
square modulus of the expansion coefficients. When the measurement
is complete, the wave-function reduction takes place, and the state
of the system is projected into the eigenstate corresponding to the
eigenvalue obtained. All this was synthesized by von Neumann in his
pioneer treatise on quantum mechanics \cite{key-4}, where he states
the two possible processes of wave function evolution: 
\begin{enumerate}
\item \emph{Reduction}, when a measurement is made; 
\item \emph{Unitary evolution} according to the Schrödinger equation, between
measurements. 
\end{enumerate}
In von Neumann's original work \cite{key-4}, the transmission of
information to the measuring apparatus is described by considering
two Hilbert spaces which we shall call $I$, the \emph{principal system}
under observation, and $II$, the \emph{measurement apparatus}. Here,
let $\hat{A}^{\left(I\right)}$ be the observable of $I$ to be measured,
$\left\{ \left|a_{n}\right\rangle \right\} $ the set of its eigenfunctions
with associated eigenvalues $a_{n}$, and let $\left|\phi^{\left(I\right)}\right\rangle $
be the initial state of the system, given as the linear combination

\begin{equation}
\left|\phi^{\left(I\right)}\right\rangle =\underset{n}{\sum}c_{n}\left|a_{n}\right\rangle .
\end{equation}
 Obviously, $\left|\left\langle a_{n}|\phi^{\left(I\right)}\right\rangle \right|^{2}$
is the probability to find $a_{n}$ as the possible result of the
measurement. Considering that the measurement results for apparatus
$II$ are shown on a scale of values, we define an observable $\hat{B}^{\left(II\right)}$
that gives us the pointer position of the eingenvalue $b_{n}$ (referring
to the eigenstate $\left|b_{n}\right\rangle $). In this case, there
is a direct correlation between $a_{n}$ and $b_{n}$: the result
of the measurement of $\hat{B}^{\left(II\right)}$ over $II$ will
give $b_{n}$ only if the result of the measurement of $\hat{A}^{\left(I\right)}$
over $I$ gives $a_{n}$.

Next, let us define the initial state $\left|\phi^{\left(I\right)}\right\rangle $
of $I$ - unknown in the sense of its being a linear combination of
eigenstates of $\hat{A}^{\left(I\right)}$ (the goal of the measurement
is to determine the state) - and the initial state $\left|\phi^{\left(II\right)}\right\rangle $
of $II$, denoted, for simplicity, by \emph{$\left|b_{0}\right\rangle $}.
The initial state of the combined system $I+II$ will then be

\begin{equation}
\left|\phi^{\left(I+II\right)}\left(0\right)\right\rangle =\left|\phi^{\left(I\right)}\right\rangle \otimes\left|\phi^{\left(II\right)}\right\rangle ,\label{psizero}
\end{equation}
 where 
\begin{equation}
\begin{cases}
\left|\phi^{\left(I\right)}\right\rangle  & =\underset{n}{\sum}c_{n}\left|a_{n}\right\rangle \\
\left|\phi^{\left(II\right)}\right\rangle  & =\left|b_{0}\right\rangle 
\end{cases}\:.\label{estsep}
\end{equation}
 For simplicity, the $\otimes$ symbol will be omitted from now on.
The transmission of information from $I$ to $II$ will be made through
the \emph{unitary operation }- i.e., by the \emph{process 2} above
- associated to the $\hat{H}_{meas}^{\left(I+II\right)}$ Hamiltonian,
which acts over the global system $I+II$ . If the measurement extends
over a time interval $\tau$, the final global state will be 
\begin{equation}
\left|\phi^{\left(I+II\right)}\left(\tau\right)\right\rangle =e^{-\frac{i}{\hbar}\hat{H}_{meas}^{\left(I+II\right)}\tau}\left|\phi^{\left(I+II\right)}\left(0\right)\right\rangle .\label{psitau}
\end{equation}
 At this point, von Neumann defines the unitary operator

\begin{equation}
\hat{\Delta}\equiv e^{-\frac{i}{\hbar}\hat{H}_{meas}^{\left(I+II\right)}\tau}\label{delta1}
\end{equation}
 to simplify his calculations. The question, then, is how to find
a form for $\hat{\Delta}$ that fits the established measurement criteria.
This is a simple task and the result is

\begin{equation}
\hat{\Delta}\underset{m,n}{\sum}x_{mn}\left|a_{m}\right\rangle \left|b_{n}\right\rangle =\underset{m,n}{\sum}x_{mn}\left|a_{m}\right\rangle \left|b_{m+n}\right\rangle .\label{delta2}
\end{equation}

It can be shown that applying (\ref{delta1}) to (\ref{psizero})
gives

\[
\left|\phi^{\left(I+II\right)}\left(\tau\right)\right\rangle =\hat{\Delta}\left|\phi^{\left(I+II\right)}\left(0\right)\right\rangle =\hat{\Delta}\underset{n}{\sum}c_{n}\left|a_{n}\right\rangle \left|b_{0}\right\rangle =\underset{n}{\sum}c_{n}\left|a_{n}\right\rangle \left|b_{n}\right\rangle .
\]
 The acquisition of information of the state is made by the process
1 above, with the results having associated probabilities $\left|c_{n}\right|^{2}$
- Born's rule.

In his work, von Neumann did not explicitly consider the time during
which the measurement apparatus interacts with the system because
the introduction of $\hat{\Delta}$ eliminates the measurement time
$\tau$ from the calculations. A way to analyze this time period within
the conventional structure of quantum mechanics as a statistical theory
(and non causal in von Neumann's point of view \cite{key-4}) consists
of considering the interaction between the system and the measurement
apparatus evolving according to the Schrödinger equation or, more
generally, the Liouville-von Neumann equation. In such an approach,
we still have an equation for the density operator and, consequently,
the probabilistic character of quantum mechanics is still present:
the populations provide the associated probabilities for the possible
results. Here, the problem of the wave-function reduction will not
be analyzed and/or refuted - there are, indeed, interpretations of
quantum mechanics where the reduction phenomenon is treated as nonexistent
\cite{key-11,key-12,key-13,key-14}.

To clarify this treatment, it is convenient to use a framework that
resembles the one by Peres \cite{key-15}, who labeled the procedure
of acquiring information regarding the physical system as {}``intervention''.
That procedure is divided into two parts: the measurement, when the
apparatus interacts with the system and acquires the information,
and the reading (or output), when the results of the intervention
become known and the reduction of the wave function occurs. In the
last step, the probabilistic information is contained in the populations
of the density operator.

As mentioned previously, a more realistic description of the measurement
apparatus implies the distinction between its microscopic and macroscopic
degrees of freedom. To each possible eigenvalue measured, there is
one and only one macroscopic value. However, this variable does not
describe completely the state of the measurement apparatus, because
there are many microscopic states corresponding to the same value
of the macroscopic variable. As in the measurement the only important
variable is the macroscopic one, we must take the partial trace over
the microscopic variables. Under the Markov approximation for the
interaction between the system and the measurement apparatus, this
procedure leads to the \emph{Lindblad equation} \cite{key-16,key-17,key-18},
where only the system coordinates appear. This equation for the density
operator has a term for \emph{unitary evolutions} (the \emph{Liouvillian},
already known) and a second term, for the \emph{non-unitary evolutions},
named \emph{Lindbladian}. It is exactly the non-unitary term that
allows us to introduce the time interval during the interaction between
the system and the measurement apparatus \cite{key-19}. In the quantum-measurement-theory
context, its derivation was sugested in \cite{key-15} and made in
a simple manner in \cite{key-27}.

Until now, we have a situation where only two systems are considered:
the system on which the measurement is made (which is the object of
our interest), and the measurement apparatus. However, in real situations,
the perfect isolation of the system is not always possible to achieve
and, consequently, the environment perturbations change the final
density operator in some way. To treat this situation, we consider
that the system interacts with both the environment and the measurement
apparatus, the effects of the environment being introduced in the
Liouvillian of the Lindblad equation. In this way, with the partial
trace over the environmental degrees of freedom, we obtain an equation
in the Lindblad form, provided the environment is assumed as Markovian.
To calculate the more general trace over the environmental degrees
of freedom appearing in the Lindblad equation, we developed a formalism
\cite{key-20} based on the superoperator algebra and the Nakajima-Zwanzig
projectors \cite{key-21,key-22}.

It should be noted that alternative methods of describing continuous
measurements that do not recur to the Lindblad superoperator exist,
and include the introduction of an extra term to the Hamiltonian to
account for the interaction with the measurement apparatus \cite{QZE},
and the use of a stochastic master equation \cite{REVIEW}. This latter
description is not incompatible with the one provided by the Lindblad
equation \cite{key-19}, which, for this choice of Lindblad operators,
is also equivalent to the master equation for continuous position
measurements derived by Barchielli et al. \cite{BARCHIELLI} Our choice
of employing the Lindblad equation as the starting point of our model,
however, has proven satisfactory for the solution of the problem of
the noisy finite-time measurement, that is, a finite-time measurement
that occurs while the system suffers errors caused by an external
environment. The investigation of this problem is what constitutes
our original contribution to the field.

In previous papers \cite{key-20,key-23,key-24}, our formalism was
applied to a two-state system in contact with an environment via phase-damping
interaction, in the case of an Ohmic spectral density. Under some
restrictions - both the natural system frequency and the environmental
temperature set to zero - chosen to simplify the problem and allow
analytical solutions, we verified that \cite{key-23}, when the measurement
does not commute with the system-environment interaction, (i) the
more intense the system-environment interaction, the more marked is
the decrease of the population - i.e., the larger is the measurement
error - and (ii) the more intense the system-measurement apparatus
interaction, the less marked is the rate of change of the populations.
From (ii), we concluded that finite-time measurements can be more
efficient than instantaneous ones, when the measurement does not commute
with the interaction between the system and the environment, because
the nature of the system-apparatus interaction reduces the rate of
change of the populations. To estimate the \emph{efficiency of a measurement},
here, we adopt the criterion that an ideal measurement would consist
of the instantaneous acquisition of information from the system before
its state was altered by the environmental noise. If the populations
of the system have been significantly altered by the time the measurement
reaches its completion, the probabilities of encountering the apparatus's
pointer in either state will be visibly different from the probabilities
associated with the initial state of the system measured, thus rendering
the process inefficient. Next, analyzing the behavior of the coherences,
which decrease with time, we established a criterion for the measurement
duration \cite{key-24}. 

In the present work, we remove the restrictions and treat the interaction
between the system and the environment as either the phase-damping
or the amplitude-damping type, aiming at verifying, in a more general
form, the influence of the environment on the final reduced density
operator. We verify that, in general, the conclusions obtained in
the previous works for state protection and system-environmental coupling
apply to both interactions (phase and amplitude damping), even in
the case of a finite environmental temperature. The addition of system
frequency induces an oscillatory behavior on the populations. For
the coherences, however, we observe that the introduction of the system
frequency causes their modulus to approach, after a certain time interval,
a constant asymptotic value, that may or may not be zero. Although
we cannot, in this case, obtain a simple expression for the duration
of the measurement, we still use the coherences to establish the time
at which the system-environment interaction can be interrupted and
the reading of the state accomplished - in this case, the time when
the coherences assume the constant asymptotic value. Obviously, in
more general cases, when the behavior of the system is not simple
(the coherences no longer show a monotonic decrease), its physical
properties must be considered to the establishment of the measurement
time. Here, we do not change the spectral density and do not increase
the number of states of the system under measurement - these extensions
are left open and can be more safely developed under the light of
the knowledge of the results presented here.

The possibility of application of this kind of measurement to improve
the quality of the results depends, however, on the practical existence
of measurements that can be performed during a finite period of time,
in accordance with the model. There is, however, a recent application
of our work - see ref. \cite{key-28}- in the context of Measurement-based
Direct quantum Feedback Control (MDFC) \cite{key-29,key-30}.

This article is structured as follows: in Sec. II, we recapitulate
our description of noisy finite-time measurements, providing the basis
for the solutions obtained in Sec. III and Sec. IV, first for the
phase-damping case, and then for the amplitude-damping one. Analyses
of the solutions, together with graphs, are presented in Sec. V, conclusions
and perspectives are given in Sec. VI. An appendix explains how this
model of finite-time measurement could be experimentally tested against
the more conservative approach that presupposes instantaneous measurements.

To present the results in a more logical and complete fashion, some
of the contents of this article have been recapitulated from our previous
works. Sec. II briefly explains the method that can be found, with
greater details, on \cite{key-20} and solutions for the Lindblad
equation for \emph{x}-component measurement without environment found
on \cite{key-23}; Sec. III shows the z-component solution and the
simplified ($T=0$ and $\omega_{0}=0$) \emph{x}-component solutions
found on \cite{key-23} for the phase-damping interaction; finally,
on Sec. V we show the formula for upper limit for the measurement
duration found on \cite{key-24}. All the other results, however,
are new.

\section{The Model of Finite-Time Measurement}

In this section, we overview the model of noisy finite-time measurement
that was developed and employed in our previous works \cite{key-23,key-24}.
This model assumes that the system interacts with the measurement
apparatus through a Markovian interaction. However, the finite nature
of this description allows for other processes to occur before the
measurement reaches its end.

In the usual treatment, when measurements are performed over the\emph{
principal system} $S$, it is assumed that $S$ is perfectly isolated
from the environment, interacting non-unitarily with the measurement
apparatus. However, as we are considering the measurement as a finite-time
process that may take a certain period to reach its completion, here
we make the assumption that the system is not isolated, but also interacting
with an \emph{environment} $B$. To deal with this situation, we describe
the additional interaction between the system and the environment
as a unitary evolution, with the Hamiltonian

\begin{equation}
\hat{H}=\hat{H}_{B}+\hat{H}_{SB}+\hat{H}_{S},
\end{equation}
 where $\hat{H}_{S}$ and $\hat{H}_{B}$ are the system and the environment
Hamiltonians, and $\hat{H}_{SB}$ is the interaction between them.

Therefore, to model the noisy measurement we describe both the measured
object and the environment by a total density operator $\hat{\rho}_{SB}$
obeying the \emph{Lindblad equation} \cite{key-16,key-17,key-18}:

\begin{equation}
\frac{\mathrm{d}}{\mathrm{d}t}\hat{\rho}_{SB}=-\frac{i}{\hbar}\left[\hat{H},\hat{\rho}_{SB}\right]+\underset{j}{\sum}\left({\color{red}{\normalcolor {\normalcolor \hat{L}}_{{\normalcolor j}}^{{\normalcolor \left(S\right)}}}}\hat{\rho}_{SB}{\color{red}{\normalcolor \hat{L}}_{{\normalcolor j}}^{{\normalcolor \left(S\right)\dagger}}}-\frac{1}{2}\left\{ {\color{red}{\normalcolor \hat{L}}_{{\normalcolor j}}^{{\normalcolor \left(S\right)\dagger}}{\normalcolor \hat{L}}_{{\normalcolor j}}^{{\normalcolor \left(S\right)}}},\hat{\rho}_{SB}\right\} \right).
\end{equation}
 Here, the first term of the right-hand side, $-\frac{i}{\hbar}\left[\hat{H},\hat{\rho}_{SB}\right]$,
is the \emph{Liouvillian }and corresponds to \emph{unitary evolutions,
}while the second term, $\underset{j}{\sum}\left(\hat{L}_{j}^{\left(S\right)}\hat{\rho}_{SB}\hat{L}_{j}^{\left(S\right)\dagger}-\frac{1}{2}\left\{ \hat{L}_{j}^{\left(S\right)\dagger}\hat{L}_{j}^{\left(S\right)},\hat{\rho}_{SB}\right\} \right)$,
is the \emph{Lindbladian} and corresponds to the \emph{non-unitary
evolutions}. The $\hat{L}_{j}^{\left(S\right)}$ operators are the
\emph{Lindblads, }which act only on the system $S$,\emph{ }and can
describe \emph{measurement processes} \cite{key-15,key-19,key-23,key-24}
if they are \emph{Hermitian}, or \emph{dissipation processes} \cite{key-17},
if \emph{non-Hermitian}. When the Lindblads are all zero, we recover
the Liouville-von Neumann equation, i. e., there is no measurement
occurring.

Here, we are only interested in the reduced density operator $\hat{\rho}_{S}$
of $S$. In our formalism \cite{key-20}, this is done with the help
of the Nakajima-Zwanzig projectors \cite{key-21,key-22} that, for
the general operator $\hat{X}\left(t\right)$ and the initial instant
$t_{0}$, act as

\begin{equation}
\hat{\hat{P}}\hat{X}\left(t\right)\equiv\hat{\rho}_{B}\left(t_{0}\right)\otimes\mathrm{Tr}_{B}\left\{ \hat{X}\left(t\right)\right\} ,
\end{equation}

\begin{equation}
\hat{\hat{Q}}\equiv1-\hat{\hat{P}},
\end{equation}
 where we are using superoperator algebra. In the present case, we
have defined superoperators referring to the contributions by the
system $S$

\begin{equation}
\hat{\hat{S}}\hat{X}=-\frac{i}{\hbar}\left[\hat{H}_{S},\hat{X}\right]+\underset{j}{\sum}\left(\hat{L}_{j}^{\left(S\right)}\hat{\rho}_{SB}\hat{L}_{j}^{\left(S\right)\dagger}-\frac{1}{2}\left\{ \hat{L}_{j}^{\left(S\right)\dagger}\hat{L}_{j}^{\left(S\right)},\hat{\rho}_{SB}\right\} \right),\label{dfS}
\end{equation}
 the environment $B:$

\begin{equation}
\hat{\hat{B}}\hat{X}=-\frac{i}{\hbar}\left[\hat{H}_{B},\hat{X}\right],\label{defB}
\end{equation}
 and to the interaction between them:

\begin{equation}
\hat{\hat{F}}\hat{X}=-\frac{i}{\hbar}\left[\hat{H}_{SB},\hat{X}\right].\label{defF}
\end{equation}
 With these definitions, we obtain the integral equation

\begin{equation}
\frac{\mathrm{d}}{\mathrm{d}t}\left[\hat{\hat{P}}\hat{\alpha}\left(t\right)\right]=\int_{0}^{t}\mathrm{d}t^{\prime}\,\left[\hat{\hat{P}}\hat{\hat{G}}\left(t\right)\hat{\hat{G}}\left(t^{\prime}\right)\hat{\hat{P}}\hat{\alpha}\left(t\right)\right],\label{eqprincipal}
\end{equation}
 where we defined the new {}``interaction-picture'' density operator

\begin{equation}
\hat{\alpha}\left(t\right)\equiv\exp\left(-\hat{\hat{S}}t-\hat{\hat{B}}t\right)\hat{\rho}_{SB}\left(t\right)
\end{equation}
 and the superoperator

\begin{equation}
\hat{\hat{G}}\left(t\right)\equiv\exp\left(-\hat{\hat{S}}t-\hat{\hat{B}}t\right)\hat{\hat{F}}\exp\left(\hat{\hat{S}}t+\hat{\hat{B}}t\right).\label{defG}
\end{equation}

One of the advantages of (\ref{eqprincipal}), as verified in \cite{key-20,key-23},
is to make the contributions referring to $S$ and $B$ easily factorized.
Both here and in our previous applications \cite{key-20,key-23,key-24},
we have considered $S$ a\emph{ 2-state system} and $B$ the \emph{environment},
with corresponding Hamiltonians

\begin{equation}
\hat{H}_{S}=\hbar\omega_{0}\hat{\sigma}_{z},
\end{equation}

\begin{equation}
\hat{H}_{B}=\hbar\underset{k}{\sum}\omega_{k}\hat{b}_{k}^{\dagger}\hat{b}_{k},
\end{equation}
 and the Lindbladian represented by a single Lindblad operator, with
the form

\begin{equation}
\hat{L}^{\left(S\right)}=\lambda\hat{\sigma}_{j},\: j=x,z,\:\lambda\in\mathbb{R},
\end{equation}
 where $\hat{b}_{k}$ and $\hat{b}_{k}^{\dagger}$ are the destruction
and creation operators of the environment $B$, $\omega_{k}$ are
the frequencies associated to each environmental mode, $\omega_{0}$
is the characteristic frequency of the system $S,$ $\lambda$ is
a constant associated with the intensity of the interaction between
the system $S$ and the measurement apparatus, and $\hat{\sigma}_{j}$
are the Pauli matrices:

\begin{equation}
\hat{\sigma}_{z}=\left(\begin{array}{cc}
1 & 0\\
0 & -1
\end{array}\right),\;\hat{\sigma}_{x}=\left(\begin{array}{cc}
0 & 1\\
1 & 0
\end{array}\right),\;\hat{\sigma}_{y}=\left(\begin{array}{cc}
0 & -i\\
i & 0
\end{array}\right).
\end{equation}
 Furthermore, we have adopted an Ohmic spectral density \cite{key-17}
given by

\begin{equation}
J\left(\omega\right)=\eta\omega e^{-\omega/\omega_{c}},\;\eta\geqslant0,\;\omega_{c}>0.\label{JOhm}
\end{equation}

When we analyse the final density operator $\hat{\rho}_{S}$, we have
to write it in terms of the basis of eigenstates of the measurement
apparatus responsible for the Lindblad $\hat{L}^{\left(S\right)}$.
Since we start our calculations using the basis corresponding to the
$\hat{\sigma}_{z}$ eigenstates, $\left\{ \left|0\right\rangle ,\left|1\right\rangle \right\} $,
the $\hat{L}^{\left(S\right)}=\lambda\hat{\sigma}_{z}$ case will
not require additional attention. However, in the $\hat{L}^{\left(S\right)}=\lambda\hat{\sigma}_{x}$
case it will be necessary the change to the basis of the $\hat{\sigma}_{x}$
eigenstates, $\left\{ \left|+\right\rangle ,\left|-\right\rangle \right\} $,
namely,

\begin{equation}
\left|\pm\right\rangle =\frac{\left|0\right\rangle \pm\left|1\right\rangle }{\sqrt{2}},
\end{equation}
 using the basis transformation matrix 
\begin{equation}
\hat{M}=\frac{1}{\sqrt{2}}\left(\begin{array}{cc}
1 & 1\\
1 & -1
\end{array}\right)=\hat{M}^{-1}.\label{M}
\end{equation}
 We observe that the basis in which the reduced density operator is
written will be specified by the superscript $\left(j\right)$, as
in $\hat{\rho}_{S}^{\left(j\right)}\left(t\right)$.

To solve the Eq. (\ref{eqprincipal}), it is often useful to adopt
the notation

\begin{equation}
\hat{R}\left(t\right)\equiv e^{-\hat{\hat{S}}t}\hat{\rho}_{S}\left(t\right),\label{defR}
\end{equation}
 so that

\begin{equation}
\hat{\hat{P}}\hat{\alpha}\left(t\right)=\hat{R}\left(t\right)\hat{\rho}_{B}.
\end{equation}
 In the end, the density operator for the system can be recovered
simply by applying the inverse transformation

\begin{equation}
\hat{\rho}_{S}\left(t\right)=e^{\hat{\hat{S}}t}\hat{R}\left(t\right).
\end{equation}

With the definition of $\hat{R}\left(t\right)$, the integrand of
Eq. (\ref{eqprincipal}) can be expanded. We illustrate how this is
done for the specific case of a phase-damping interaction,

\begin{equation}
\hat{H}_{SB}=\hat{\sigma}_{z}{\color{red}{\normalcolor \sum_{k}\hbar\left(g_{k}b_{k}^{\dagger}+g_{k}^{*}b_{k}\right)}},
\end{equation}
 where the $g_{k}$ are the associated coefficients of the spectral
density. In this case, $\hat{\hat{G}}\left(t\right)\hat{\hat{G}}\left(t^{\prime}\right)\hat{R}\left(t\right)\hat{\rho}_{B}$
from Eq. (\ref{defG}) yields

\begin{eqnarray}
\hat{\hat{G}}\left(t\right)\hat{\hat{G}}\left(t^{\prime}\right)\hat{R}\left(t\right)\hat{\rho}_{B} & = & ie^{-\hat{\hat{S}}t}e^{-\hat{\hat{B}}t}\hat{\hat{F}}\left\{ e^{\hat{\hat{S}}\left(t-t'\right)}\left[\left(e^{\hat{\hat{S}}t'}\hat{R}\left(t\right)\right)\hat{\sigma}_{z}\right]\right\} \nonumber \\
 & \times & \left\{ e^{\hat{\hat{B}}\left(t-t'\right)}\left[\left(e^{\hat{\hat{B}}t'}\hat{\rho}_{B}\right){\color{red}{\normalcolor \sum_{k}\hbar\left(g_{k}b_{k}^{\dagger}+g_{k}^{*}b_{k}\right)}}\right]\right\} \nonumber \\
 &  & -ie^{-\hat{\hat{S}}t}e^{-\hat{\hat{B}}t}\hat{\hat{F}}\left\{ e^{\hat{\hat{S}}\left(t-t'\right)}\left[\hat{\sigma}_{z}\left(e^{\hat{\hat{S}}t'}\hat{R}\left(t\right)\right)\right]\right\} \nonumber \\
 & \times & \left\{ e^{\hat{\hat{B}}\left(t-t'\right)}\left[{\color{red}{\normalcolor \sum_{k}\hbar\left(g_{k}b_{k}^{\dagger}+g_{k}^{*}b_{k}\right)}}\left(e^{\hat{\hat{B}}t'}\hat{\rho}_{B}\right)\right]\right\} .\label{GGro}
\end{eqnarray}

The following procedure is to replace the superoperators in Eq. (\ref{GGro})
with the definitions from (\ref{dfS}), (\ref{defB}) and (\ref{defF}).
To do this, it is necessary to clarify the effect of the temporal
exponentials of the superoperators, $e^{\hat{\hat{S}}t}$ and $e^{\hat{\hat{B}}t}$.
For example, defining $\hat{X}'$ as the result of applying $e^{\hat{\hat{B}}t}$
on an operator $\hat{X}$:

\begin{equation}
\hat{X}'=e^{\hat{\hat{B}}t}\hat{X}
\end{equation}
 and taking the time derivative on both sides, we have

\begin{equation}
\frac{\mathrm{d}}{\mathrm{d}t}\hat{X}'=\hat{\hat{B}}e^{\hat{\hat{B}}t}\hat{X}=\hat{\hat{B}}\hat{X}'\label{dX}
\end{equation}
 but, by the definition (\ref{defB}), Eq. (\ref{dX}) will be

\begin{equation}
\frac{\mathrm{d}}{\mathrm{d}t}\hat{X}'=-\frac{i}{\hbar}\left[\hat{H}_{B},\hat{X}'\right].
\end{equation}
 This is the Liouville-von Neumann equation, whose solution, for a
time-independent Hamiltonian $\hat{H}_{B}$, will be, remembering
that $\hat{X}'\left(0\right)=\hat{X}$,

\begin{equation}
\hat{X}'=e^{\hat{\hat{B}}t}\hat{X}=e^{-i\frac{\hat{H}_{B}}{\hbar}t}\hat{X}e^{i\frac{\hat{H}_{B}}{\hbar}t}.
\end{equation}

Equivalently, the action of $e^{\hat{\hat{S}}t}$ over the density
operator elements $\rho_{ij}$ ($i,j=1,2$) will be determined by
Eq. (\ref{dfS}), being the solution of the Lindblad superoperator
with no environment. For $\hat{L}^{\left(S\right)}=\lambda\hat{\sigma}_{z}$,
the solution will be given by \cite{key-19}

\begin{equation}
\rho_{11}^{\left(z\right)}\left(t\right)=\rho_{11}^{\left(z\right)}\left(0\right),
\end{equation}

\begin{equation}
\rho_{12}^{\left(z\right)}\left(t\right)=\rho_{12}^{\left(z\right)}\left(0\right)e^{-2\lambda^{2}t}e^{-i2\omega_{0}t}.
\end{equation}

However, for $\hat{L}^{\left(S\right)}=\lambda\hat{\sigma}_{x}$,
the solution is more complicated and will be given, on the $\hat{\sigma}_{z}$
eigenstates basis, by \cite{key-23}

\begin{equation}
\begin{cases}
\rho_{11}^{\left(z\right)}\left(t\right) & =\frac{1}{2}+\frac{2\rho_{11}^{\left(z\right)}\left(0\right)-1}{2}e^{-2\lambda^{2}t},\\
\\
\rho_{12}^{\left(z\right)}\left(t\right) & =e^{-\lambda^{2}t}\left\{ \rho_{12}^{\left(z\right)}\left(0\right)\cosh\left(\sqrt{\lambda^{4}-4\omega_{0}^{2}}t\right)\right.\\
 & -\rho_{12}^{\left(z\right)}\left(0\right)\frac{i2\omega_{0}}{\sqrt{\lambda^{4}-4\omega_{0}^{2}}}\sinh\left(\sqrt{\lambda^{4}-4\omega_{0}^{2}}t\right)\\
 & +\left.\frac{\lambda^{2}}{\sqrt{\lambda^{4}-4\omega_{0}^{2}}}\rho_{12}^{\left(z\right)*}\left(0\right)\sinh\left(\sqrt{\lambda^{4}-4\omega_{0}^{2}}t\right)\right\} .
\end{cases}\label{solfx}
\end{equation}
 Finally, it is necessary to change the basis to the eigenstates of
the measurement apparatus, furnishing: 
\begin{equation}
\begin{cases}
\rho_{11}^{\left(x\right)}\left(t\right)= & \frac{1}{2}+\mathrm{Re}\left\{ \rho_{12}^{\left(z\right)}\left(t\right)\right\} ,\\
\rho_{12}^{\left(x\right)}\left(t\right)= & -\frac{1}{2}+\rho_{11}^{\left(z\right)}\left(t\right)-i\mathrm{Im}\left\{ \rho_{12}^{\left(z\right)}\left(t\right)\right\} .
\end{cases}
\end{equation}

In possession of these solutions, it is possible to solve the master
equation under certain circumstances, which is done in the next two
sections.

In our previous works \cite{key-20,key-23,key-24}, to obtain analytical
solutions, we had considered only the system-environment interaction
in the phase-damping format and, in the case when the Lindblad does
not commute with that interaction, we had considered the particular
situation when the frequency $\omega_{0}$ and the environment temperature
$T$ are both zero. In the following sections, we will present numerical
implementations without these restrictions and with the use of both
kinds of interaction. For the environment, we adopt the initial thermal
state:

\begin{equation}
\hat{\rho}_{B}=\frac{1}{Z_{B}}\underset{p}{\prod}e^{-\hbar\beta\omega_{p}\hat{b}_{p}^{\dagger}\hat{b}_{p}},Z_{B}=\underset{l}{\prod}\frac{1}{1-e^{-\hbar\beta\omega_{l}}},
\end{equation}
 with $\beta=\frac{1}{k_{B}T}$, where $k_{B}$ is the \emph{Boltzmann
constant} and $T$ is the \emph{environmental temperature}.

\section{Phase-damping interaction}

The case of phase-damping interaction was solved in two different
manners: first, using the master equation given in the section above,
valid for weak system-environment interaction. This is presented in
the first subsection below. Afterward, we explain how the algorithm
to numerically solve the Lindblad equation was written.

\subsection{Solutions of the master equation}

For the case of the phase-damping interaction,

\begin{equation}
\hat{H}_{SB}=\hbar\hat{\sigma}_{z}\underset{k}{\sum}\left(g_{k}\hat{b}_{k}^{\dagger}+g_{k}^{*}\hat{b}_{k}\right)
\end{equation}
 the treatment of the environment degrees of freedom was shown on
\cite{key-23}, furnishing

\begin{eqnarray}
\hat{\hat{P}}\hat{\hat{G}}\left(t\right)\hat{\hat{G}}\left(t'\right)\hat{\hat{P}}\hat{\alpha}\left(t\right) & = & \eta\int_{0}^{\infty}\mathrm{d}\omega\omega e^{-\frac{\omega}{\omega_{c}}}\left\{ \coth\left(\frac{\hbar\beta\omega}{2}\right)\cos\left[\omega\left(t-t'\right)\right]+i\sin\left[\omega\left(t-t'\right)\right]\right\} \nonumber \\
 &  & \qquad\times e^{-\hat{\hat{S}}t}\left[\hat{\sigma}_{z},\left\{ e^{\hat{\hat{S}}\left(t-t'\right)}\left[\left(e^{\hat{\hat{S}}t'}\hat{R}\left(t\right)\right)\hat{\sigma}_{z}\right]\right\} \right]\otimes\hat{\rho}_{B}\nonumber \\
 &  & +\eta\int_{0}^{\infty}\mathrm{d}\omega\omega e^{-\frac{\omega}{\omega_{c}}}\left\{ \coth\left(\frac{\hbar\beta\omega}{2}\right)\cos\left[\omega\left(t-t'\right)\right]-i\sin\left[\omega\left(t-t'\right)\right]\right\} \nonumber \\
 &  & \qquad\times e^{-\hat{\hat{S}}t}\left[\left\{ e^{\hat{\hat{S}}\left(t-t'\right)}\left[\hat{\sigma}_{z}\left(e^{\hat{\hat{S}}t'}\hat{R}\left(t\right)\right)\right]\right\} ,\hat{\sigma}_{z}\right]\otimes\hat{\rho}_{B}\label{eq:ambcont}
\end{eqnarray}
 where we took the limit to the continuum by defining the spectral
density function 
\begin{equation}
J\left(\omega\right)=\underset{l}{\sum}\left|g_{l}\right|^{2}\delta\left(\omega-\omega_{l}\right),\label{eq:Spectraldensity}
\end{equation}
 and employed Ohmic spectral density (\ref{JOhm}).

\subsubsection{The z-component measurement}

The measurement of the $z$ component, i.e., $\hat{L}^{\left(S\right)}=\lambda\hat{\sigma}_{z}$
will furnish the (relatively) simple solutions \cite{key-23}:

\begin{equation}
\begin{cases}
\rho_{11}\left(t\right) & =\rho_{11}\left(0\right),\\
\rho_{12}\left(t\right) & =\rho_{12}\left(0\right)\left[\frac{\Gamma\left(\frac{1}{\omega_{c}\beta\hbar}+i\frac{t}{\beta\hbar}\right)\Gamma\left(\frac{1}{\omega_{c}\beta\hbar}-i\frac{t}{\beta\hbar}\right)}{\Gamma^{2}\left(\frac{1}{\omega_{c}\beta\hbar}\right)}\frac{\Gamma\left(\frac{1}{\omega_{c}\beta\hbar}+1+i\frac{t}{\beta\hbar}\right)\Gamma\left(\frac{1}{\omega_{c}\beta\hbar}+1-i\frac{t}{\beta\hbar}\right)}{\Gamma^{2}\left(\frac{1}{\omega_{c}\beta\hbar}+1\right)}\right]^{2\eta}e^{-2\lambda^{2}t}e^{i2\omega_{0}t},
\end{cases}\label{eq:solzZZ}
\end{equation}

\subsubsection{The x-component measurement}

In the $\hat{\sigma}_{z}$ eigenbasis, as obtained on \cite{key-23},
the populations for $\hat{L}^{\left(S\right)}=\lambda\hat{\sigma}_{x}$
will be 
\begin{equation}
\begin{cases}
\rho_{11}^{\left(z\right)}\left(t\right)= & \frac{2\rho_{11}^{\left(z\right)}\left(0\right)-1}{2}e^{-2\lambda^{2}t}+\frac{1}{2}\\
\rho_{12}^{\left(z\right)}\left(t\right)= & \frac{e^{-\lambda^{2}t}}{\Omega}\left\{ \left[\Omega\cosh\left(\Omega t\right)-i2\omega_{0}\sinh\left(\Omega t\right)\right]r_{12}\left(t\right)+\lambda^{2}\sinh\left(\Omega t\right)r_{21}\left(t\right)\right\} 
\end{cases}\label{solzPD}
\end{equation}
 
\begin{equation}
\Omega=\sqrt{\lambda^{4}-4\omega_{0}^{2}}
\end{equation}
 where $r_{12}\left(t\right)$ and $r{}_{21}\left(t\right)$ are both
solutions of the system

\begin{equation}
\begin{cases}
\frac{\mathrm{d}}{\mathrm{d}t}r_{12}\left(t\right)= & -4\frac{\eta}{\Omega^{3}}\int_{0}^{t}\mathrm{d}t'\int_{0}^{\infty}\mathrm{d}\omega\,\omega e^{-\frac{\omega}{\omega_{c}}}\cos\left[\omega\left(t-t'\right)\right]\coth\left(\frac{\beta\hbar\omega}{2}\right)\\
 & \times\left[Q_{1}\left(t,t'\right)r_{12}\left(t\right)+Q_{2}\left(t,t'\right)r_{21}\left(t\right)\right]\\
\frac{\mathrm{d}}{\mathrm{d}t}r_{12}\left(t\right)= & -4\frac{\eta}{\Omega^{3}}\int_{0}^{t}\mathrm{d}t'\int_{0}^{\infty}\mathrm{d}\omega\,\omega e^{-\frac{\omega}{\omega_{c}}}\cos\left[\omega\left(t-t'\right)\right]\coth\left(\frac{\beta\hbar\omega}{2}\right)\\
 & \times\left[Q_{2}^{*}\left(t,t'\right)r_{12}\left(t\right)+Q_{1}^{*}\left(t,t'\right)r_{21}\left(t\right)\right]
\end{cases}\label{sistphasex}
\end{equation}
 with

\begin{equation}
\begin{cases}
Q_{1}\left(t,t'\right)\equiv & K_{1}\left(t\right)\left[K_{1}^{*}\left(t-t'\right)K_{1}^{*}\left(t'\right)-K_{2}\left(t-t'\right)K_{2}\left(t'\right)\right]\\
 & +K_{2}\left(t\right)\left[K_{2}\left(t-t'\right)K_{1}^{*}\left(t'\right)-K_{1}\left(t-t'\right)K_{2}\left(t'\right)\right]\\
Q_{2}\left(t,t'\right)\equiv & K_{1}\left(t\right)\left[K_{1}^{*}\left(t-t'\right)K_{2}\left(t'\right)-K_{2}\left(t-t'\right)K_{1}\left(t'\right)\right]\\
 & +K_{2}\left(t\right)\left[K_{2}\left(t-t'\right)K_{2}\left(t'\right)-K_{1}\left(t-t'\right)K_{1}\left(t'\right)\right]
\end{cases}
\end{equation}
 
\begin{equation}
\begin{cases}
K_{1}\left(t\right)\equiv & \Omega\cosh\left(\Omega t\right)+i2\omega_{0}\sinh\left(\Omega t\right)\\
K_{2}\left(t\right)\equiv & \lambda^{2}\sinh\left(\Omega t\right)
\end{cases}
\end{equation}

\subsection{Numerical solution of the Lindblad equation}

To compare the semi-analytical results of the hybrid master equation
with the numerical simulations of the Lindblad equation, we employed
the same superoperator-splitting method described in a previous article
\cite{key-23}. A few adaptations were necessary for the inclusion
of non-vanishing $\omega_{0}$and temperature $T$. We shall approach
the introduction of these two parameters separately in the subsections
that follow.

\subsubsection{Introduction of $\omega_{0}$}

As stated in Eq. (41) of \cite{key-23}, the final state of the system
subject to phase noise and a perpendicular measurement can be calculated
using the superoperator-splitting method through the algorithm:

\begin{equation}
\left(\begin{array}{c}
\rho_{12}^{\left(z\right)}\left(t\right)\\
\rho_{21}^{\left(z\right)}\left(t\right)
\end{array}\right)=e^{-\lambda^{2}t}\sum_{q_{1}\in\left\{ -1,1\right\} }\ldots\sum_{q_{N}\in\left\{ -1,1\right\} }\prod_{n=1}^{N}\left[A_{q_{n}}\left(\Delta t\right)\right]\mathrm{Tr}_{B}\left\{ \prod_{n=1}^{N}\left[\hat{\hat{K}}_{q_{n}}\left(\Delta t\right)\right]\hat{\rho}_{B}\right\} \left(\begin{array}{c}
\rho_{12}^{\left(z\right)}\left(0\right)\\
\rho_{21}^{\left(z\right)}\left(0\right)
\end{array}\right),\label{eq:algorithm}
\end{equation}
 where $t=N\Delta t$ and the superoperators $\hat{\hat{K}}_{q}\left(\Delta t\right)$
satisfy, when the noise is along the $\hat{\mathbf{z}}$ direction,

\[
\hat{\hat{K}}_{q}\left(\Delta t\right)\hat{X}\equiv e^{-i\sum_{k}\omega_{k}\left(\hat{b}_{k}+qg_{k}/\omega_{k}\right)^{\dagger}\left(\hat{b}_{k}+qg_{k}/\omega_{k}\right)\Delta t}\hat{X}e^{i\sum_{k}\omega_{k}\left(\hat{b}_{k}-qg_{k}/\omega_{k}\right)^{\dagger}\left(\hat{b}_{k}-qg_{k}/\omega_{k}\right)\Delta t},
\]
 and the $A_{q_{n}}\left(\Delta t\right)$ are $2\times2$ matrices
that depend on the solution of the noiseless Lindblad equation, i.
e.

\[
\frac{\mathrm{d}}{\mathrm{d}t}\rho\left(t\right)=-i\omega_{0}\left[\sigma_{z},\rho\left(t\right)\right]+\lambda^{2}\left[\sigma_{x}\rho\left(t\right)\sigma_{x}-\rho\left(t\right)\right].
\]

In our previous article we chose $\omega_{0}=0$, thus rendering the
solution of the equation in the eigenbasis of $\sigma_{z}$ simply

\[
\left(\begin{array}{c}
\rho_{12}^{\left(z\right)}\left(\Delta t\right)\\
\rho_{21}^{\left(z\right)}\left(\Delta t\right)
\end{array}\right)=e^{-\lambda^{2}\Delta t}A_{+}^{\left(0\right)}\left(\Delta t\right)\left(\begin{array}{c}
\rho_{12}^{\left(z\right)}\left(0\right)\\
\rho_{21}^{\left(z\right)}\left(0\right)
\end{array}\right)+e^{-\lambda^{2}\Delta t}A_{-}^{\left(0\right)}\left(\Delta t\right)\left(\begin{array}{c}
\rho_{12}^{\left(z\right)}\left(0\right)\\
\rho_{21}^{\left(z\right)}\left(0\right)
\end{array}\right),
\]
 where the matrices $A_{\pm}^{\left(0\right)}\left(\Delta t\right)$
are

\[
A_{+}^{\left(0\right)}\left(\Delta t\right)=\left(\begin{array}{cc}
\cosh\left(\lambda^{2}\Delta t\right) & \sinh\left(\lambda^{2}\Delta t\right)\\
0 & 0
\end{array}\right),\quad A_{-}^{\left(0\right)}\left(\Delta t\right)=\left(\begin{array}{cc}
0 & 0\\
\sinh\left(\lambda^{2}\Delta t\right) & \cosh\left(\lambda^{2}\Delta t\right)
\end{array}\right).
\]

In this new situation where $\omega_{0}\ne0$, the solution for the
coherences will require these matrices instead:

\begin{eqnarray*}
A_{+}\left(\Delta t\right) & = & \frac{1}{\Omega}\left(\begin{array}{cc}
\Omega\cosh\left(\Omega\Delta t\right)-2i\omega_{0}\sinh\left(\Omega\Delta t\right) & \lambda^{2}\sinh\left(\Omega\Delta t\right)\\
0 & 0
\end{array}\right),\\
A_{-}\left(\Delta t\right) & = & \frac{1}{\Omega}\left(\begin{array}{cc}
0 & 0\\
\lambda^{2}\sinh\left(\Omega\Delta t\right) & \Omega\cosh\left(\Omega\Delta t\right)+2i\omega_{0}\sinh\left(\Omega\Delta t\right)
\end{array}\right),
\end{eqnarray*}
 where $\Omega\equiv\sqrt{\lambda^{4}-4\omega_{0}^{2}}$. It is easy
to verify that, when $\omega_{0}=0$, $\Omega=\lambda^{2}$, the original
matrices $A_{\pm}^{\left(0\right)}\left(\Delta t\right)$ are recovered.

The full expression of the algorithm in Eq. (\ref{eq:algorithm})
requires a product of $N$ such matrices for each term of the $N$
summations. We can reduce the computational overhead by simplifying
analytically the results before writing the program. Following the
method employed in the previous article, we first define the functions

\begin{eqnarray*}
c_{+}\left(\Delta t\right) & \equiv & \cosh\left(\Omega\Delta t\right)-2i\omega_{0}\frac{1}{\Omega}\sinh\left(\Omega\Delta t\right),\\
c_{-}\left(\Delta t\right) & \equiv & \lambda^{2}\frac{1}{\Omega}\sinh\left(\Omega\Delta t\right),
\end{eqnarray*}
 so that the products of these matrices can be written as

\[
A_{q_{n}}\left(\Delta t\right)A_{q_{n+1}}\left(\Delta t\right)=c_{q_{n}q_{n+1}}^{q_{n}}\left(\Delta t\right)\left(\sigma_{x}\right)^{\left(1-q_{n}q_{n+1}\right)/2}A_{q_{n+1}}\left(\Delta t\right),
\]
 where a superior negative index in the $c\left(\Delta t\right)$
represents taking its complex conjugate ($c^{+}=c$,$c^{-}=c^{*}$).

Proceeding iteratively with all the $N$ products, we find

\[
\prod_{n=1}^{N}\left[A_{q_{n}}\left(\Delta t\right)\right]=\prod_{n=1}^{N-1}c_{q_{n}q_{n+1}}^{q_{n}}\left(\Delta t\right)\left(\sigma_{x}\right)^{\sum_{n=1}^{N-1}\left(1-q_{n}q_{n+1}\right)/2}A_{q_{N}}\left(\Delta t\right).
\]
 Noting additionally that $\left(\sigma_{x}\right)^{\left(1-q_{1}q_{2}\right)/2}\left(\sigma_{x}\right)^{\left(1-q_{2}q_{3}\right)/2}=\left(\sigma_{x}\right)^{1-q_{2}\left(q_{1}+q_{3}\right)/2}=\left(\sigma_{x}\right)^{\left(1-q_{1}q_{3}\right)/2}$,
the product simplifies even further. Back to Eq. (\ref{eq:algorithm}),

\begin{eqnarray*}
\left(\begin{array}{c}
\rho_{12}\left(N\Delta t\right)\\
\rho_{21}\left(N\Delta t\right)
\end{array}\right) & = & e^{-\lambda^{2}\left(N\Delta t\right)}\sum_{q_{1}\in\left\{ -1,1\right\} }\ldots\sum_{q_{N}\in\left\{ -1,1\right\} }\left[\prod_{n=1}^{N-1}c_{q_{n}q_{n+1}}^{q_{n}}\left(\Delta t\right)\right]\left(\sigma_{x}\right)^{\left(1-q_{1}q_{N}\right)/2}\\
 &  & A_{q_{N}}\left(\Delta t\right)\prod_{m,n=1}^{N}\left\{ 1+\frac{2\left(\omega_{c}\Delta t\right)^{-2}+\left(1-2\left|m-n\right|^{2}\right)}{\left[\left(\omega_{c}\Delta t\right)^{-2}+\left|m-n\right|^{2}\right]^{2}}\right\} ^{-\eta q_{m}q_{n}}\left(\begin{array}{c}
\rho_{12}\left(0\right)\\
\rho_{21}\left(0\right)
\end{array}\right),
\end{eqnarray*}
 where we have already replaced the trace for its the result at zero
temperature, given in Appendix G of \cite{key-23}. In the next section,
we will see how this trace is modified for the case when $T>0$.

\subsubsection{Introduction of $T$}

In the case when the system has some temperature $T>0$, the trace
we have to take is

\[
\mathrm{Tr}_{B}\left\{ \prod_{n=1}^{N}\left[\hat{\hat{K}}_{q_{n}}\left(\Delta t\right)\right]\frac{e^{-\beta\hbar\sum_{k}\omega_{k}\hat{b_{k}}^{\dagger}\hat{b_{k}}}}{Z}\right\} .
\]

We will employ once again the basis of coherent states to take the
trace. When acting upon the superoperator $\hat{\hat{K}}_{q_{n}}\left(\Delta t\right)$,
coherent-state bras and kets result in:

\begin{eqnarray*}
\bigotimes_{k}\left\langle \alpha_{k}^{\prime}\right|\hat{\hat{K}}_{q}\left(\Delta t\right)\hat{X}\bigotimes_{k}\left|\alpha_{k}\right\rangle  & = & \bigotimes_{k}\left\langle e^{i\omega_{k}\Delta t}\alpha_{k}^{\prime}+\left(e^{i\omega_{k}\Delta t}-1\right)G_{k}\right|\hat{X}\bigotimes_{k}\left|e^{i\omega_{k}\Delta t}\alpha_{k}-\left(e^{i\omega_{k}\Delta t}-1\right)G_{k}\right\rangle \\
 &  & \times e^{\sum_{k}\left(G_{k}^{*}\left(1-e^{i\omega_{k}\Delta t}\right)\left(\alpha_{k}+\alpha_{k}^{\prime}\right)-G_{k}\left(1-e^{-i\omega_{k}\Delta t}\right)\left(\alpha_{k}+\alpha_{k}^{\prime}\right)^{*}\right)/2},
\end{eqnarray*}
 where we have defined

\[
G_{k}\equiv q\frac{g_{k}}{\omega_{k}}.
\]

Repeating the procedure $N$ times, we find:

\begin{eqnarray*}
\bigotimes_{k}\left\langle \alpha_{k}\right|\prod_{n=1}^{N}\left[\hat{\hat{K}}_{q_{n}}\left(\Delta t\right)\right]\rho_{B}\left(0\right)\bigotimes_{k}\left|\alpha_{k}\right\rangle  & = & \bigotimes_{k}\left\langle e^{iN\omega_{k}\Delta t}\left[\alpha_{k}+\sum_{n=1}^{N}\left(e^{i\omega_{k}\Delta t}-1\right)G_{n,k}\right]\right|\\
 &  & \frac{e^{-\beta\hbar\sum_{k}\omega_{k}\hat{b_{k}}^{\dagger}\hat{b_{k}}}}{Z}\bigotimes_{k}\left|e^{iN\omega_{k}\Delta t}\left[\alpha_{k}-\sum_{n=1}^{N}\left(e^{i\omega_{k}\Delta t}-1\right)G_{n,k}\right]\right\rangle \\
 &  & \times e^{\sum_{n,k}\left(G_{n,k}^{*}e^{-i\omega_{k}\Delta t}\left(1-e^{i\omega_{k}\Delta t}\right)\alpha_{k}-G_{n,k}e^{i\omega_{k}\Delta t}\left(1-e^{-i\omega_{k}\Delta t}\right)\alpha_{k}^{*}\right)},
\end{eqnarray*}
 where

\[
G_{n,k}\equiv e^{-in\omega_{k}\Delta t}q_{n}\frac{g_{k}}{\omega_{k}}
\]
 and the $\hat{X}$ operator has been replaced by the initial density
matrix of the environment. Its action on the coherent state yields:

\begin{eqnarray*}
e^{-\beta\hbar\omega_{k}\hat{b_{k}}^{\dagger}\hat{b_{k}}}\left|e^{iN\omega_{k}\Delta t}\left[\alpha_{k}-\sum_{n=1}^{N}\left(e^{i\omega_{k}\Delta t}-1\right)G_{n,k}\right]\right\rangle  & = & \left|e^{-\beta\hbar\omega_{k}}e^{iN\omega_{k}\Delta t}\left[\alpha_{k}-\sum_{n=1}^{N}\left(e^{i\omega_{k}\Delta t}-1\right)G_{n,k}\right]\right\rangle \\
 &  & \times e^{-\frac{1}{2}\left(1-e^{-2\beta\hbar\omega_{k}}\right)\left|\alpha_{k}-\sum_{n=1}^{N}\left(e^{i\omega_{k}\Delta t}-1\right)G_{n,k}\right|^{2}}.
\end{eqnarray*}

Therefore,

\begin{eqnarray*}
\bigotimes_{k}\left\langle \alpha_{k}\right|\prod_{n=1}^{N}\left[\hat{\hat{K}}_{q_{n}}\left(\Delta t\right)\right]\rho_{B}\left(0\right)\bigotimes_{k}\left|\alpha_{k}\right\rangle  & = & \prod_{k}\left(\frac{1}{1-e^{-\beta\hbar\omega_{k}}}\right)^{-1}\exp\left\{ -\left(1-e^{-\beta\hbar\omega_{k}}\right)\left|\alpha_{k}\right|^{2}\right\} \\
 &  & \times\exp\left\{ -\left(1+e^{-\beta\hbar\omega_{k}}\right)\sum_{n=1}^{N}2i\mathrm{Im}\left[G_{n,k}^{*}\left(1-e^{-i\omega_{k}\Delta t}\right)\alpha_{k}\right]\right\} \\
 &  & \times\exp\left\{ -\left(1+e^{-\beta\hbar\omega_{k}}\right)\left|\sum_{n=1}^{N}\left(e^{i\omega_{k}\Delta t}-1\right)G_{n,k}\right|^{2}\right\} 
\end{eqnarray*}

Taking the trace, we find ourselves integrating over the complex plane:

\begin{eqnarray*}
\mathrm{Tr}\ldots & = & \prod_{k}\left(\frac{1}{1-e^{-\beta\hbar\omega_{k}}}\right)^{-1}\frac{1}{\pi}\\
 &  & \times\int_{-\infty}^{\infty}\mathrm{d}x\exp\left\{ -\left(1-e^{-\beta\hbar\omega_{k}}\right)x^{2}\right\} \exp\left\{ -\left(1+e^{-\beta\hbar\omega_{k}}\right)2i\mathrm{Im}\left[\sum_{n=1}^{N}G_{n,k}^{*}\left(1-e^{-i\omega_{k}\Delta t}\right)\right]x\right\} \\
 &  & \times\int_{-\infty}^{\infty}\mathrm{d}y\exp\left\{ -\left(1-e^{-\beta\hbar\omega_{k}}\right)y^{2}\right\} \exp\left\{ -\left(1+e^{-\beta\hbar\omega_{k}}\right)2i\mathrm{Re}\left[\sum_{n=1}^{N}G_{n,k}^{*}\left(1-e^{-i\omega_{k}\Delta t}\right)\right]y\right\} \\
 &  & \times\exp\left\{ -\left(1+e^{-\beta\hbar\omega_{k}}\right)\left|\sum_{n=1}^{N}\left(e^{i\omega_{k}\Delta t}-1\right)G_{n,k}\right|^{2}\right\} .
\end{eqnarray*}

These are integrals of Gaussians, which can be solved by:

\[
\int_{-\infty}^{\infty}\mathrm{d}x\; e^{-ax^{2}}e^{-2i\mathrm{Im}\left(b\right)x}\int_{-\infty}^{\infty}\mathrm{d}y\; e^{-ay^{2}}e^{-2i\mathrm{Re}\left(b\right)y}=\frac{\pi}{a}e^{-\left|b\right|^{2}/a},
\]
 yielding the result of, after taking the continuous limit of frequencies,
and Ohmic spectral density (\ref{JOhm}):

\[
\mathrm{Tr}\ldots=\exp\left\{ -8\eta\sum_{m,n=1}^{N}q_{m}q_{n}\int_{0}^{\infty}\mathrm{d}\omega e^{-\omega/\omega_{c}}\frac{e^{-i\left(m-n\right)\omega\Delta t}}{\omega}\coth\left(\frac{\hbar\beta\omega}{2}\right)\sin^{2}\left(\frac{\omega t}{2}\right)\right\} .
\]

This final expression clearly recovers the integral given in Appendix
G of our previous article \cite{key-23} in the limit when $T\to0$,
because

\[
\lim_{\beta\to\infty}\coth\left(\frac{\hbar\beta\omega}{2}\right)=1.
\]

Moreover, it reduces to the expression of decoherence in a finite-temperature
(Eq. 18 of Ref. \cite{key-25}) environment when we take just one
step ($N=1$):

\[
\exp\left\{ -8\eta\int_{0}^{\infty}\mathrm{d}\omega\frac{e^{-\omega/\omega_{c}}}{\omega}\coth\left(\frac{\hbar\beta\omega}{2}\right)\sin^{2}\left(\frac{\omega t}{2}\right)\right\} .
\]

\section{Amplitude-damping interaction}

The amplitude-damping interaction - which, here, is only solved using
the master equation - is described by the Jaynes-Cummings interaction
Hamiltonian:

\begin{equation}
\hat{H}_{SB}=\hbar\underset{k}{\sum}\left(g_{k}\hat{b}_{k}\hat{\sigma}_{+}+g_{k}^{*}\hat{b}_{k}^{\dagger}\hat{\sigma}_{-}\right)
\end{equation}
 where

\begin{equation}
\hat{\sigma}_{+}=\left(\begin{array}{cc}
0 & 1\\
0 & 0
\end{array}\right),\:\hat{\sigma}_{-}=\left(\begin{array}{cc}
0 & 0\\
1 & 0
\end{array}\right).
\end{equation}

The partial trace for the environmental degrees of freedom, following
the procedures already shown in \cite{key-20,key-23}, furnish:

\begin{eqnarray}
\hat{\hat{P}}\hat{\hat{G}}\left(t\right)\hat{\hat{G}}\left(t'\right)\hat{\hat{P}}\hat{\alpha}\left(t\right) & = & \eta\int_{0}^{\infty}\mathrm{d}\omega\omega e^{-\frac{\omega}{\omega_{c}}}\frac{e^{-i\omega\left(t-t'\right)}}{e^{\hbar\beta\omega}-1}e^{-\hat{\hat{S}}t}\left[\hat{\sigma}_{+},\left\{ e^{\hat{\hat{S}}\left(t-t'\right)}\left[\left(e^{\hat{\hat{S}}t'}\hat{R}\left(t\right)\right)\hat{\sigma}_{-}\right]\right\} \right]\hat{\rho}_{B}\nonumber \\
 &  & +\eta\int_{0}^{\infty}\mathrm{d}\omega\omega e^{-\frac{\omega}{\omega_{c}}}\frac{e^{i\omega\left(t-t'\right)}}{1-e^{-\hbar\beta\omega}}e^{-\hat{\hat{S}}t}\left[\hat{\sigma}_{-},\left\{ e^{\hat{\hat{S}}\left(t-t'\right)}\left[\left(e^{\hat{\hat{S}}t'}\hat{R}\left(t\right)\right)\hat{\sigma}_{+}\right]\right\} \right]\hat{\rho}_{B}\nonumber \\
 &  & +\eta\int_{0}^{\infty}\mathrm{d}\omega\omega e^{-\frac{\omega}{\omega_{c}}}\frac{e^{i\omega\left(t-t'\right)}}{e^{\hbar\beta\omega}-1}e^{-\hat{\hat{S}}t}\left[\left\{ e^{\hat{\hat{S}}\left(t-t'\right)}\left[\hat{\sigma}_{+}\left(e^{\hat{\hat{S}}t'}\hat{R}\left(t\right)\right)\right]\right\} ,\hat{\sigma}_{-}\right]\hat{\rho}_{B}\nonumber \\
 &  & +\eta\int_{0}^{\infty}\mathrm{d}\omega\omega e^{-\frac{\omega}{\omega_{c}}}\frac{e^{-i\omega\left(t-t'\right)}}{1-e^{-\hbar\beta\omega}}e^{-\hat{\hat{S}}t}\left[\left\{ e^{\hat{\hat{S}}\left(t-t'\right)}\left[\hat{\sigma}_{-}\left(e^{\hat{\hat{S}}t'}\hat{R}\left(t\right)\right)\right]\right\} ,\hat{\sigma}_{+}\right]\hat{\rho}_{B}.\label{ambJCz}
\end{eqnarray}
 where we are employing the Ohmic spectral density (\ref{JOhm}).

\subsection{The z-component measurement}

For $\hat{L}^{\left(S\right)}=\lambda\hat{\sigma}_{z}$ , the treatment
of (\ref{ambJCz}) and its substitution on (\ref{eqprincipal}) will
furnish, for $\hat{R}\left(t\right)$,

\begin{equation}
\begin{cases}
\frac{\mathrm{d}}{\mathrm{d}t}R_{11}= & 2R_{22}\left(t\right)\eta\int_{0}^{t}\mathrm{d}\tau\int_{0}^{\infty}\mathrm{d}\omega\frac{\omega e^{-\frac{\omega}{\omega_{c}}}}{e^{\hbar\beta\omega}-1}e^{-2\lambda^{2}\tau}\cos\left[\left(2\omega_{0}-\omega\right)\tau\right]\\
 & -2R_{11}\left(t\right)\eta\int_{0}^{t}\mathrm{d}\tau\int_{0}^{\infty}\mathrm{d}\omega\frac{\omega e^{-\frac{\omega}{\omega_{c}}}}{1-e^{-\hbar\beta\omega}}e^{-2\lambda^{2}\tau}\cos\left[\left(2\omega_{0}-\omega\right)\tau\right]\\
\\
R_{12}\left(t\right)= & R_{12}\left(0\right)\exp\left\{ -\eta\int_{0}^{t}\mathrm{d}t'\int_{0}^{\infty}\mathrm{d}\omega\omega e^{-\frac{\omega}{\omega_{c}}}\frac{e^{2\lambda^{2}t'}e^{i\left(2\omega_{0}-\omega\right)t'}-1}{2\lambda^{2}+i\left(2\omega_{0}-\omega\right)}\coth\left(\frac{\hbar\beta\omega}{2}\right)\right\} 
\end{cases}\label{sistzAD}
\end{equation}
 remembering that $R_{11}\left(t\right)+R_{22}\left(t\right)=1$ and
$R_{12}\left(t\right)=R_{21}^{*}\left(t\right)$.

\subsection{The x-component measurement}

For $\hat{L}^{\left(S\right)}=\lambda\hat{\sigma}_{x}$ , as expected,
the expressions will be more complex, furnishing, 
\begin{equation}
\begin{cases}
\frac{\mathrm{d}}{\mathrm{d}t}R_{11} & =2R_{11}\left(t\right)e^{2\lambda^{2}t}\int_{0}^{t}\mathrm{d}t'\int_{0}^{\infty}\mathrm{d}\omega J\left(\omega\right)\mathrm{Re}\left\{ a_{2}\left(t'\right)b_{1}^{*}\left(t-t'\right)\frac{e^{-i\omega\left(t-t'\right)}}{e^{\hbar\beta\omega}-1}-a_{1}\left(t'\right)b_{1}\left(t-t'\right)\frac{e^{i\omega\left(t-t'\right)}}{1-e^{-\hbar\beta\omega}}\right\} \\
 & +2R_{22}\left(t\right)e^{2\lambda^{2}t}\int_{0}^{t}\mathrm{d}t'\int_{0}^{\infty}\mathrm{d}\omega J\left(\omega\right)\mathrm{Re}\left\{ a_{1}\left(t'\right)b_{1}^{*}\left(t-t'\right)\frac{e^{-i\omega\left(t-t'\right)}}{e^{\hbar\beta\omega}-1}-a_{2}\left(t'\right)b_{1}\left(t-t'\right)\frac{e^{i\omega\left(t-t'\right)}}{1-e^{-\hbar\beta\omega}}\right\} \\
\\
\frac{\mathrm{d}}{\mathrm{d}t}R_{12} & =-R_{12}\left(t\right)\int_{0}^{t}\mathrm{d}t'\int_{0}^{\infty}\mathrm{d}\omega J\left(\omega\right)e^{-2\lambda^{2}\left(t-t'\right)}\coth\left(\frac{\hbar\beta\omega}{2}\right)\\
 & \qquad\times\left[b_{1}\left(-t\right)b_{1}\left(t'\right)e^{-i\omega\left(t-t'\right)}+b_{2}\left(-t\right)b_{2}\left(t'\right)e^{i\omega\left(t-t'\right)}\right]\\
 & -R_{21}\left(t\right)\int_{0}^{t}\mathrm{d}t'\int_{0}^{\infty}\mathrm{d}\omega J\left(\omega\right)e^{-2\lambda^{2}\left(t-t'\right)}\coth\left(\frac{\hbar\beta\omega}{2}\right)\\
 & \qquad\times\left[b_{1}\left(-t\right)b_{2}\left(t'\right)e^{-i\omega\left(t-t'\right)}+b_{1}^{*}\left(t'\right)b_{2}\left(-t\right)e^{i\omega\left(t-t'\right)}\right]
\end{cases}
\end{equation}
 where

\begin{equation}
\begin{cases}
a_{1}\left(t\right) & =\frac{1+e^{-2\lambda^{2}t}}{2}\\
a_{2}\left(t\right) & =\frac{1-e^{-2\lambda^{2}t}}{2}\\
b_{1}\left(t\right) & =\frac{e^{-\lambda^{2}t}}{\Omega}\left[\Omega\cosh\left(\Omega t\right)-i2\omega_{0}\sinh\left(\Omega t\right)\right]\\
b_{2}\left(t\right) & =\frac{\lambda^{2}}{\Omega}e^{-\lambda^{2}t}\sinh\left(\Omega t\right)
\end{cases}
\end{equation}
 remembering that $R_{11}\left(t\right)+R_{22}\left(t\right)=1$ e
$R_{12}\left(t\right)=R_{21}^{*}\left(t\right)$. After the $R$ have
been found, it is still necessary a change of basis to find the density
operator in the eigenbasis of $\hat{\sigma}_{x}$, as in the discussion
of Eq. (\ref{sistphasex}).

\section{Results and Analysis}

After laying down the methods for obtaining the solutions, we now
can observe what they look like in graphs. There are four parameters
to be analysed: 
\begin{itemize}
\item $\lambda$, the coupling strength between the system and the measuring
apparatus; 
\item $\eta,$ the coupling strength between the system and the environment; 
\item $\omega_{0}$, the splitting, in frequency units, between the two
energy levels of the principal system, henceforth called the natural
frequency of the system; 
\item $T$, the environment temperature, also expressed in terms of $\beta=1/k_{B}T$,
where $k_{B}$ is the Boltzmann constant. Note that $T=0$ corresponds
to the limit in which $\beta\rightarrow\infty$. 
\end{itemize}
In this section, we observe how the measurement, represented by $\lambda$,
affects the evolution of the system for different values of the other
three parameters $\eta$, $\omega_{0}$, and $\beta$.

It may be noted that in some cases the natural frequency of the system
is taken as zero ($\omega_{0}=0$). This is not simply a manner of
simplifying the calculations, despite certainly having this advantage.
This is equivalent to turning off the natural evolution of the two-level
system, as in the case of a memory qubit. It is, therefore, a situation
of relevance in the context of quantum computation, because quantum
memories are necessary to the proper synchronization of the internal
processes of a quantum computer \cite{Q_MEMORY}.

\subsection{Phase-damping}

The phase damping channel merely causes the loss of coherence in the
basis of $\hat{\sigma}_{z}$, thus having an effect similar to a measurement
of this observable. It does not affect the populations in this basis
and, therefore, the outcome of the $\hat{\sigma}_{z}$ measurement
remains unchanged. However, as it makes the coherences disappear,
it does affect the outcomes of the measurement of $\hat{\sigma}_{x}$,
for example. In this case, as the coherences in the original basis
vanish, the populations in the basis of $\hat{\sigma}_{x}$ approach
$1/2$, thus destroying any information that we initially could obtain
from a measurement of that observable.

\subsubsection{The z-component measurement}

Eq. (\ref{eq:solzZZ}) is the complete analytical solution for this
case, and it includes all the parameters of our problem. For this
case, we can see \cite{key-23} that: 
\begin{itemize}
\item the populations are constant in time, for any value of the parameters
$\omega_{0}$, $\lambda$, $\eta$, and $T$; 
\item the moduli of the coherences are independent of $\omega_{0}$ and
an increase in the other parameters just makes the decay more intense. 
\end{itemize}
These results are expected, because the frequency $\omega_{0}$ only
affects the phases of the coherences, and both the measuring apparatus
and the environment are responsible for measuring the system in the
eigenbasis of $\hat{\sigma}_{z}$, thus causing decoherence. The only
new result is that the environmental temperature also helps to increase
the decoherence, an effect not unlike those found in other works that
did not involve the measuring apparatus \cite{key-25}.

\subsubsection{The x-component measurement}

While in \cite{key-23} we had to make the restrictions $\omega_{0}=0$
and $T=\frac{1}{k_{B}\beta}=0$ in (\ref{sistphasex}) to obtain analytical
solutions, here the equation is solved numerically to analyse the
influence of all parameters.

\begin{figure}
\includegraphics[width=0.5\textwidth]{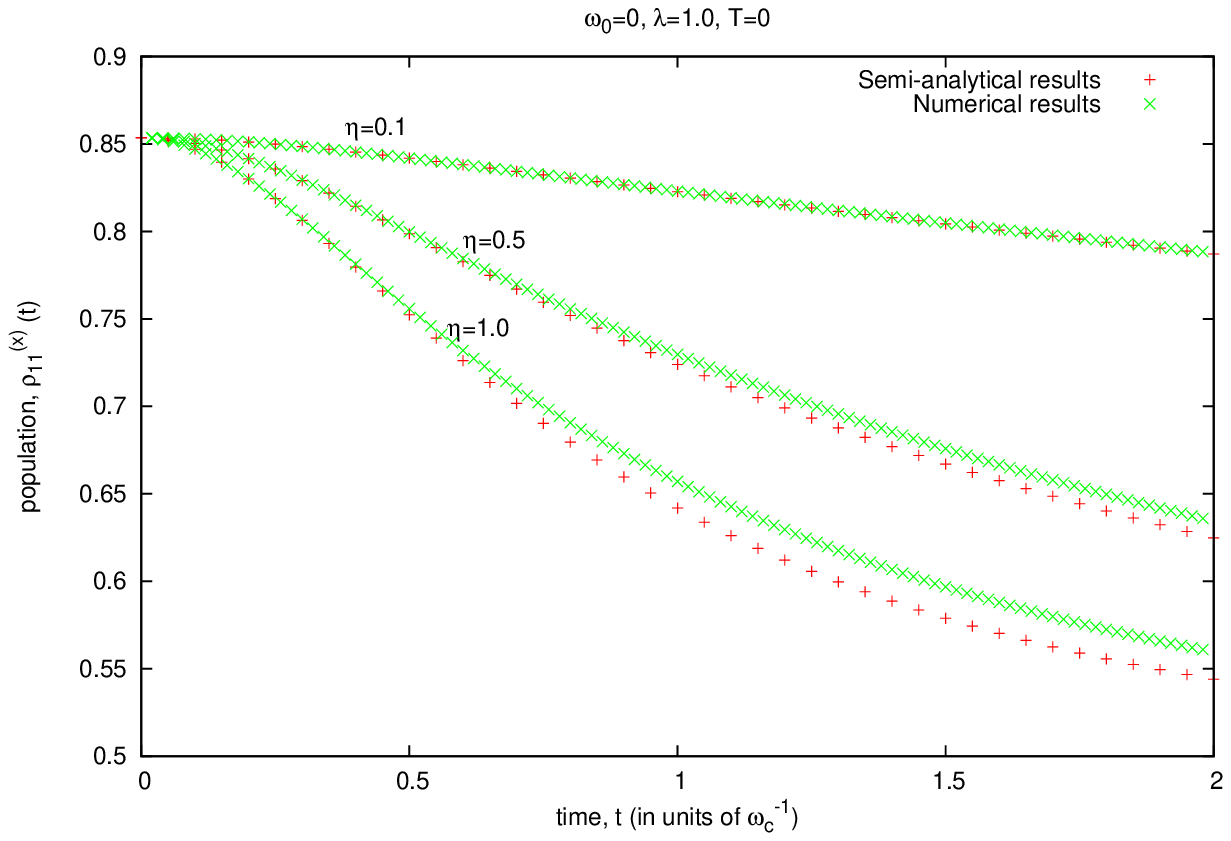}\includegraphics[width=0.5\textwidth]{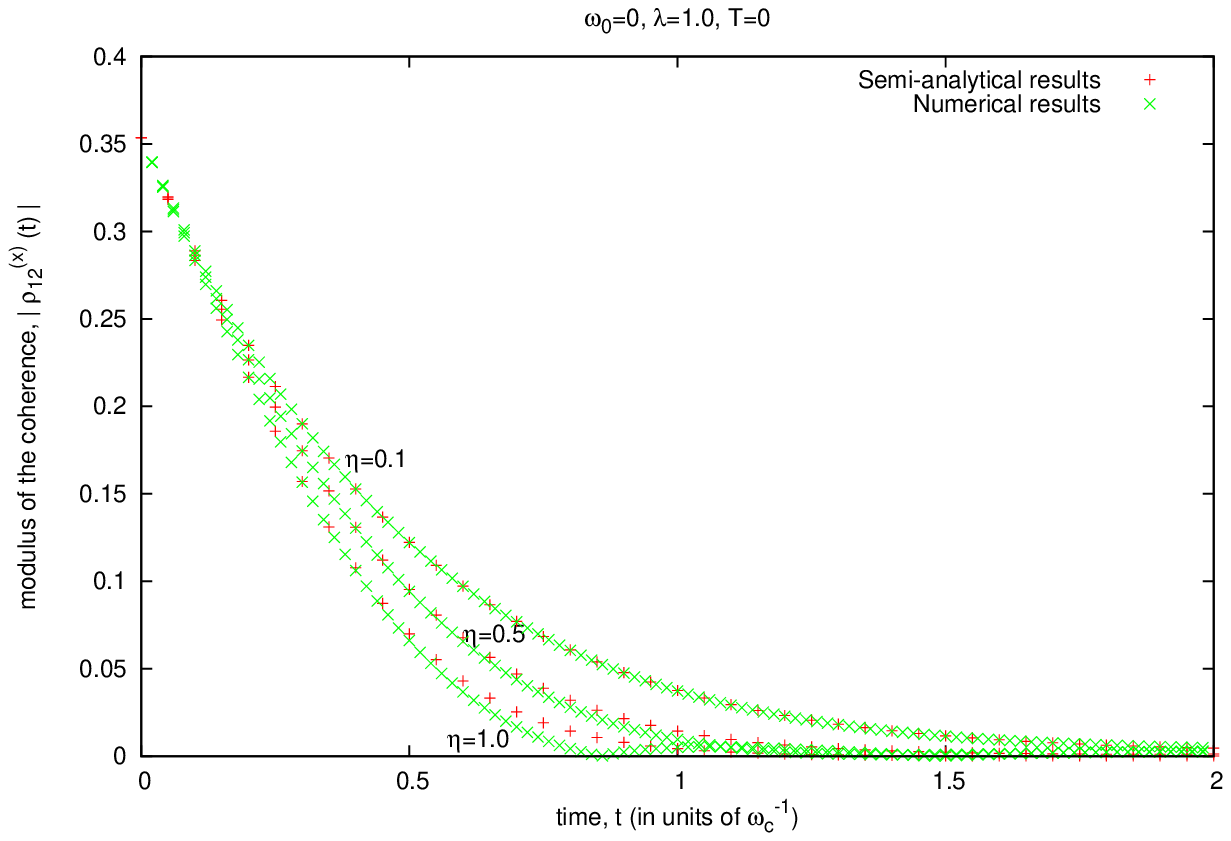}

\includegraphics[width=0.5\textwidth]{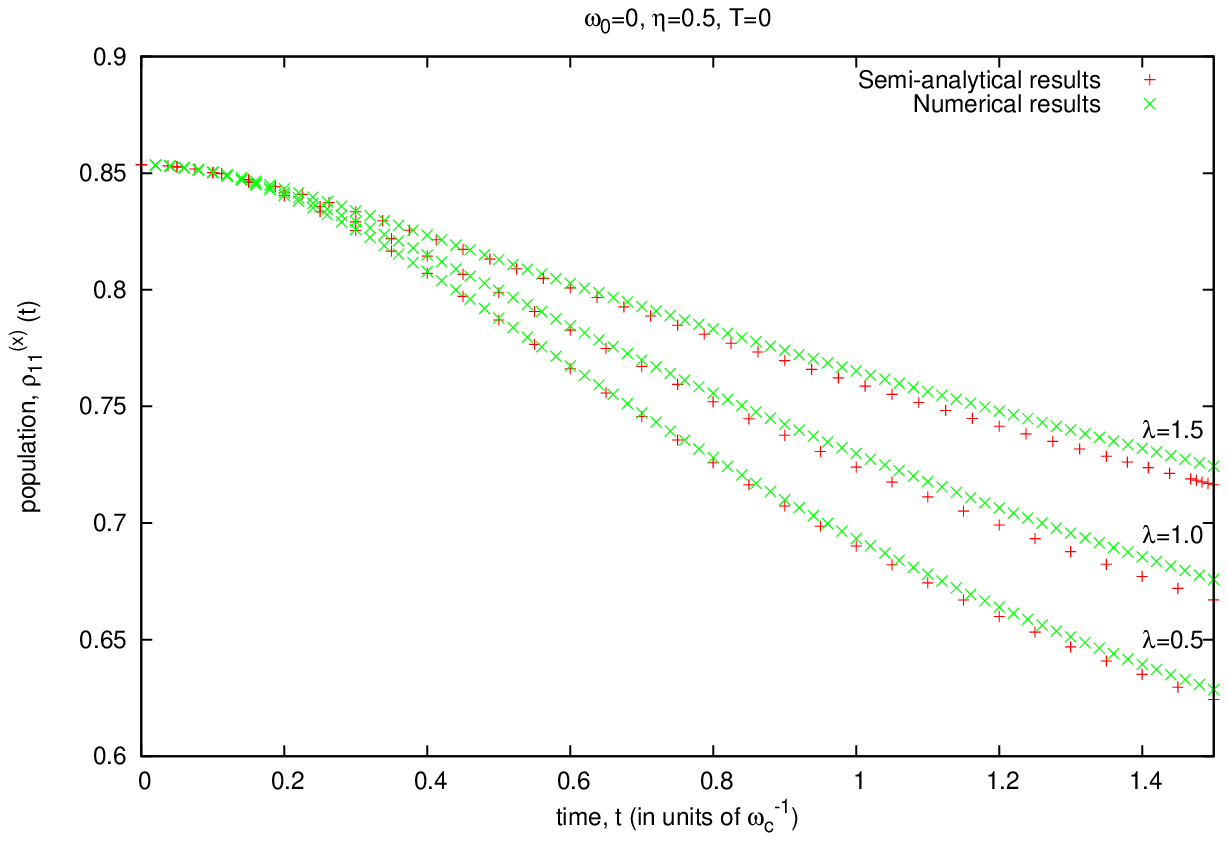}\includegraphics[width=0.5\textwidth]{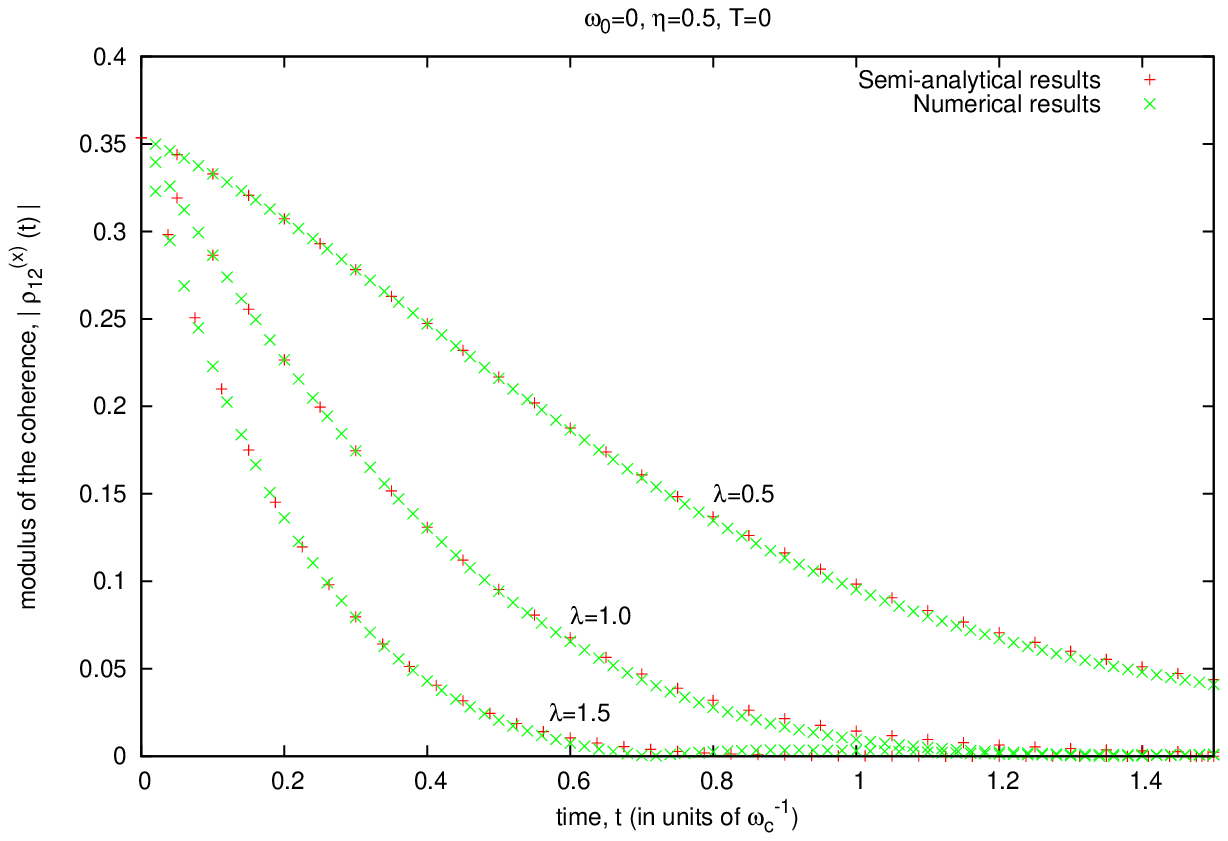}

\caption{Populations and coherences in the eigenbasis of the measured observable
for $\omega_{0}=0$ and $T=0$. In this case the conclusions are the
same as those presented in reference \cite{key-23}.}
\end{figure}

Figure 1 reproduces the analytical results of \cite{key-23,key-24}
for $\omega_{0}=0$ and $T=0$, namely, that the change in value of
the population, caused by the interaction with the environment and
therefore increased with $\eta$, is attenuated by the system-apparatus
coupling $\lambda$. Regarding the coherences, which tend to zero
in absolute value, we are able to establish a time for the measurement
process to reach completion. In \cite{key-24} we have found, for
this specific situation, an upper limit for the measurement duration:

\begin{equation}
t_{M}=-\frac{1}{2\lambda^{2}}\ln f
\end{equation}
 where $f$ is an arbitrary fraction of the initial value of the modulus
of the coherences ($0<f<1$).

Moreover, the comparison between the solution of the master equation
and the numerical results obtained from the superoperator-splitting
procedure are in good agreement for low values of $\eta$, but the
two graphs depart more and more from each other as the strength of
the coupling increases. This is due to the fact that the master equation
is a lower-order approximation valid only in the limit of small $\eta$.
For higher values of $\eta$, the approximation fails.

After reproducing the numerical results known to the analytical approach,
we add a new parameter: the system frequency, $\omega_{0}$. Figure
2 is equivalent to Fig. 1, now considering a non-zero value for $\omega_{0}$.

\begin{figure}
\includegraphics[width=0.5\textwidth]{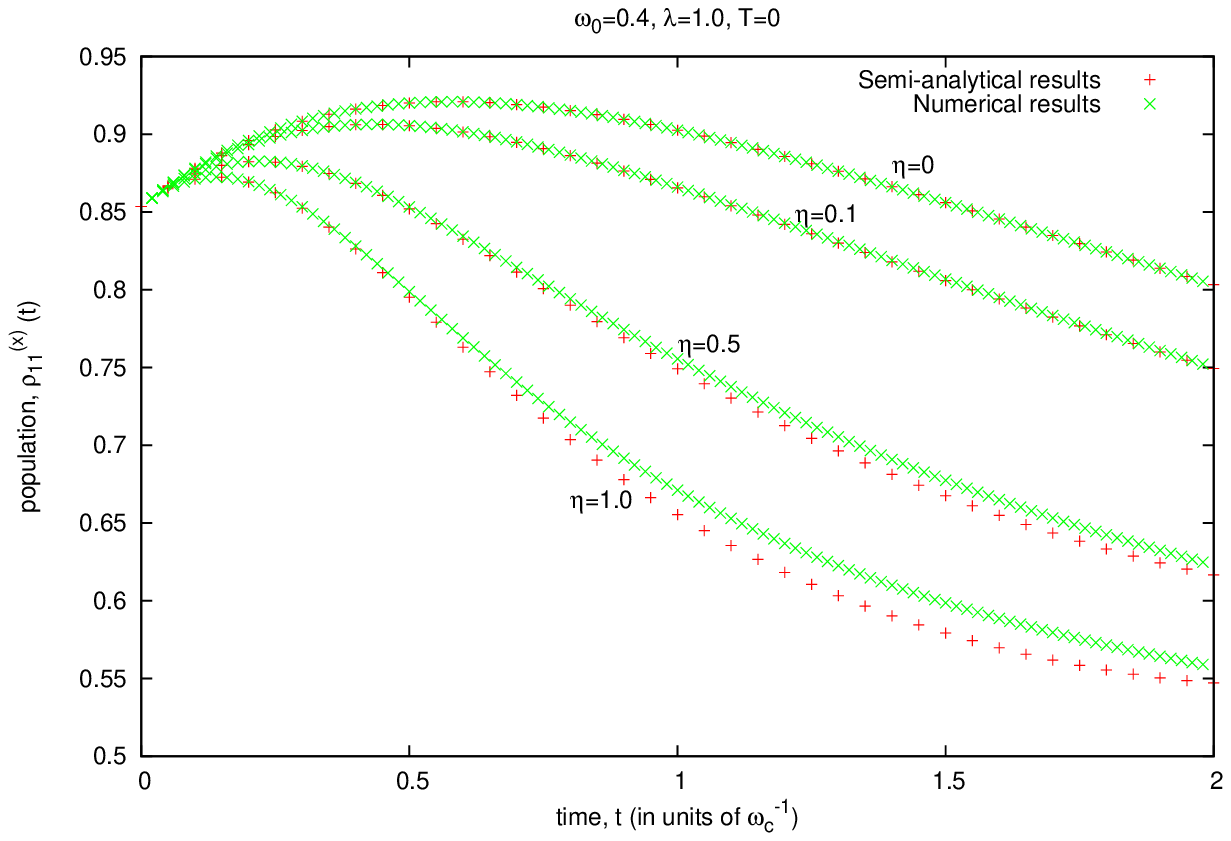}\includegraphics[width=0.5\textwidth]{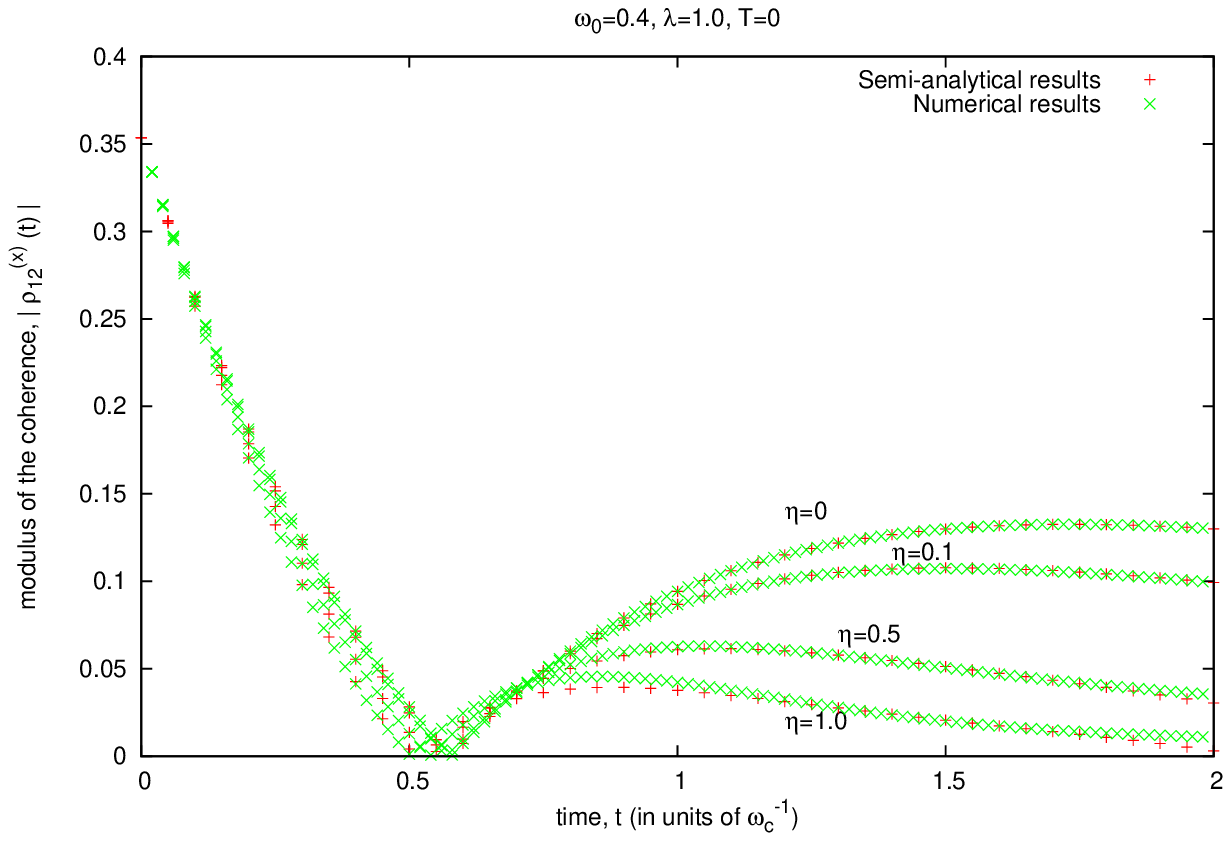}

\includegraphics[width=0.5\textwidth]{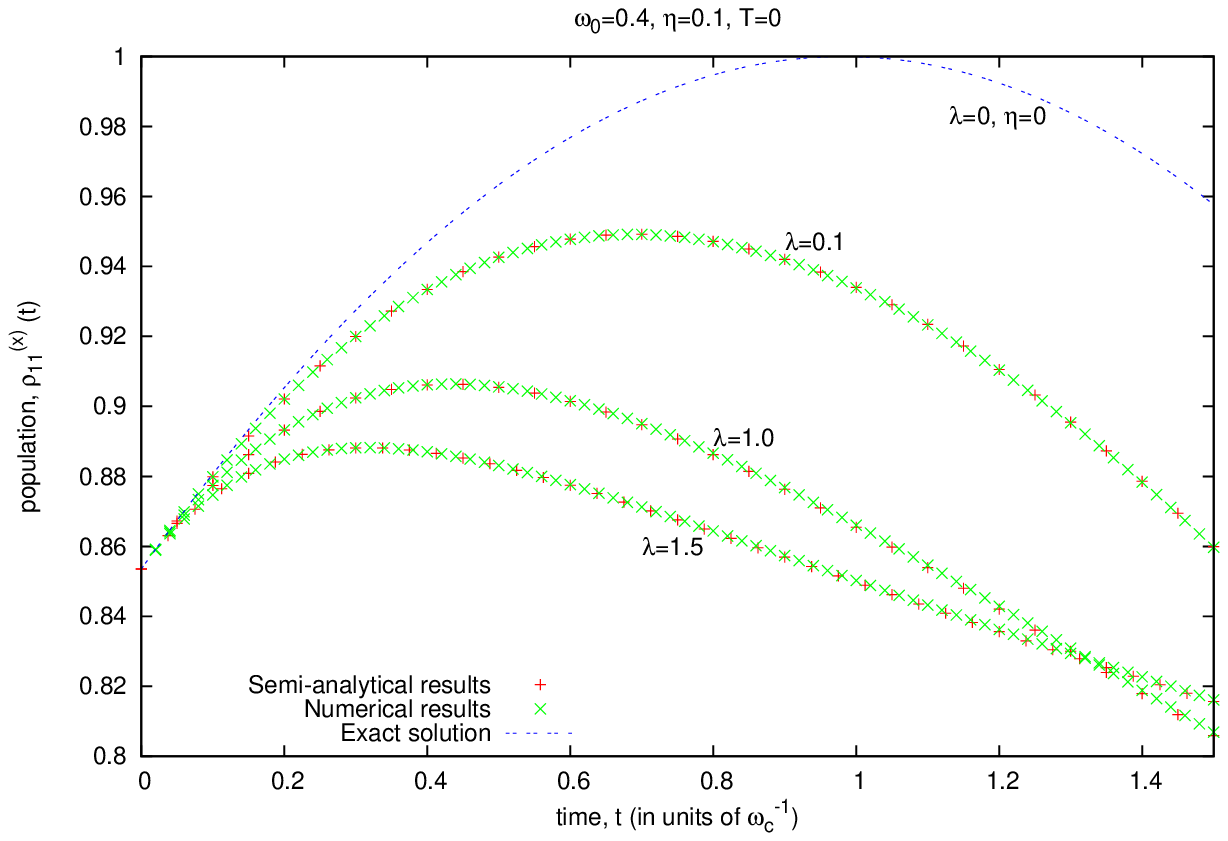}\includegraphics[width=0.5\textwidth]{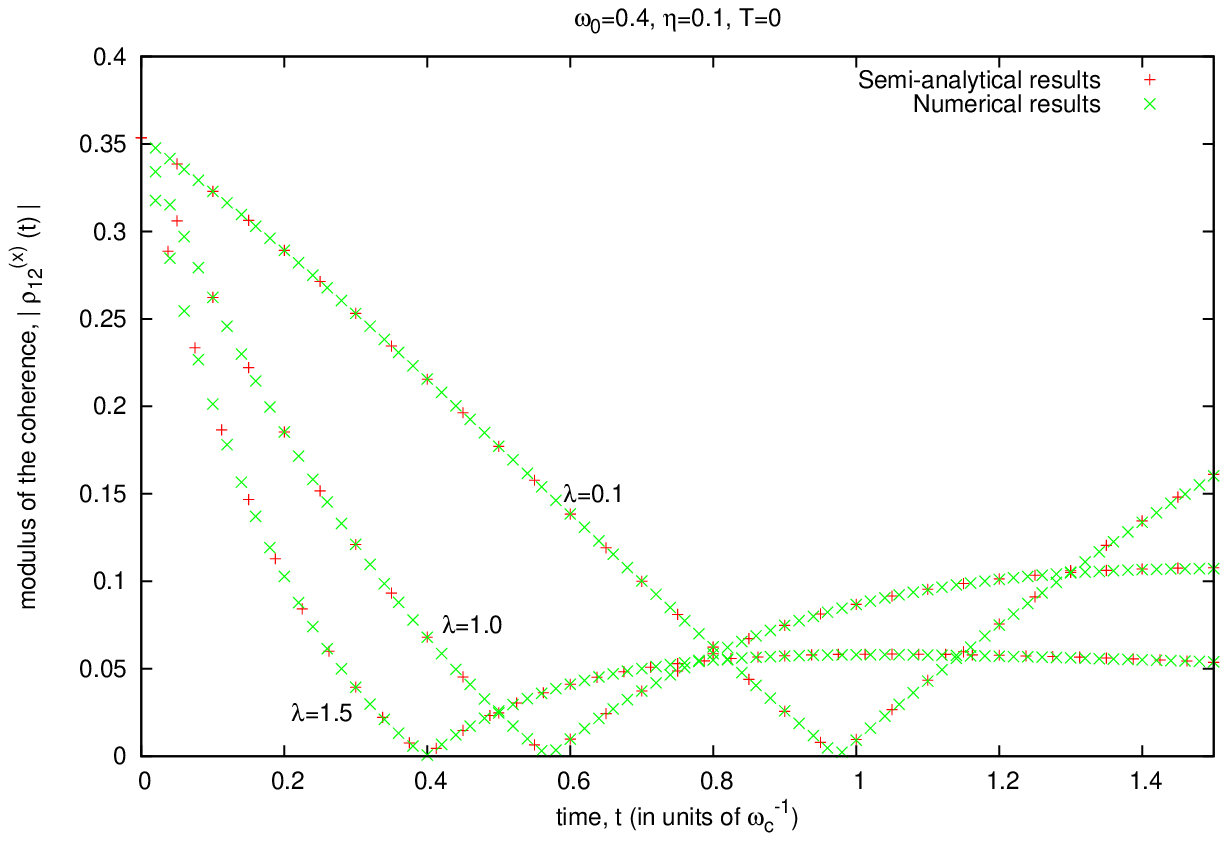}

\caption{Curves referring to the population in the eigenbasis of the measured
observable, considering many values of $\lambda$ and $\eta$, and
a non-zero value of $\omega_{0}$, while keeping $T=0$.}
\end{figure}

From Fig. 2, we note that the increase of the system frequency changes
the behavior of the populations, which is no longer monotonic. In
the time interval considered here, we are able to observe that the
populations oscillate. From Fig. 2(c), we observe that, the lower
the value of $\lambda$, the closer is the evolution of the populations
to the unperturbed case (dashed lines). This might mislead us into
thinking that the finite-time measurement is less efficient than the
instantaneous in this case. This notion is false, however, because
the state we want to measure is the one at the instant $t=0$, when
the measurement apparatus is turned on. In general, the curve with
the highest $\lambda$ is still the one that conserves the population
closest to its initial value for most of the time, meaning that the
measurement protects the system both from the environment and prevents
its inherent evolution, as in the Zeno effect.

For the modulus of the coherences, the monotonic behavior is likewise
eliminated. However, while for the populations we found a point of
maximum, here we find also a point of minimum at zero, where the coherences
change sign and their modulus bounces upwards. As the system-environment
coupling strength $\eta$ becomes more intense, the time when the
magnitude of the coherences is zero becomes smaller, while keeping
the modulus in general closer to zero even after the point of minimum.
A similar effect is observed for $\lambda$, which is expected from
the fact that both the apparatus and the environment interactions
are contributing to the decoherence.

Now, we will consider the effects of another parameter neglected in
our previous works \cite{key-23,key-24}: the temperature of the environment.
Fig. 3 shows the behavior of populations and coherences for a fixed
value of the strength of the coupling between the system and the environment
when the system has zero frequency $\omega_{0}$, in the case in which
we vary $\lambda$ and $T$. It is possible to observe, for populations,
that, while an increase of the temperature makes the deleterious effects
of the environment more pronounced, a similar increase in $\lambda$
is capable of compensating for these effects, thus attenuating the
rate of population change. Therefore, we can conclude that the simple
addition of the environment temperature to the model does not change
our previously published conclusions about state protection and measurement
error \cite{key-23,key-24}.

\begin{figure}
\includegraphics[width=0.5\textwidth]{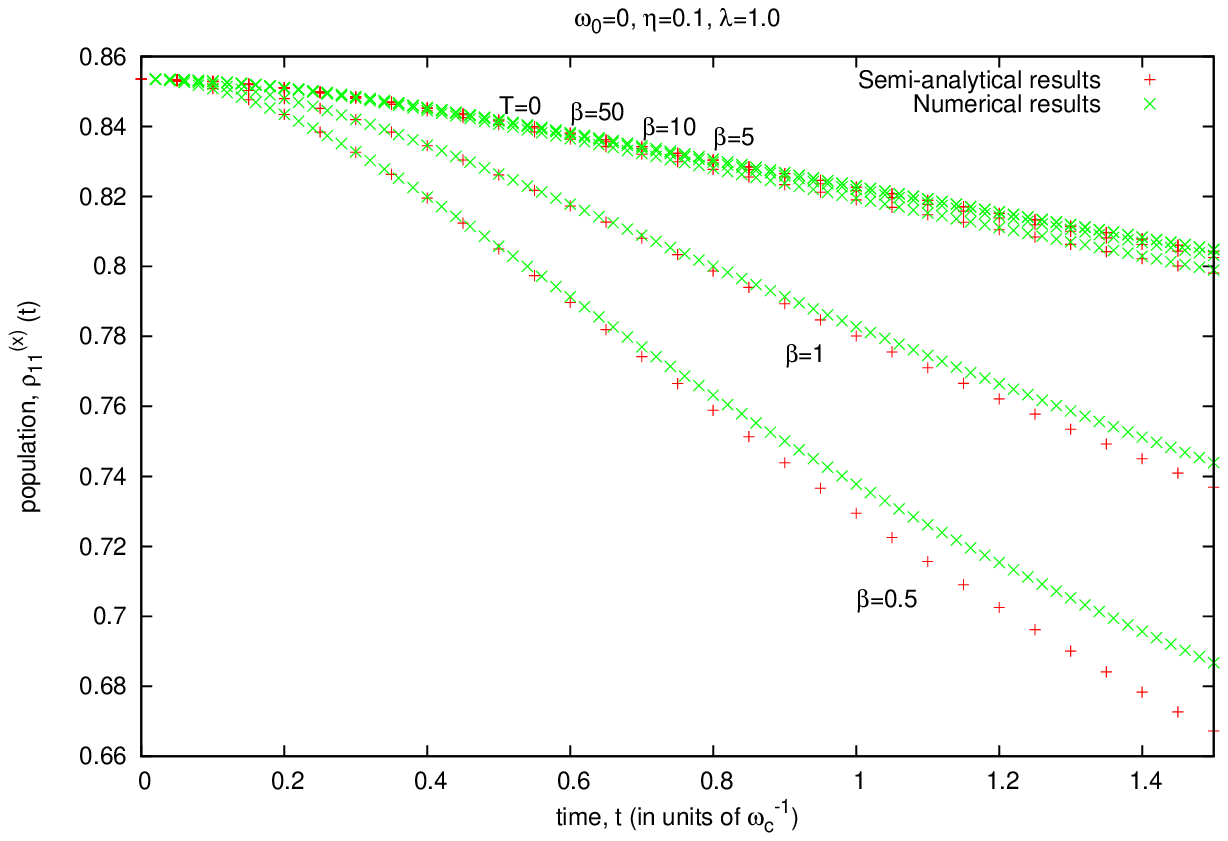}\includegraphics[width=0.5\textwidth]{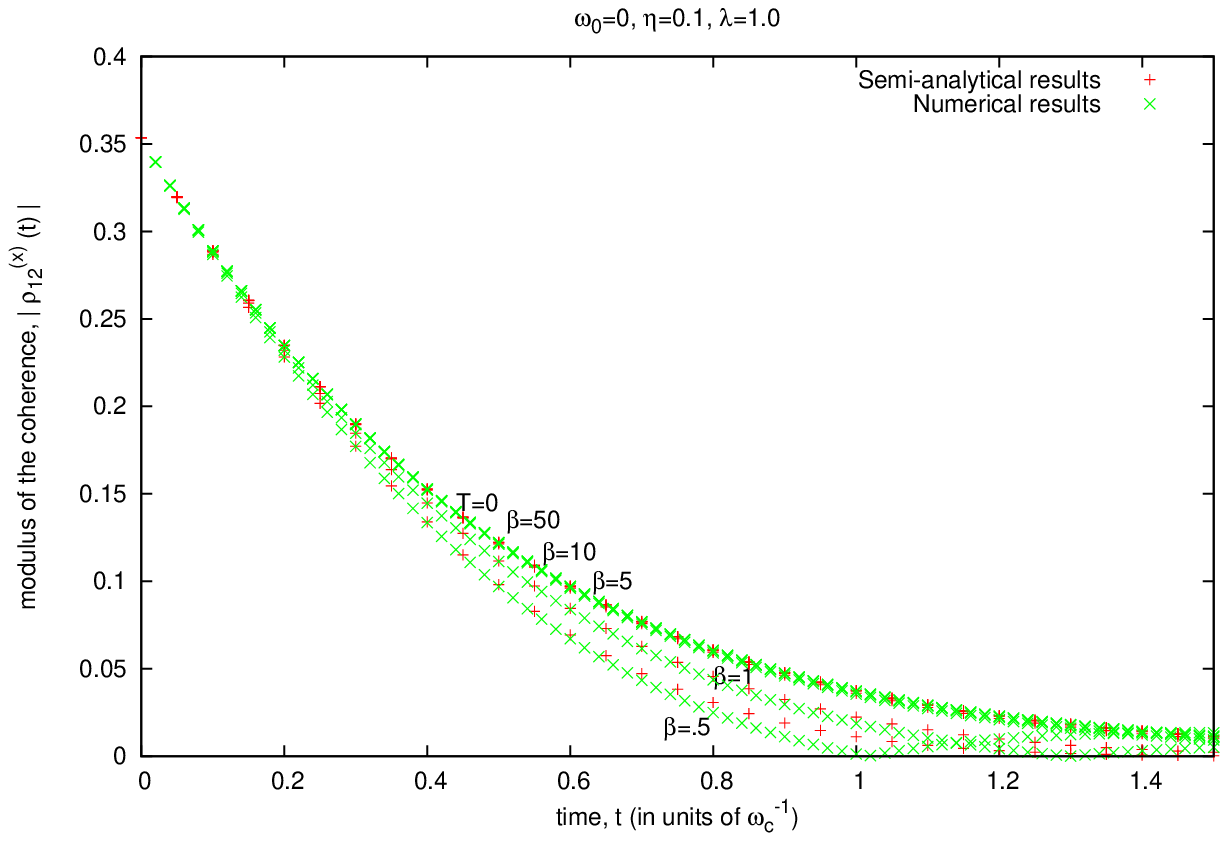}

\includegraphics[width=0.5\textwidth]{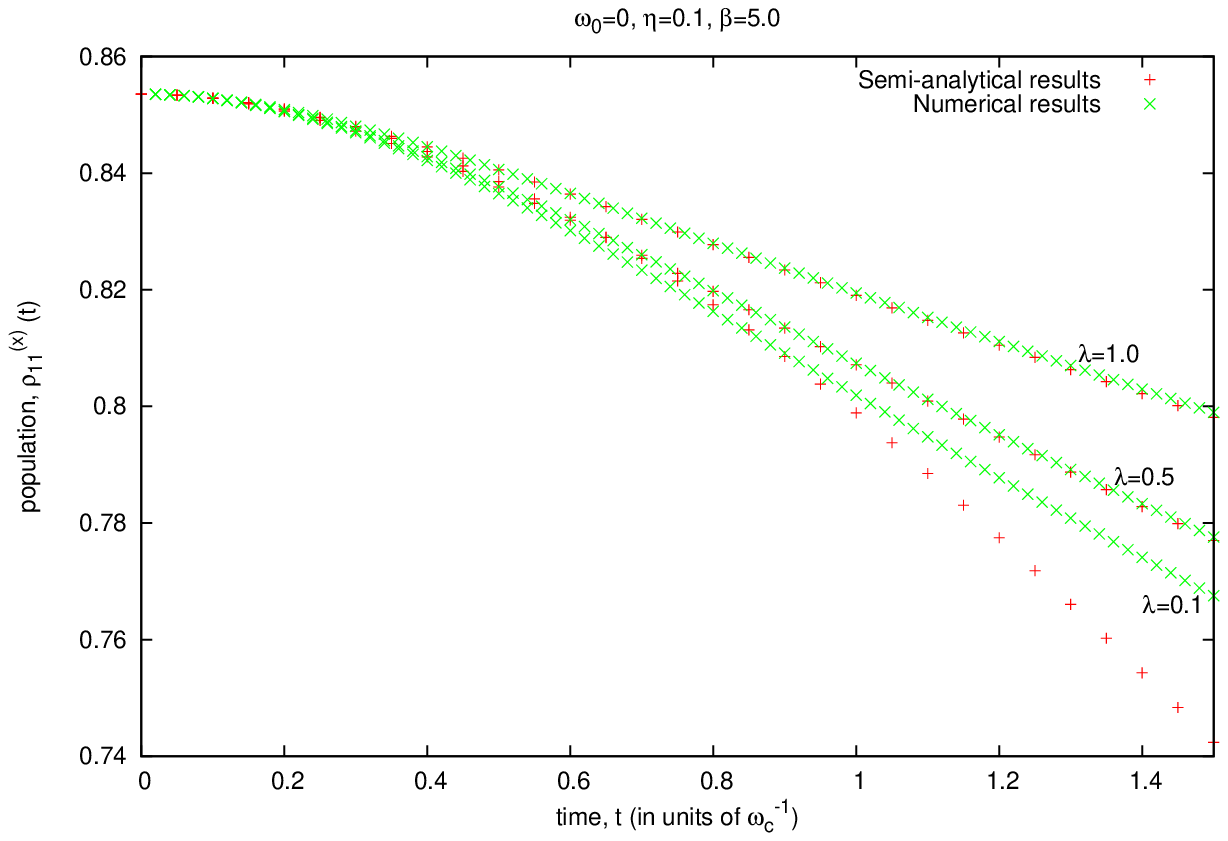}\includegraphics[width=0.5\textwidth]{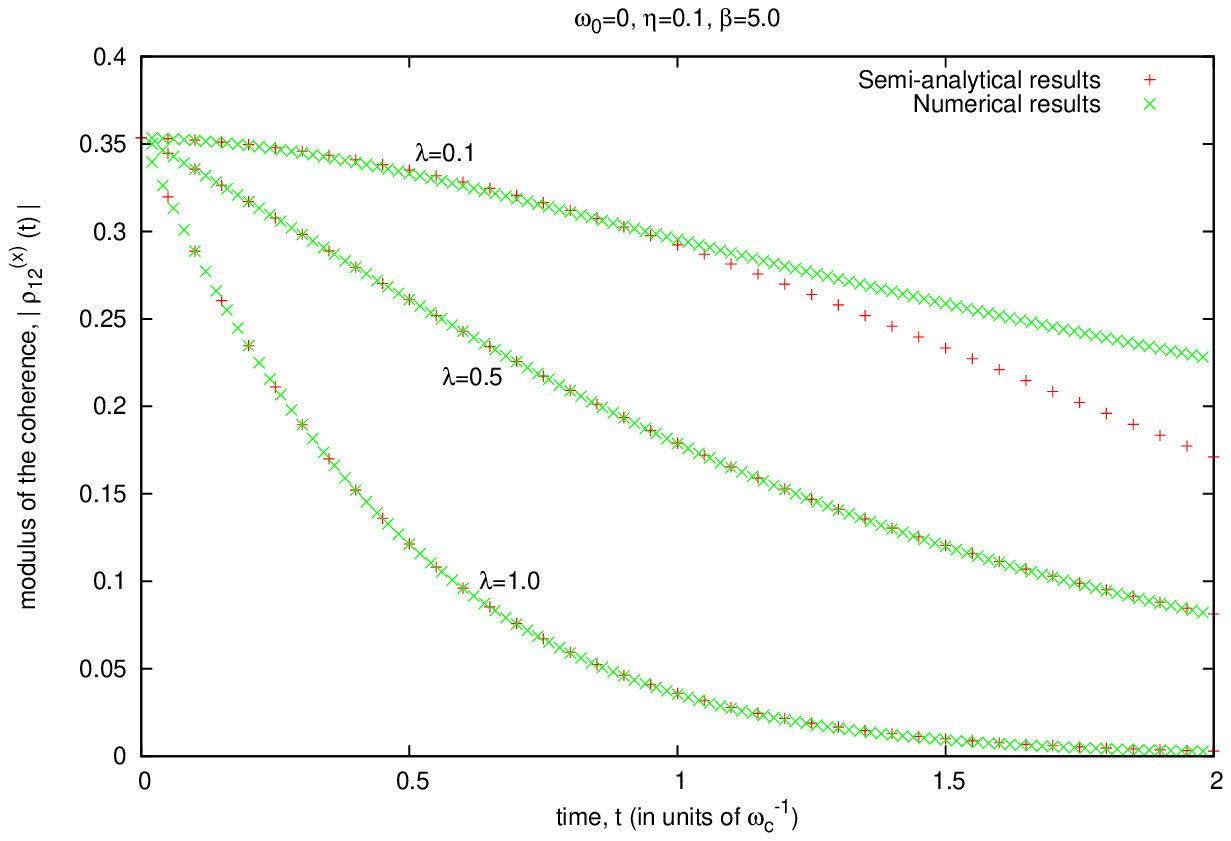}

\caption{Curves for the analysis of the effects of the environmental temperature
for $\omega_{0}=0$.}
\end{figure}

For the modulus of the coherences, we still have the same effects
of faster decrease for higher $\lambda$, as seen in Fig. 3(d). The
only new fact is that the environmental finite temperature has a similar
effect to $\eta$ in making the coherences approach zero.

Ending the analysis of phase-damping interaction, Fig. 4 shows several
curves with fixed and non-zero environment parameters $\eta$ and
$T$, varying $\lambda$. As in the previous cases, here we once again
verify that the measurement has a twofold effect on the system: it
attenuates the changes in the populations and increases the velocity
with which the coherences vanish, confirming, so far, the conclusions
of our previous works \cite{key-23,key-24}.

\begin{figure}
\includegraphics[width=0.5\textwidth]{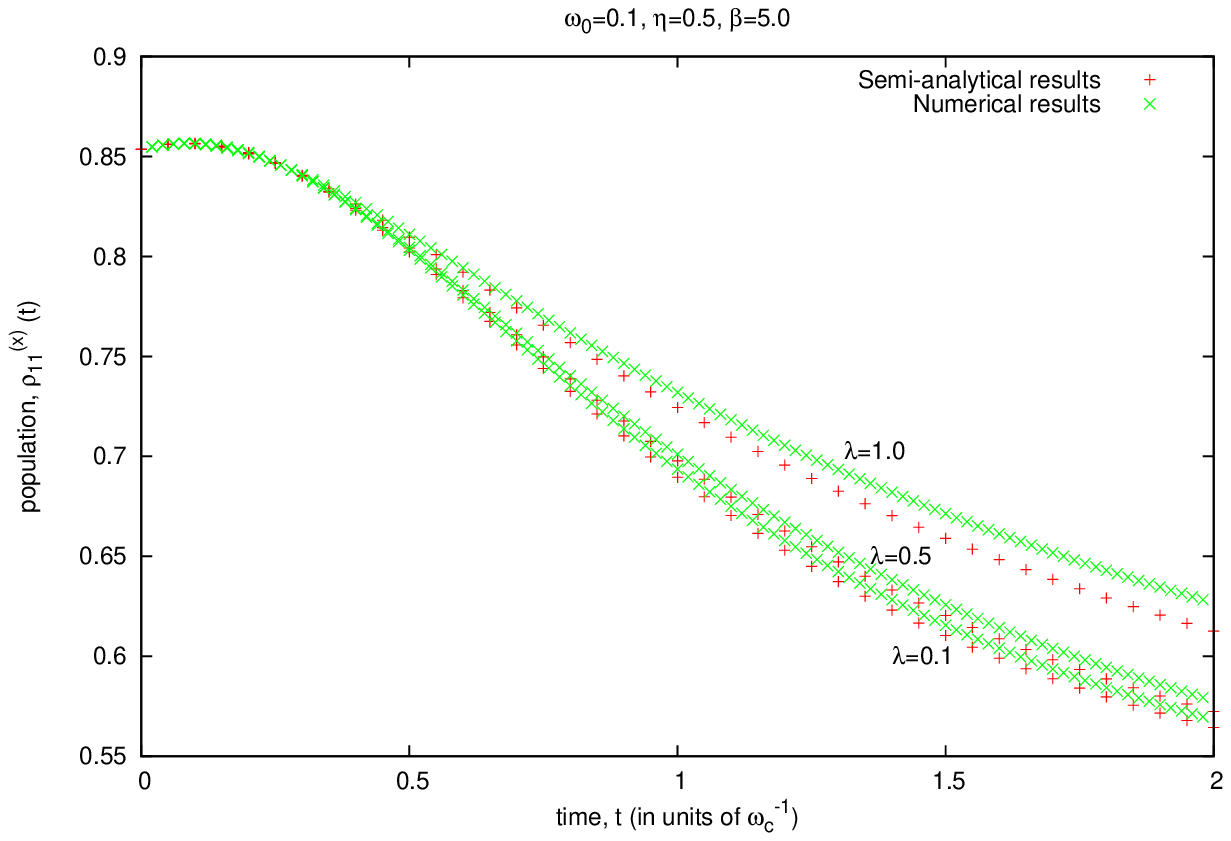}\includegraphics[width=0.5\textwidth]{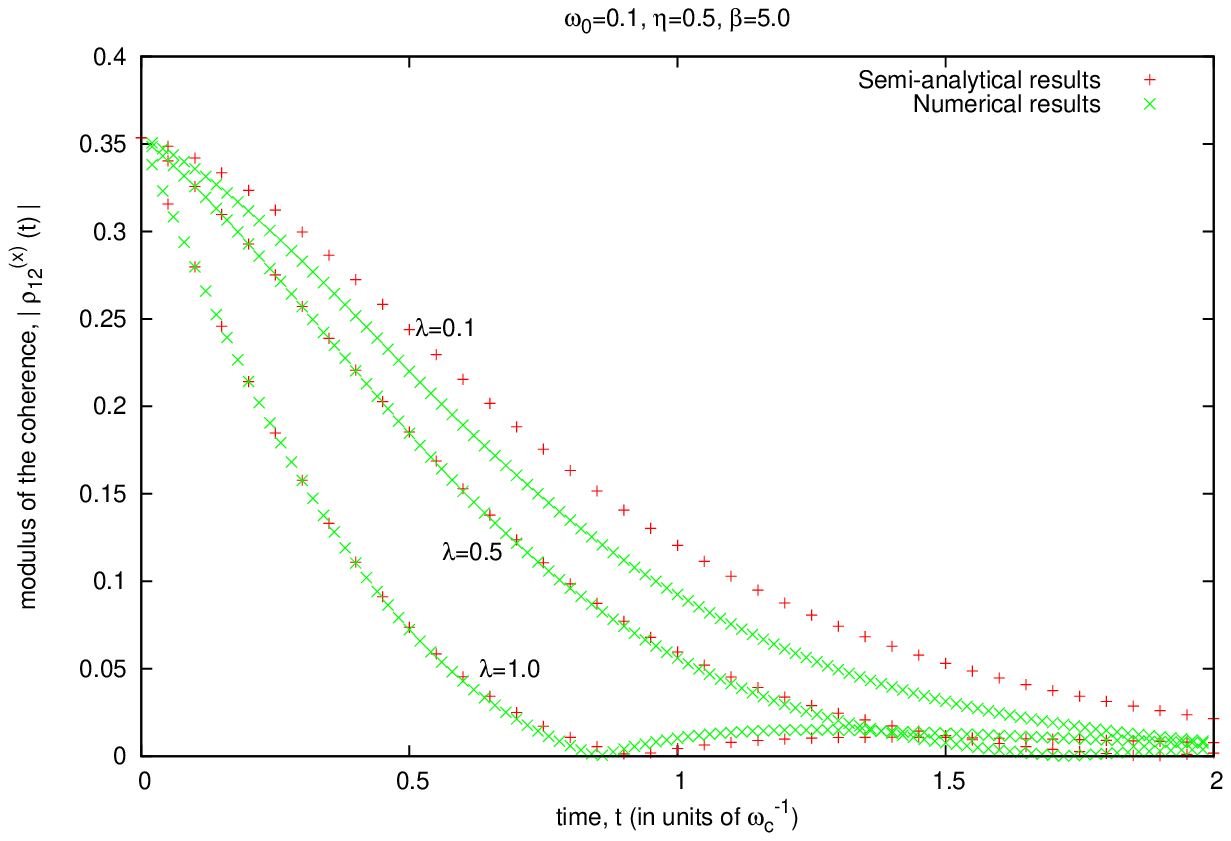}

\caption{Simultaneous influence of the frequency of the system and the environmental
temperature.}
\end{figure}

\subsection{Amplitude-damping}

The amplitude-damping system-environment interaction differs from
phase-damping by not simply being an agent of decoherence. This kind
of interaction causes the decay of the excited state $\left|1\right\rangle $
to the ground state $\left|2\right\rangle $, thus resulting in the
gradual decrease of the population $\rho_{11}\left(t\right)$ until
its complete disappearance.

However, if the temperature is not zero, as in many cases we are treating
here, the repopulation of the excited state by the environment will
result in a final equilibrium population that lies above zero. This
value, proportional to the Boltzmann weight $e^{-\hbar\beta\omega_{0}}$,
increases with temperature (with more energy available, the environment
will more easily repopulate the excited state) but decreases with
$\omega_{0}$ (the higher the energy difference, the more difficult
it is to repopulate). For $T>0$ and $\omega_{0}=0$, both populations
assume a final value of $0.5$, as will be seen below.

\subsubsection{The z-component measurement}

Contrary to the phase-damping case, amplitude-damping changes the
value of the populations in the eigenbasis of $\hat{\sigma}_{z}$,
allowing us to observe interesting effects even in this case. Figure
5 shows curves for the population and modulus of the coherence considering
several values of $\eta$ and $\lambda$, but still with temperature
$T$ and frequency $\omega_{0}$ turned off.

\begin{figure}
\includegraphics[width=0.5\textwidth]{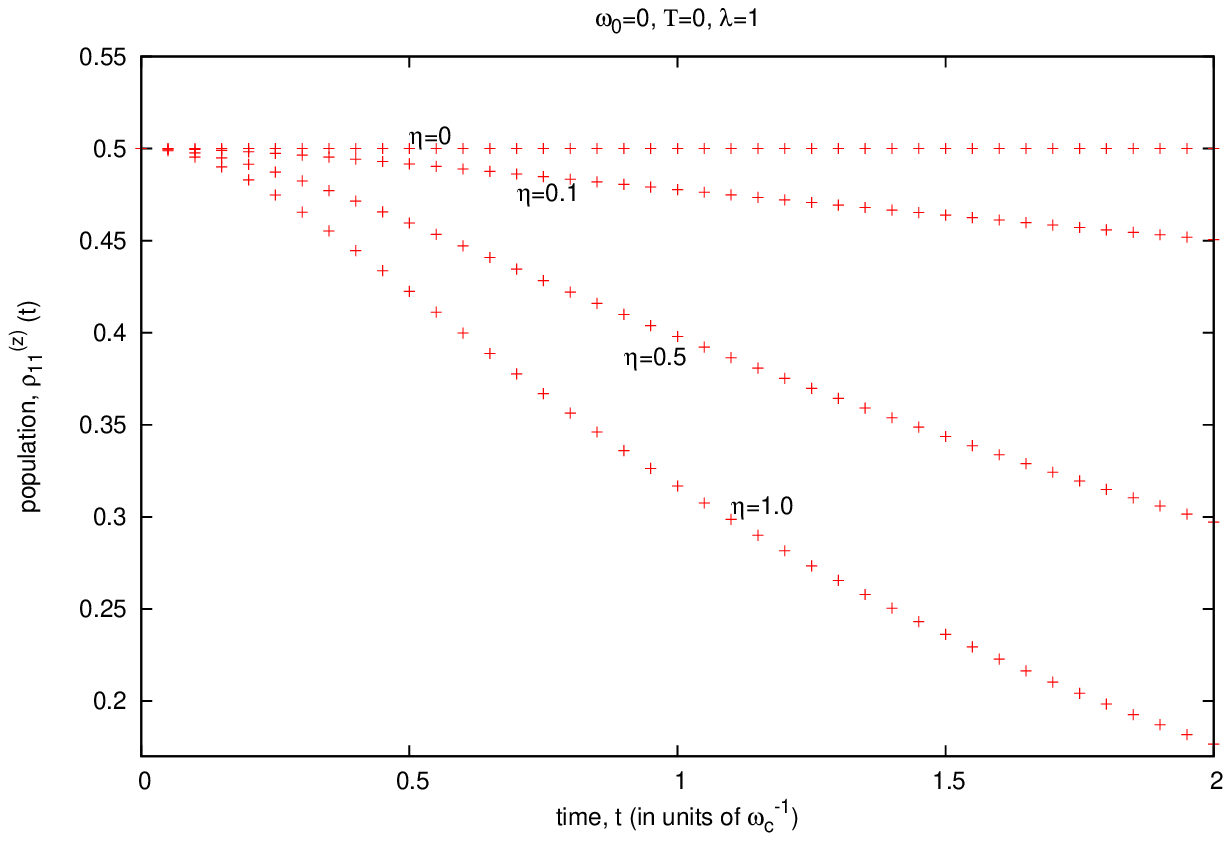}\includegraphics[width=0.5\textwidth]{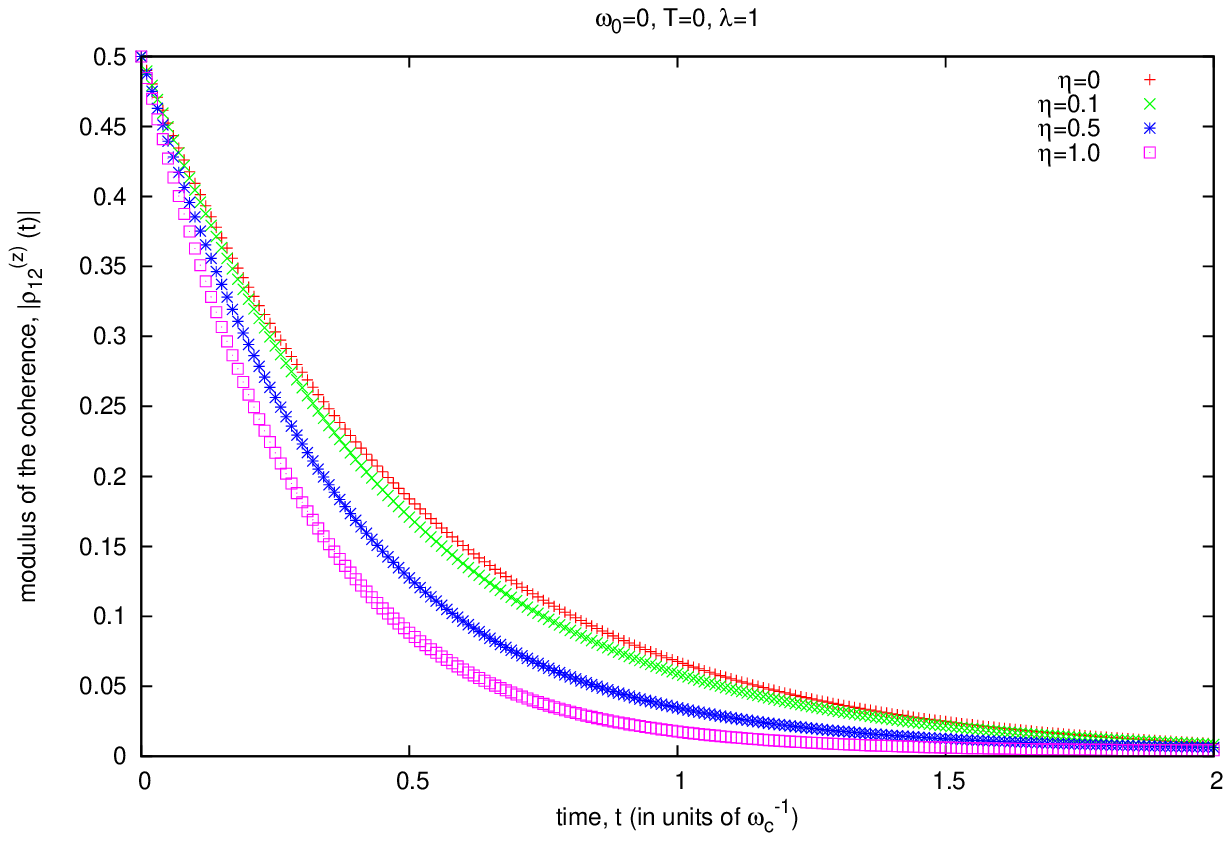}

\includegraphics[width=0.5\textwidth]{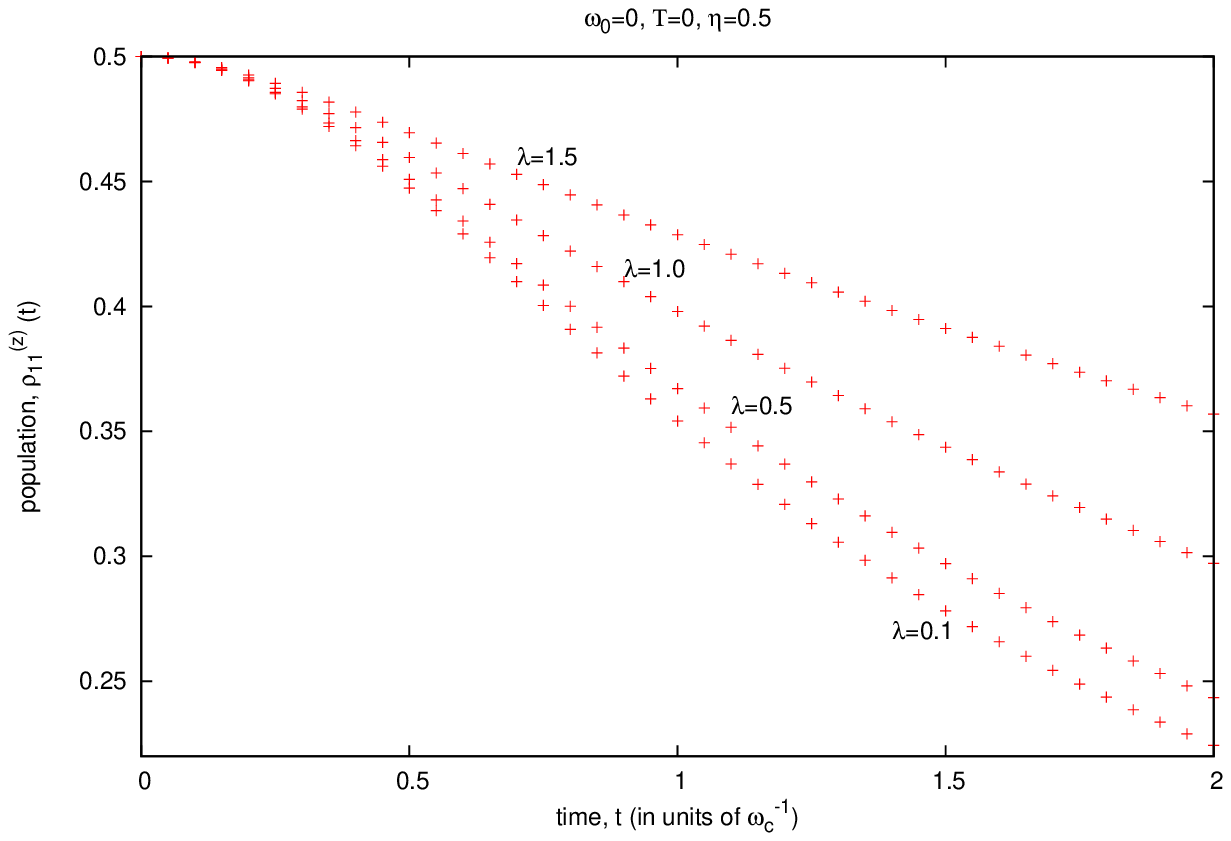}\includegraphics[width=0.5\textwidth]{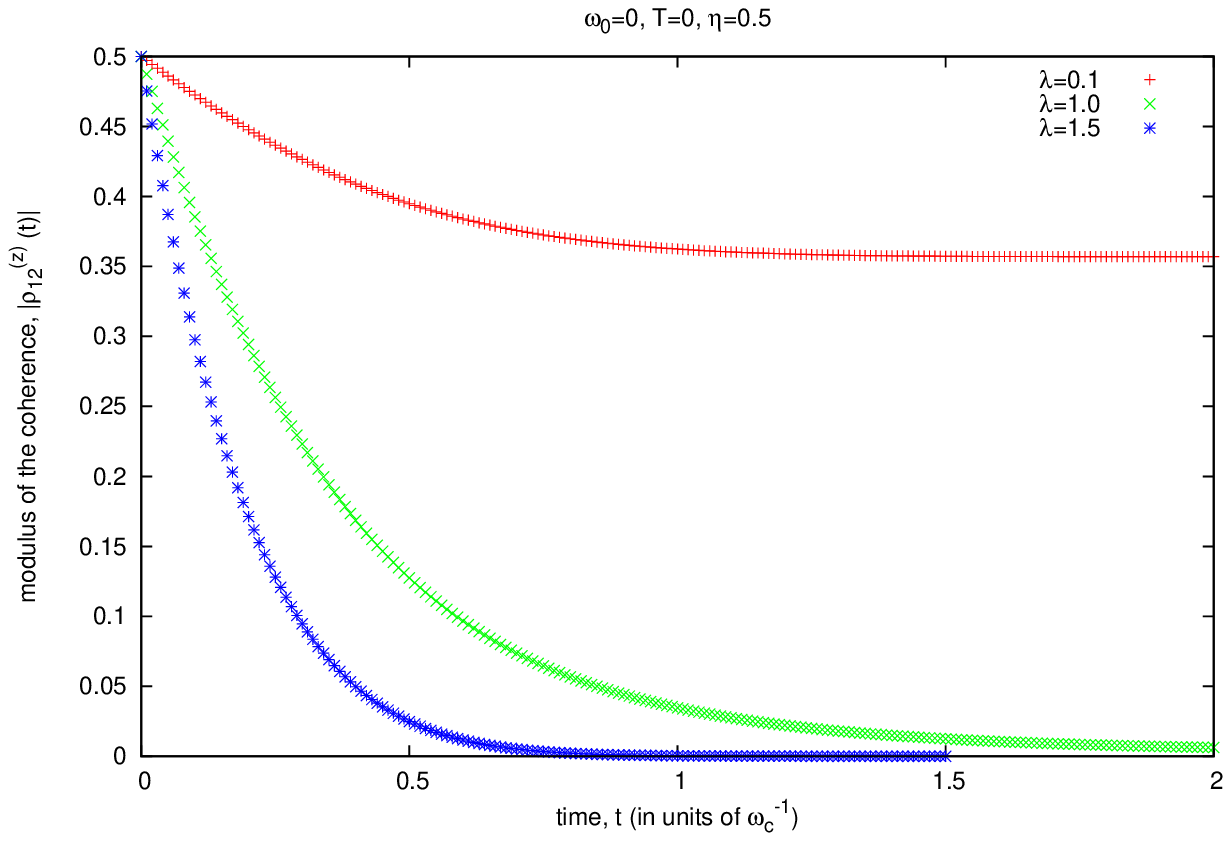}

\caption{Curves for the amplitude-damping interaction, during a measurement
of the $z$-component, considering different values of $\lambda$
and $\eta$, but with $T=\omega_{0}=0$.}
\end{figure}

The strengthening of the system-environment coupling $\eta$ causes
a more intense decrease of the population that can be compensated
by an increase in the system-measurement apparatus coupling $\lambda$.
This causes a reduction of the rate of decay analogous to the phase-damping
interaction with an $x$-component measurement. For the coherence,
the increase of $\eta$ or $\lambda$ causes a more intense decrease
of the modulus. Therefore, the conclusions from \cite{key-23,key-24}
and the previous section of this article remain valid here, despite
our having changed not simply the system-environment interaction but
the measured component too.

Let us introduce the system frequency, $\omega_{0}$. From Fig. 6,
we note this parameter does not change significantly the population
and coherence behavior and, then, we have the same conclusions about
the effects of $\lambda$ and $\eta$.

\begin{figure}
\includegraphics[width=0.5\textwidth]{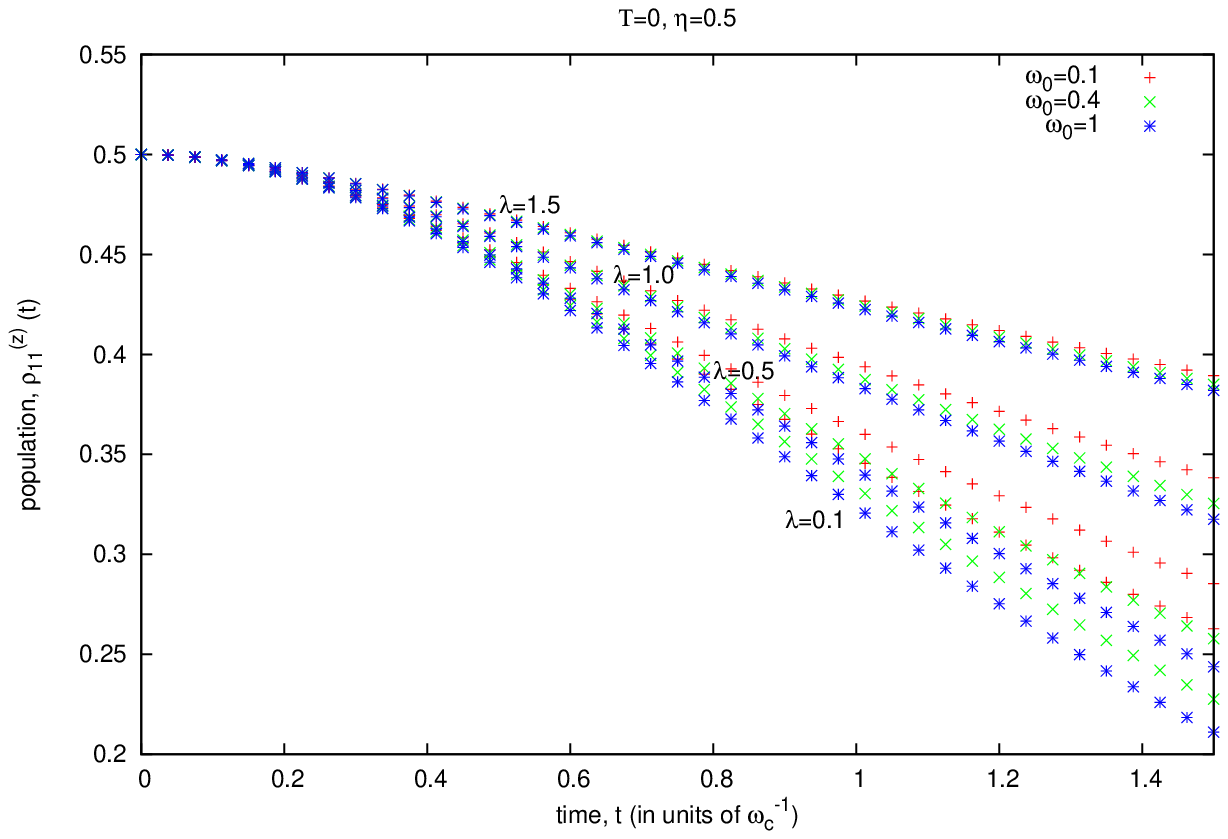}\includegraphics[width=0.5\textwidth]{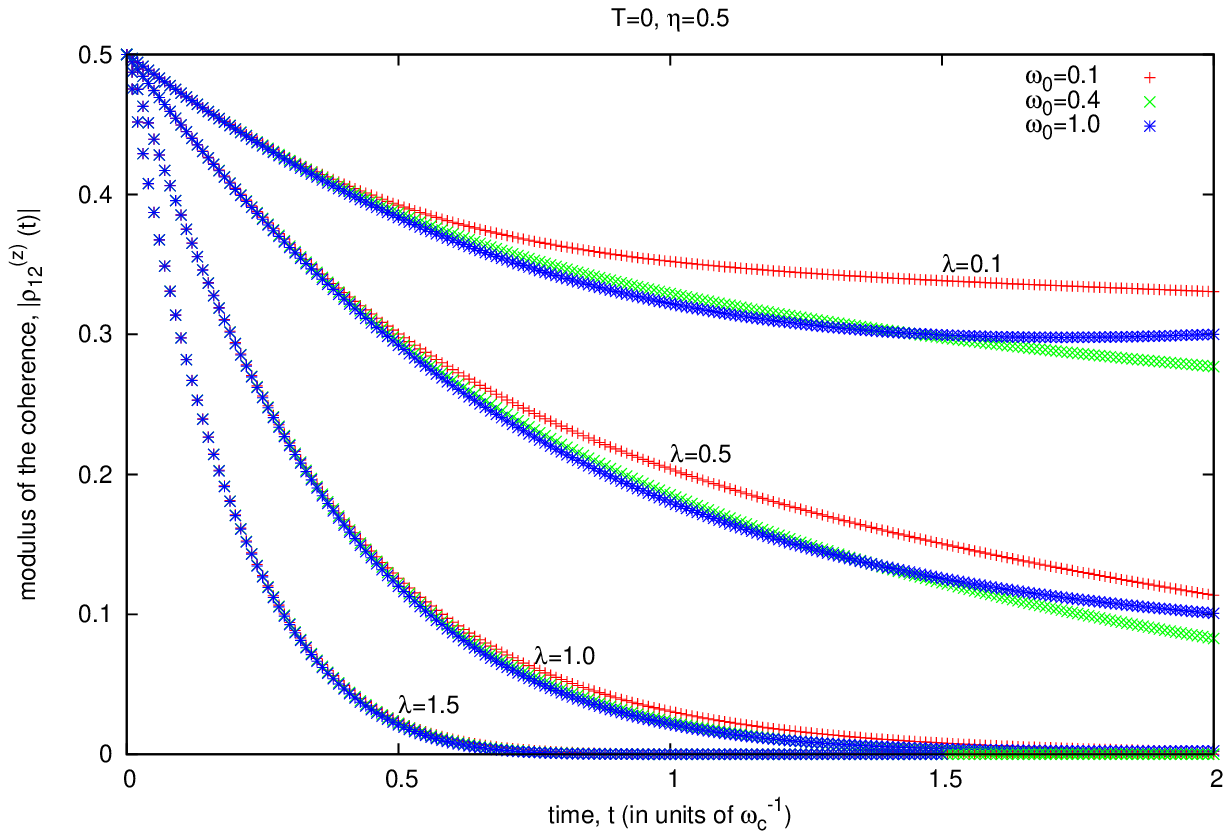}

\caption{Curves for the population and modulus of the coherence considering
different values of $\lambda$, with non-zero $\omega_{0}$.}
\end{figure}

The coherences have their behavior governed mainly by $\lambda$,
which, as always, makes the modulus go faster to zero while keeping
the population closer to its original value. For both variables, a
larger system frequency $\omega_{0}$ results in a more intense decrease.
For the populations, this fact can be easily explained because an
increase in the difference between the energy levels makes repopulation
more difficult. Then, $\omega_{0}$ and $\lambda$ have compensating
effects.

With the introduction of the environmental temperature $T$ (Fig.
7), we note that, as expected, if we keep $T$ constant, we have the
well-known protecting effect of the finite-time measurement. On the
other hand, the measurement apparatus has its already-observed effect
of increasing the reduction rate of the coherences.

\begin{figure}
\includegraphics[width=0.5\textwidth]{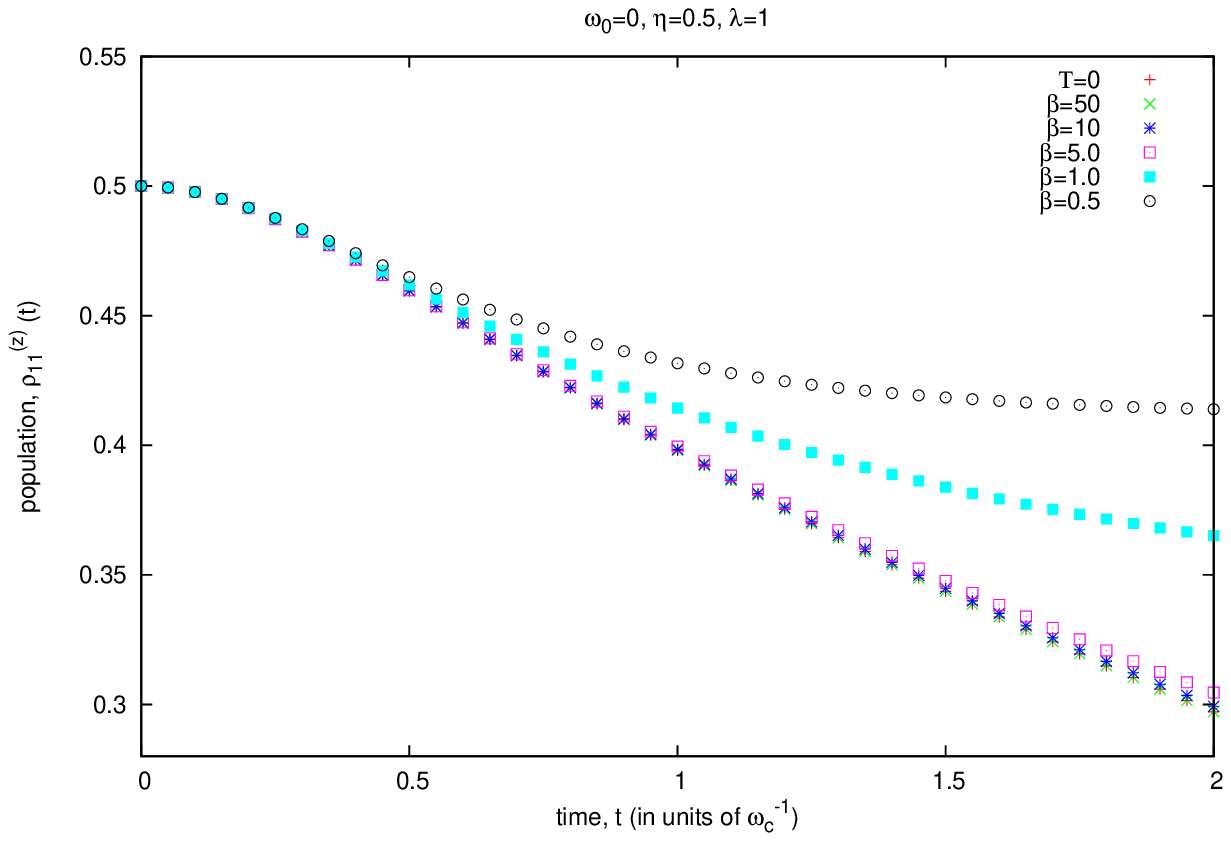}\includegraphics[width=0.5\textwidth]{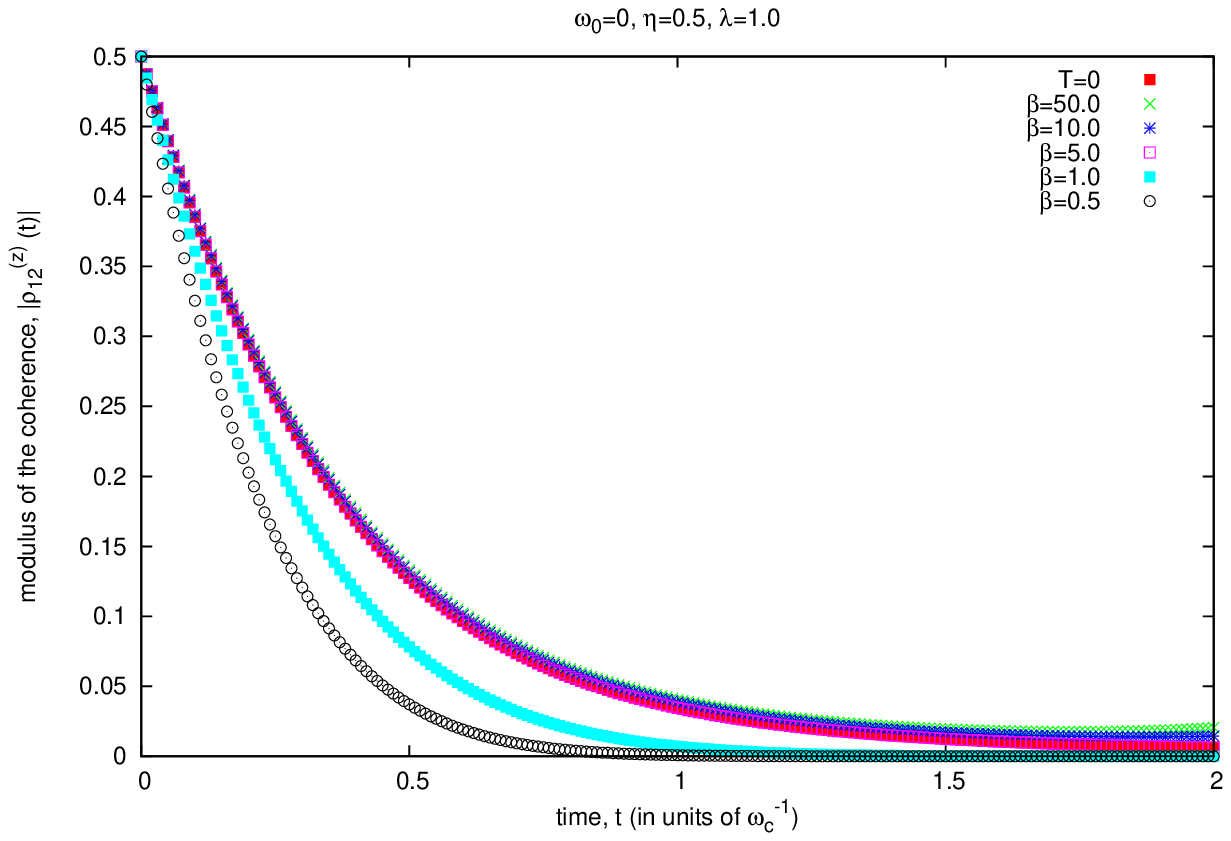}

\includegraphics[width=0.5\textwidth]{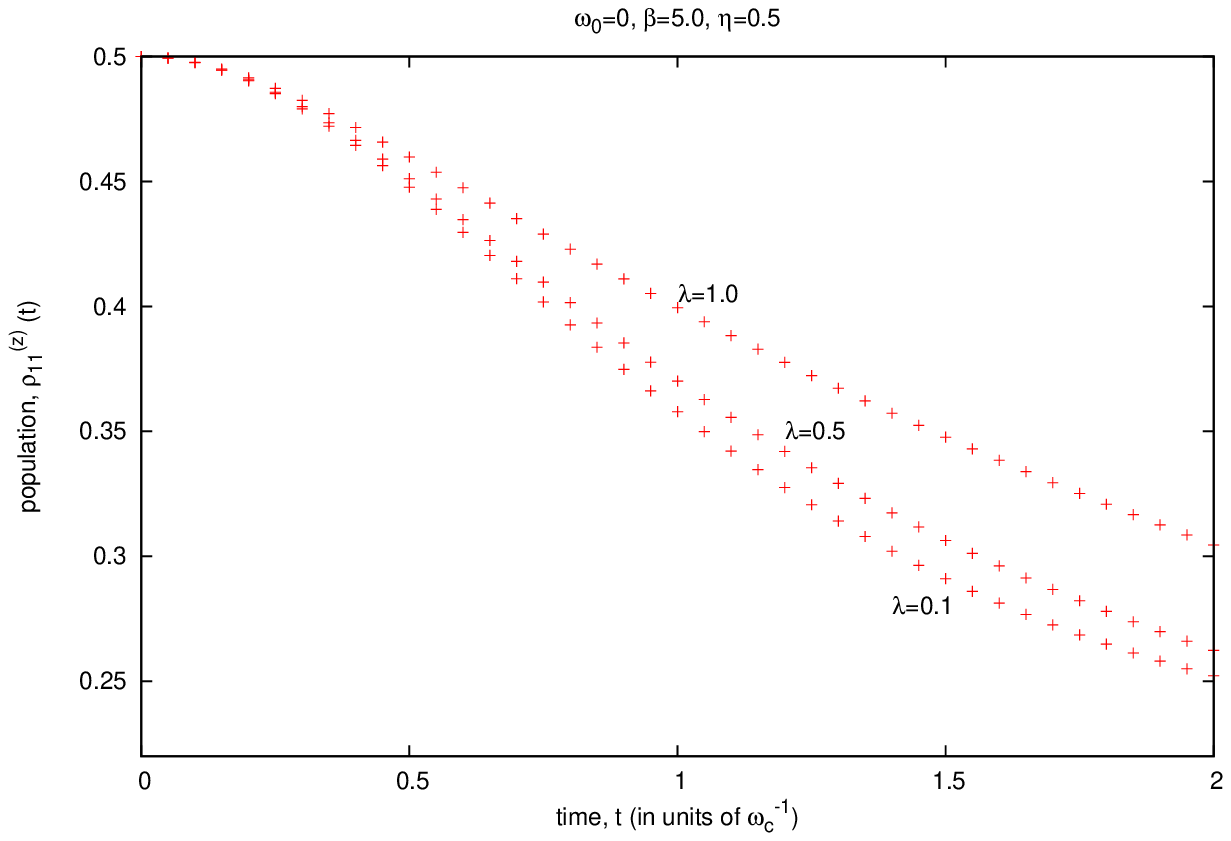}\includegraphics[width=0.5\textwidth]{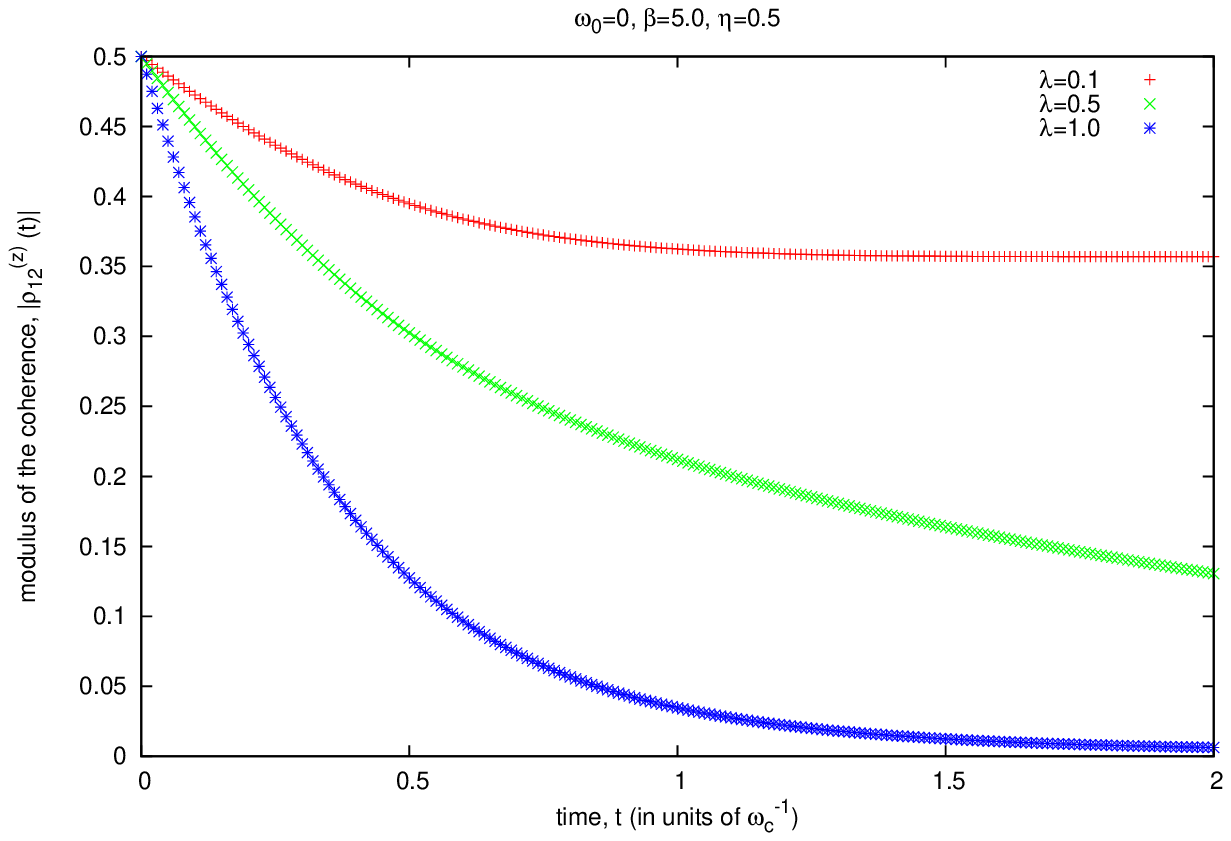}

\caption{Introduction of temperature of the environment for the amplitude-damping
interaction.}
\end{figure}

The coherence has its rate of decay sharpened as the environment temperature
rises. From the graphs, we can also see that as the environmental
temperature rises the population has its decrease rate attenuated,
i.e., for a higher temperature, we have a lower measurement error.
Hence we see that the environment can have contradictory effects over
the measuring process: the coupling intensity $\eta$ acts to increase
the error associated to the measurement process and decrease the measurement
time; however, the temperature $T$ acts to reinforce the effects
of the system-measurement apparatus coupling $\lambda$, preventing
the decay and increasing the measurement time.

This curious effect is a fortuitous coincidence caused by our choice
of $\omega_{0}$ and initial state. With $\omega_{0}=0$, the thermal
energy from the environment will eventually lead to a state of thermal
equilibrium between the two systems that will stabilize both populations
at the same value. Starting at $\rho_{11}\left(0\right)=0.5$, our
initial state is already the expected final state. An increase in
temperature, in this case, only makes faster the approach of the equilibrium
situation.

\begin{figure}
\includegraphics{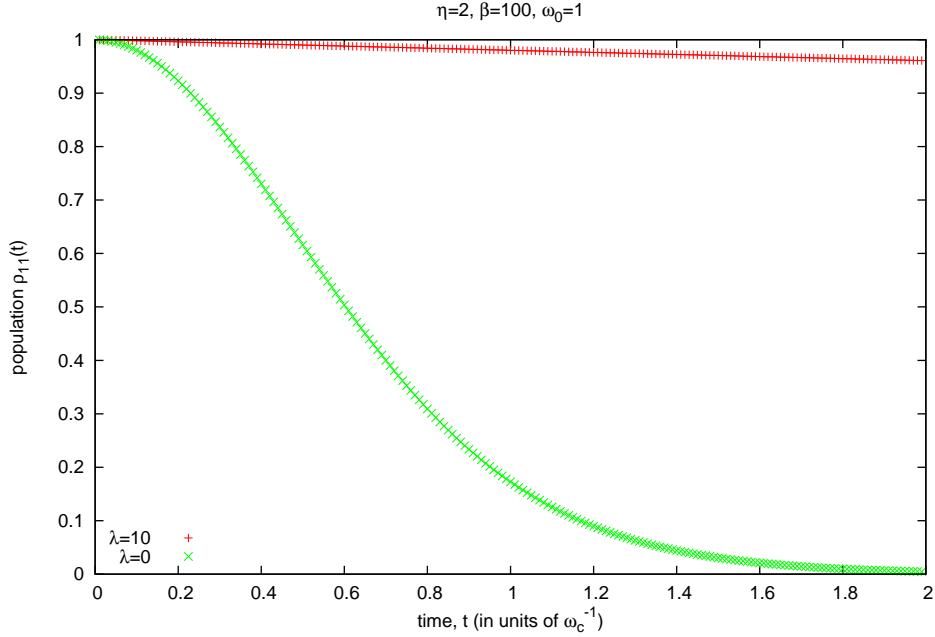}

\caption{Attenuated decay caused by the finite-time measurement of the excited
state of the system. In green, the curve corresponding to the free
evolution of the population; in red, the one indicating its evolution
during measurement.}
\end{figure}

The maintenance of the system in the excited state, however, can be
verified even with the $\omega_{0}$ and $T$ parameters above zero.
As illustrated in Fig. 8, we can, with a strong enough choice of the
measurement coupling $\lambda$, keep an ensemble of initially excited
systems as close as possible to their original states even at times
when they would normally have completely decayed by themselves. This
is not simply a more efficient way of performing a measurement of
their initial state, but may be a means of observing the quantum Zeno
effect \cite{key-1} (i. e., keeping a system in its excited state
by observing it) for finite-time measurements.

Finally, Fig. 9 shows curves for non-zero environmental temperature,
considering constant system-environment coupling and changing the
frequency of system and its coupling with the measurement apparatus.
We see that the introduction of $\omega_{0}$ does not change the
evolution of the modulus of the coherences, but intensifies the population
decrease for a fixed $\lambda$.

\begin{figure}
\includegraphics[width=0.5\textwidth]{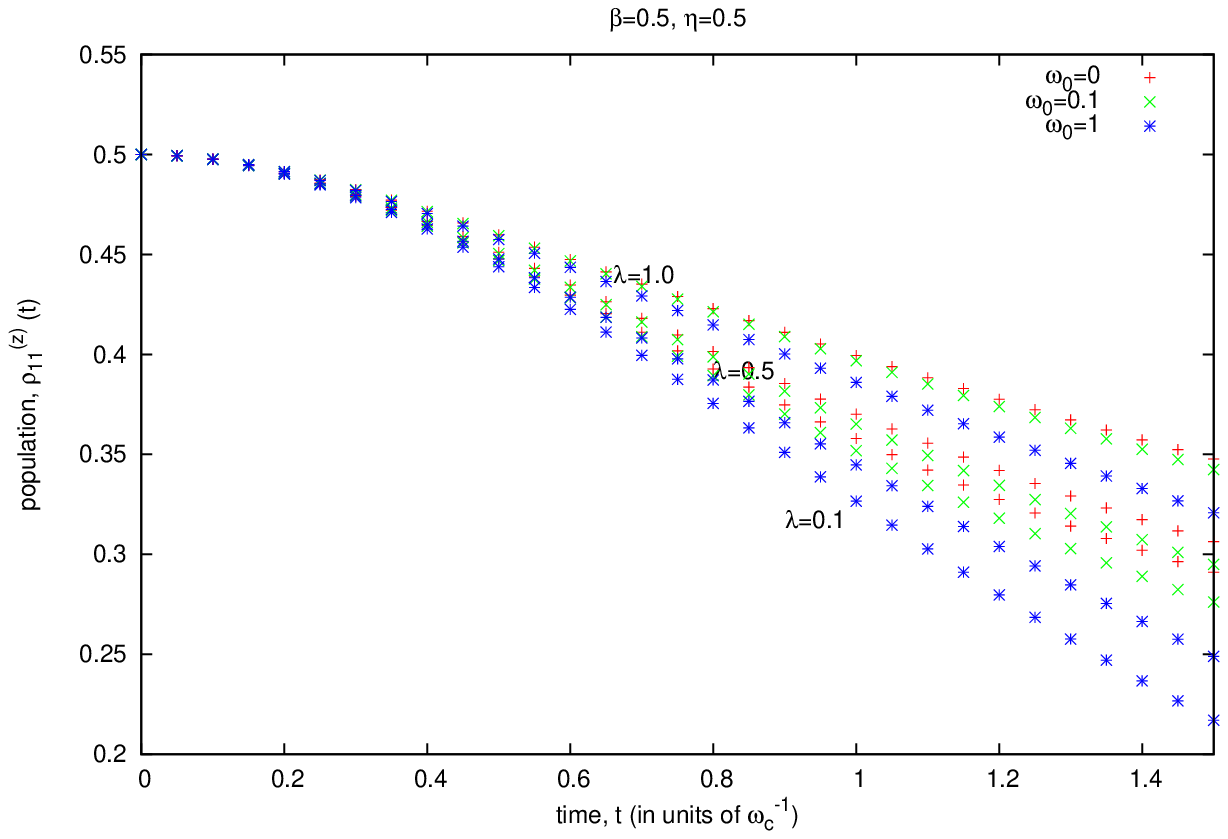}\includegraphics[width=0.5\textwidth]{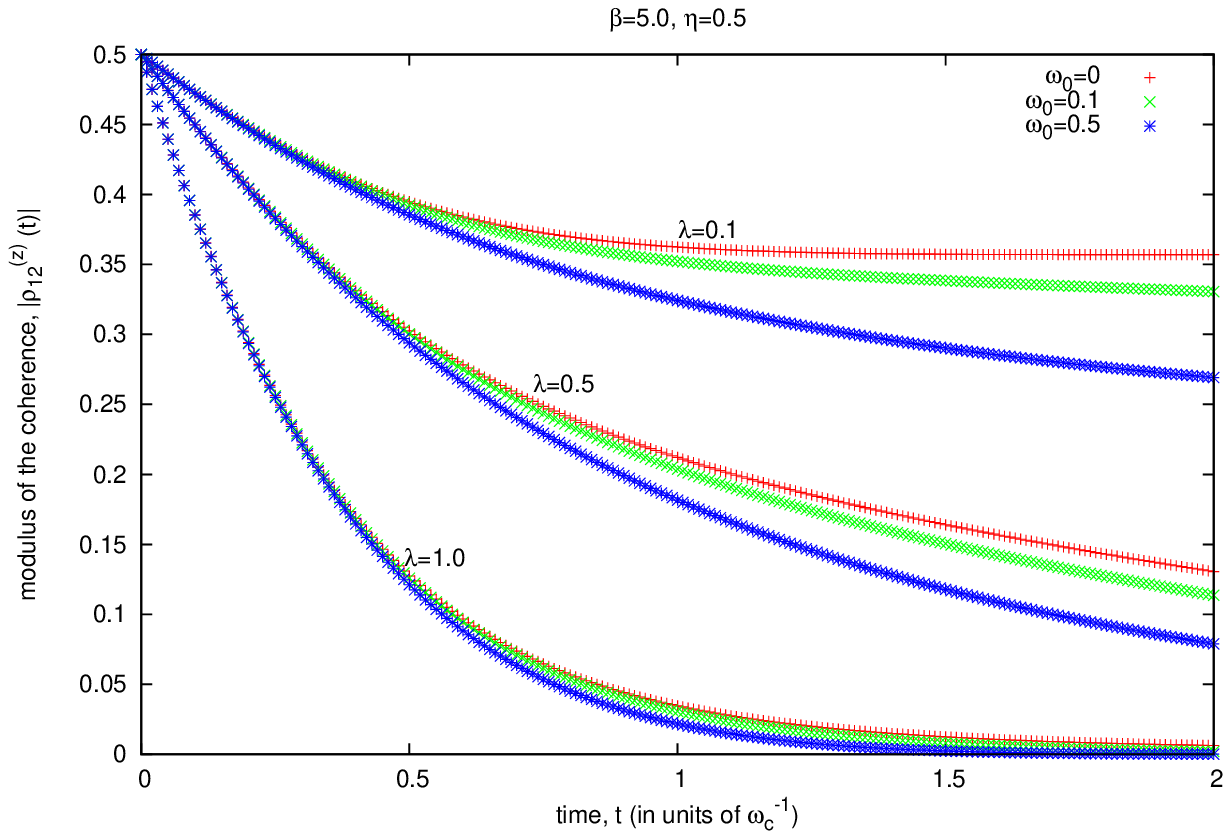}

\caption{Populations and modulus of the coherences, and their dependence on
$\lambda$ and $T$.}
\end{figure}

Contrary to the case with phase-damping interaction, here we verified
some curious effects when we change the system-environment interaction
to amplitude-damping: 
\begin{itemize}
\item both system-measurement apparatus coupling $\lambda$ and environment
temperature $T$ can have the effect of preserving the system in the
excited state; the first effect is universal, while the second only
occurs for certain values of the initial population and temperature,
and is a direct consequence of the repopulation of the excited state
caused by the thermal energy of the environment; 
\item both the system frequency $\omega_{0}$ and the system-environment
coupling $\eta$ intensify the decrease of the population; 
\item in general, the conclusions obtained by the measurement of the $x$-component
with phase-damping interaction are valid in the present case, except,
perhaps, the formula for the upper limit of the measurement duration. 
\end{itemize}

\subsubsection{The x-component measurement}

Figure 9 shows population and coherence curves for several values
of $\lambda$ and $\eta$, while keeping $\omega_{0}=T=0$. As previously
observed, the system-environment interaction acts in the sense of
accentuating the population decrease and, consequently, the error
associated to the measurement process. However, the measurement apparatus
still keeps the populations close to their original values.

\begin{figure}
\includegraphics[width=0.5\textwidth]{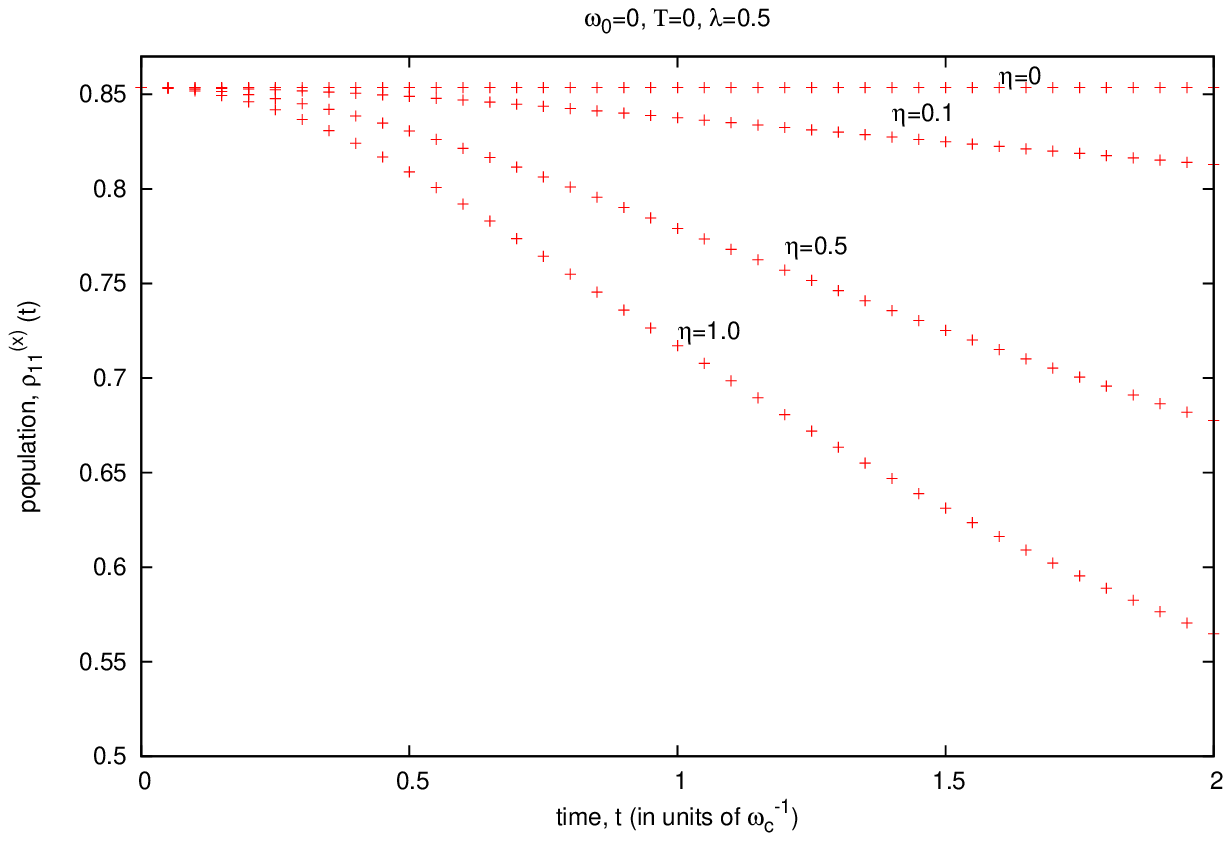}\includegraphics[width=0.5\textwidth]{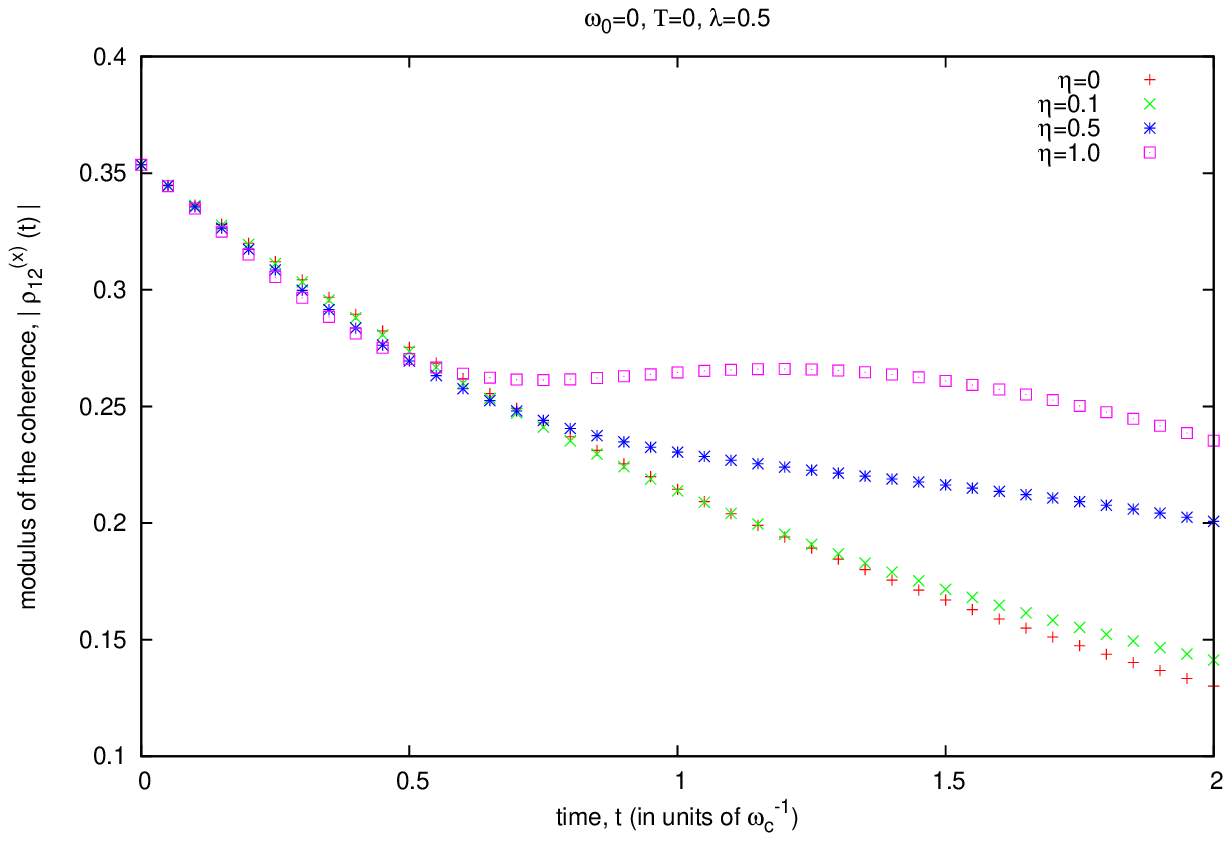}

\includegraphics[width=0.5\textwidth]{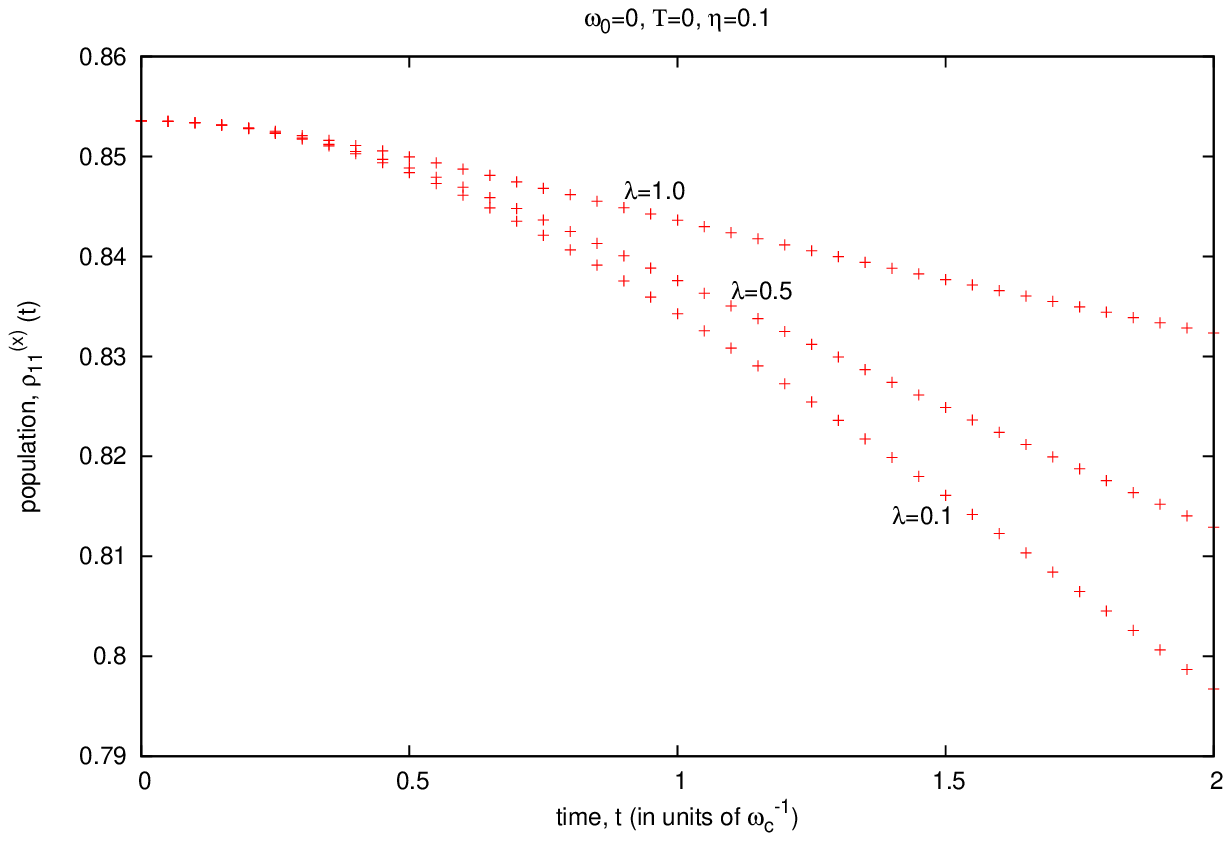}\includegraphics[width=0.5\textwidth]{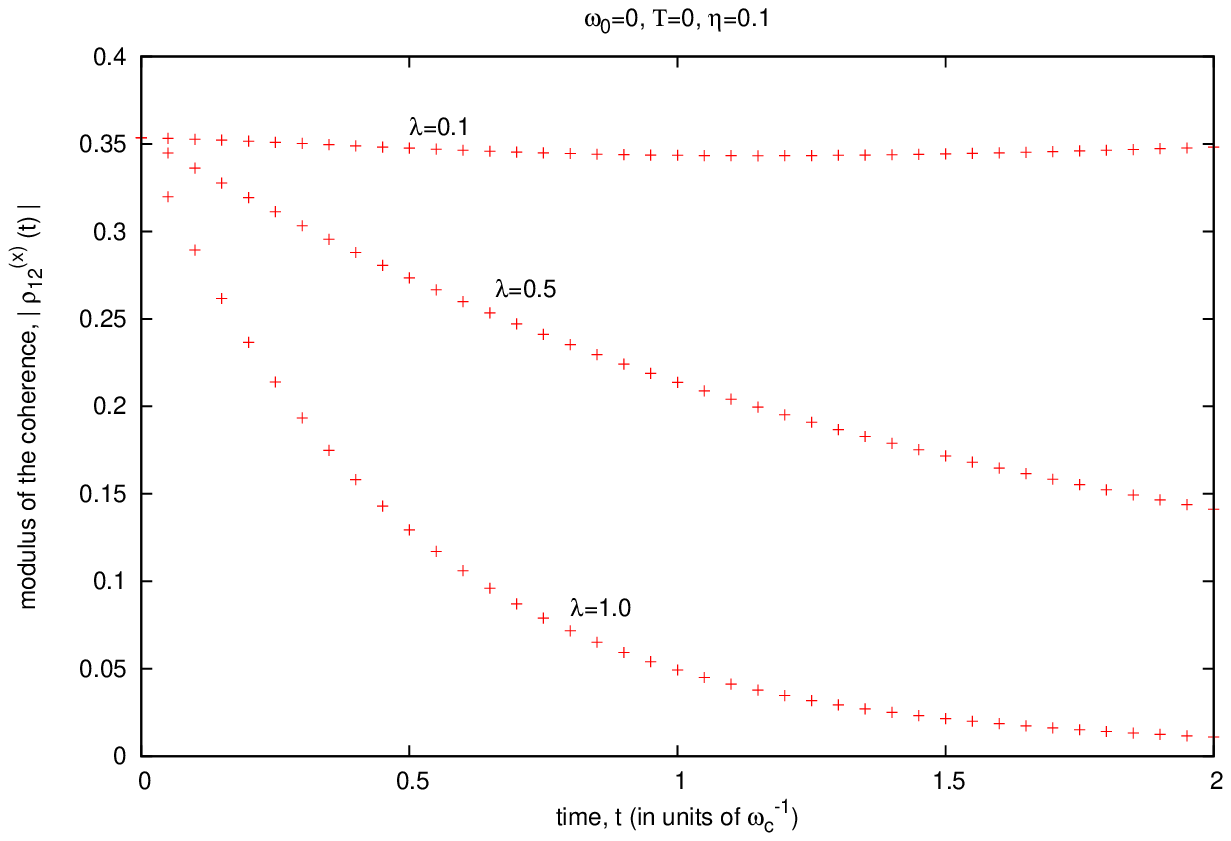}

\caption{Populations and coherences for two values of $\lambda$ and different
values of $\eta$.}
\end{figure}

Although we cannot, in this case, obtain a simple expression for the
duration of the measurement, we still use the coherences to establish
the time at which the system-environment interaction can be interrupted
and the reading of the state accomplished - in this case, the time
when the coherences assume the constant asymptotic value. Obviously,
in more general cases, when the behavior of the system is not simple
(the coherences no longer display a monotonic decrease), its physical
properties must be considered to the establishment of the measurement
time. In these non-monotonous cases, there are at least two possible
criteria that could be taken to determine the end of the measurement:
the instant when the modulus of the coherences reaches zero, or when
it assumes an asymptotically constant value.

The next step is to introduce a system frequency $\omega_{0}$. In
this case, we observe a similarity with the behavior of the measurement
of the $x$-component for phase-damping interaction: the populations
do not exhibit monotonic behavior anymore, instead displaying a maximum
point. As the system-environment coupling increases, both its intensity
and the time when it occurs decreases. On the other hand, increasing
the system-measurement apparatus coupling diminishes the maximum intensity.

Figure 11 shows that the element that determines the qualitative format
of the curves is the frequency of the system $\omega_{0}$. One should
note, while observing these results, that all of them were plotted
in the basis of the measurement, that is, $\hat{\sigma}_{x}$. These
populations depend on the real part of the coherences in the basis
of $\hat{\sigma}_{z}$, which is expected to oscillate when $\omega_{0}\ne0$.
For example, in the case of a spin in a $z$-oriented field, the expectation
values of $\hat{\sigma}_{x}$ and $\hat{\sigma}_{y}$ change while
the spin rotates around the $z$-axis. This explains the non-monotonous
behavior of the populations, a phenomenon observed previously in Fig.
2(a).

\begin{figure}
\includegraphics[width=0.5\textwidth]{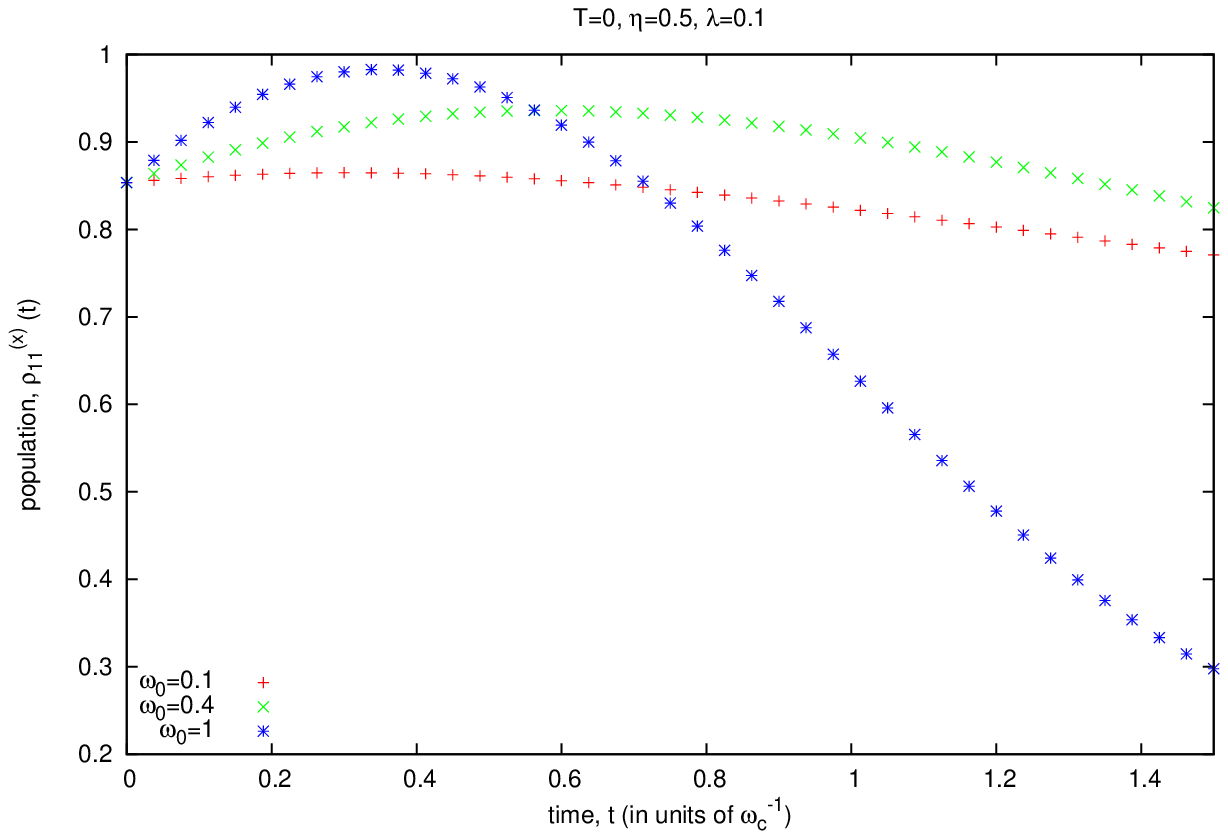}\includegraphics[width=0.5\textwidth]{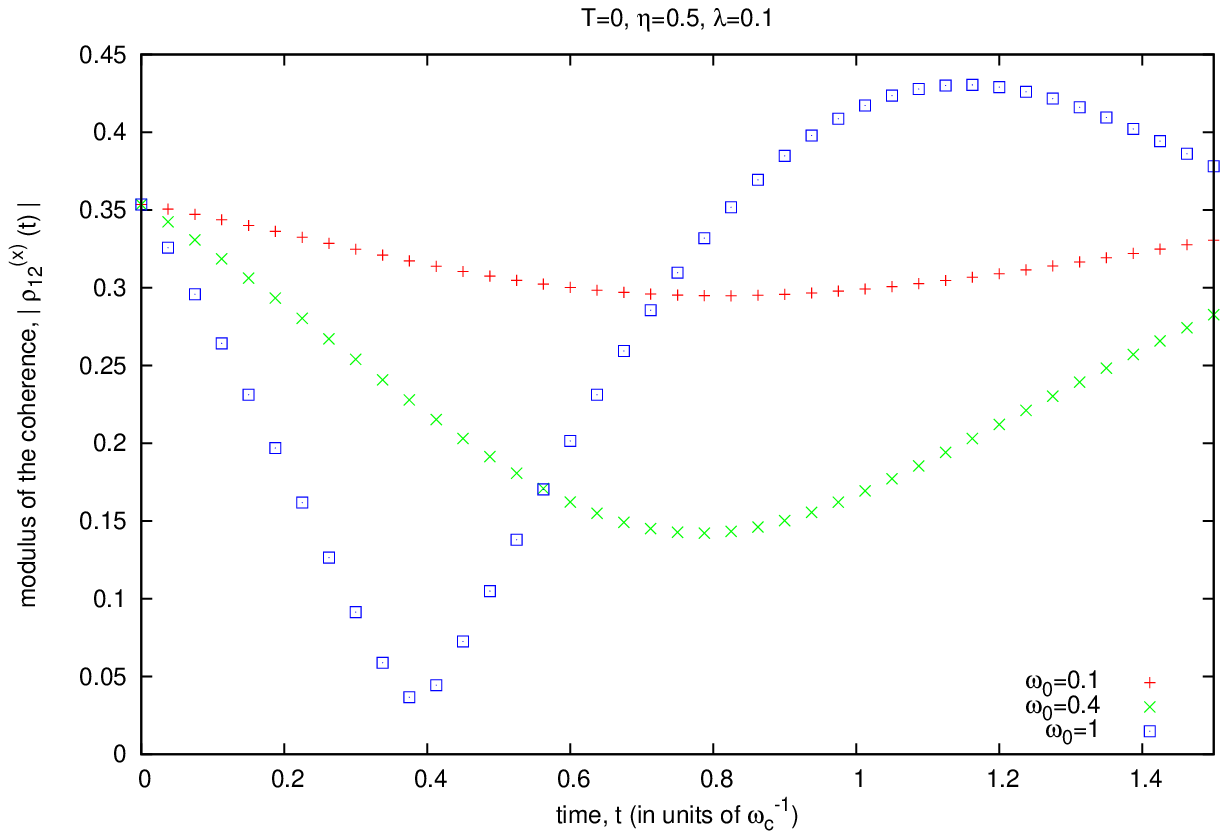}

\includegraphics[width=0.5\textwidth]{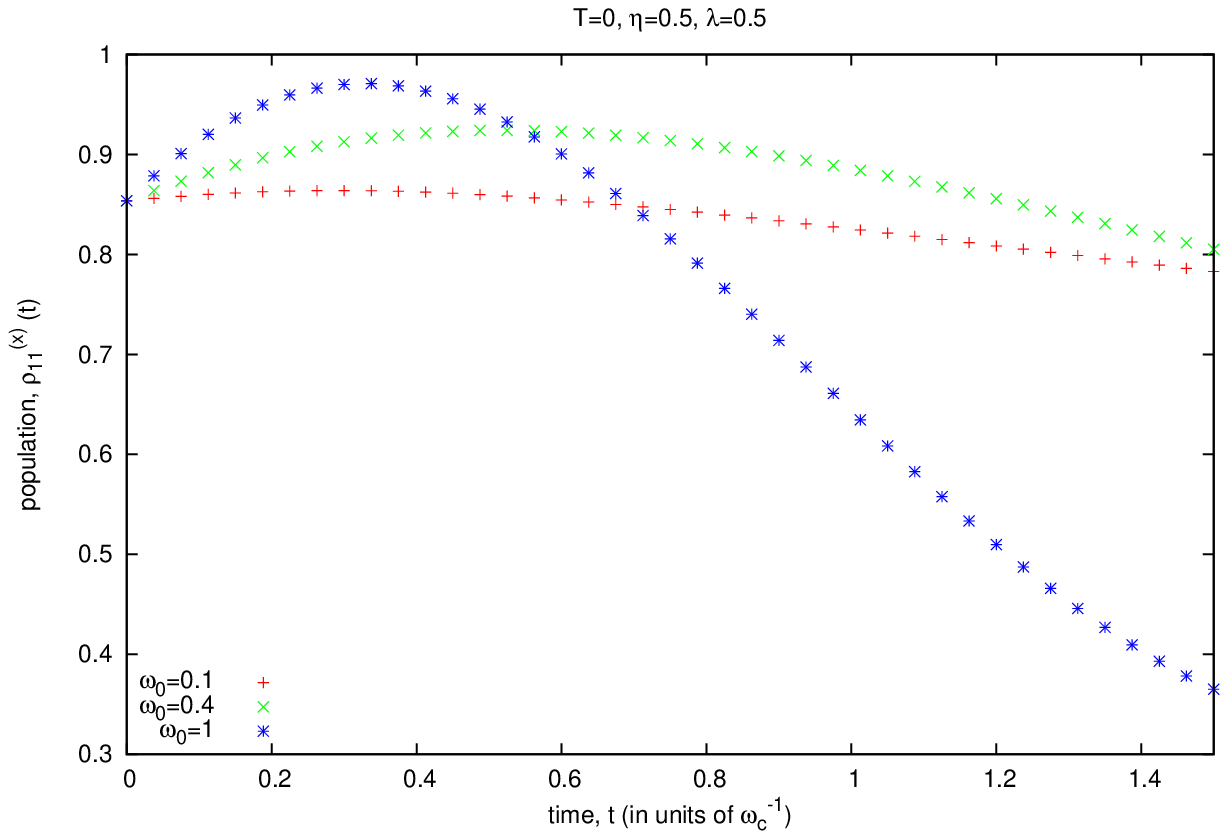}\includegraphics[width=0.5\textwidth]{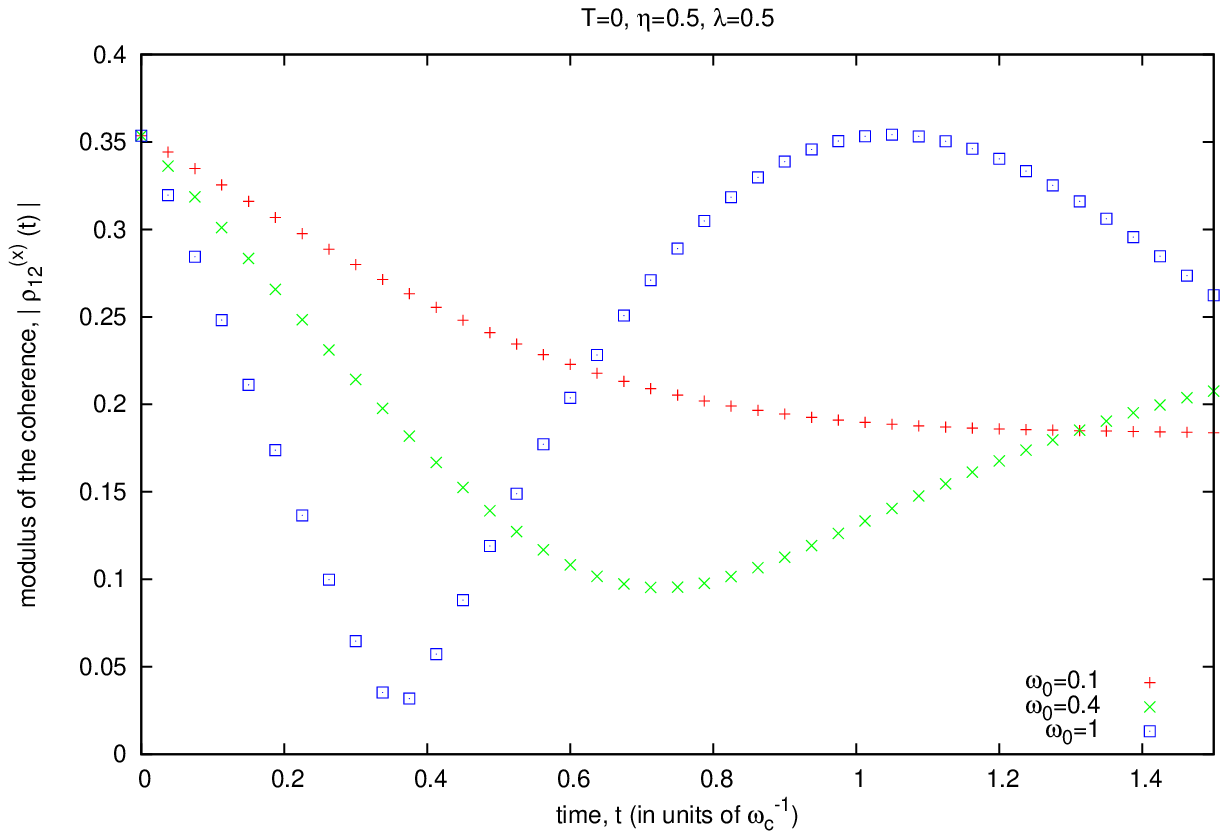}

\caption{Populations and coherences obtained for a fixed $\eta$, varying $\lambda$
and $\omega_{0}$.}
\end{figure}

Lastly, we analyse the effects of the environmental temperature. For
the present case, contrary to the previous section, we verified that,
for a larger environmental temperature, the rate of change of the
populations is increased. Then, the environment - through $T$ - acts
to increase the measurement error, reinforcing the fact that the temperature-originated
protection found in the previous section was merely fortuitous.

Figure 12 shows several curves for different environment temperatures
and $\omega_{0}=0$ and, after that, curves where all relevant parameters
are present. Once again, the qualitative behavior is determined by
$\omega_{0}$.

\begin{figure}
\includegraphics[width=0.5\textwidth]{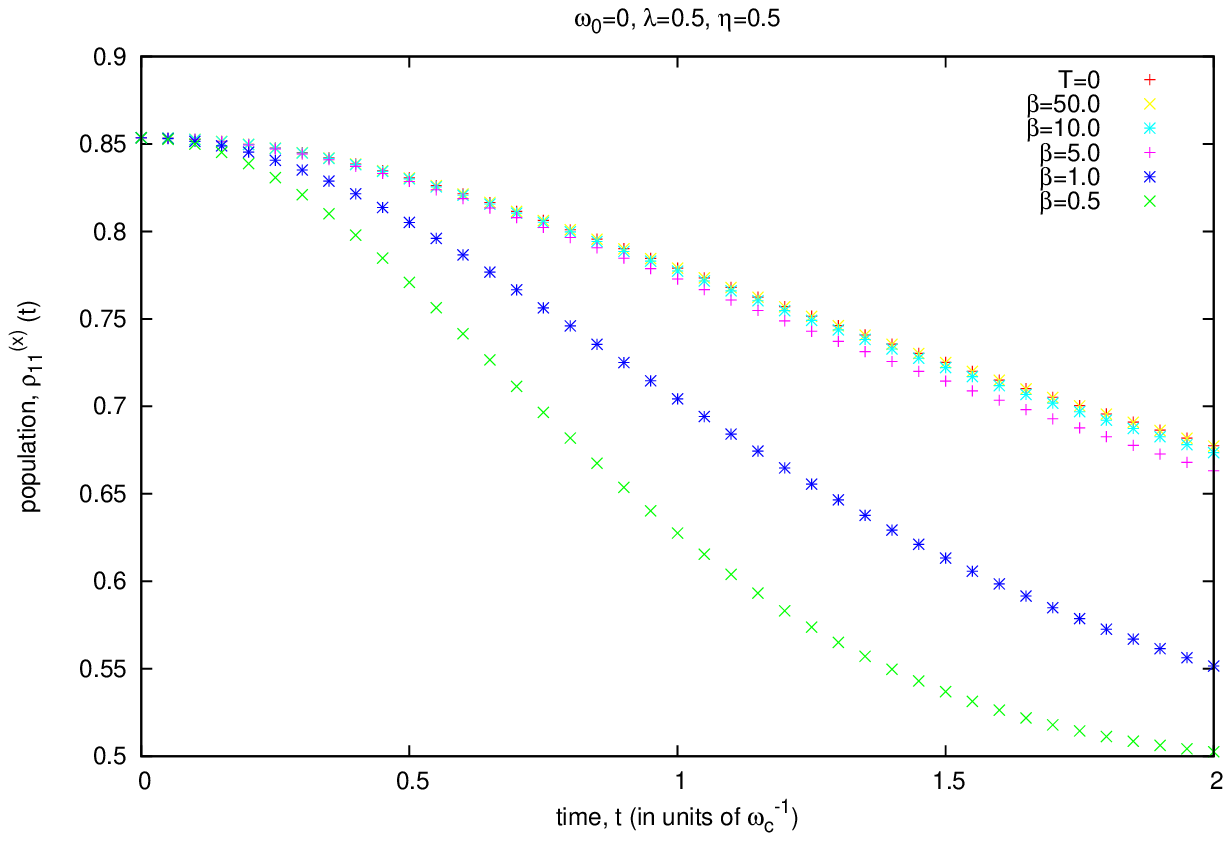}\includegraphics[width=0.5\textwidth]{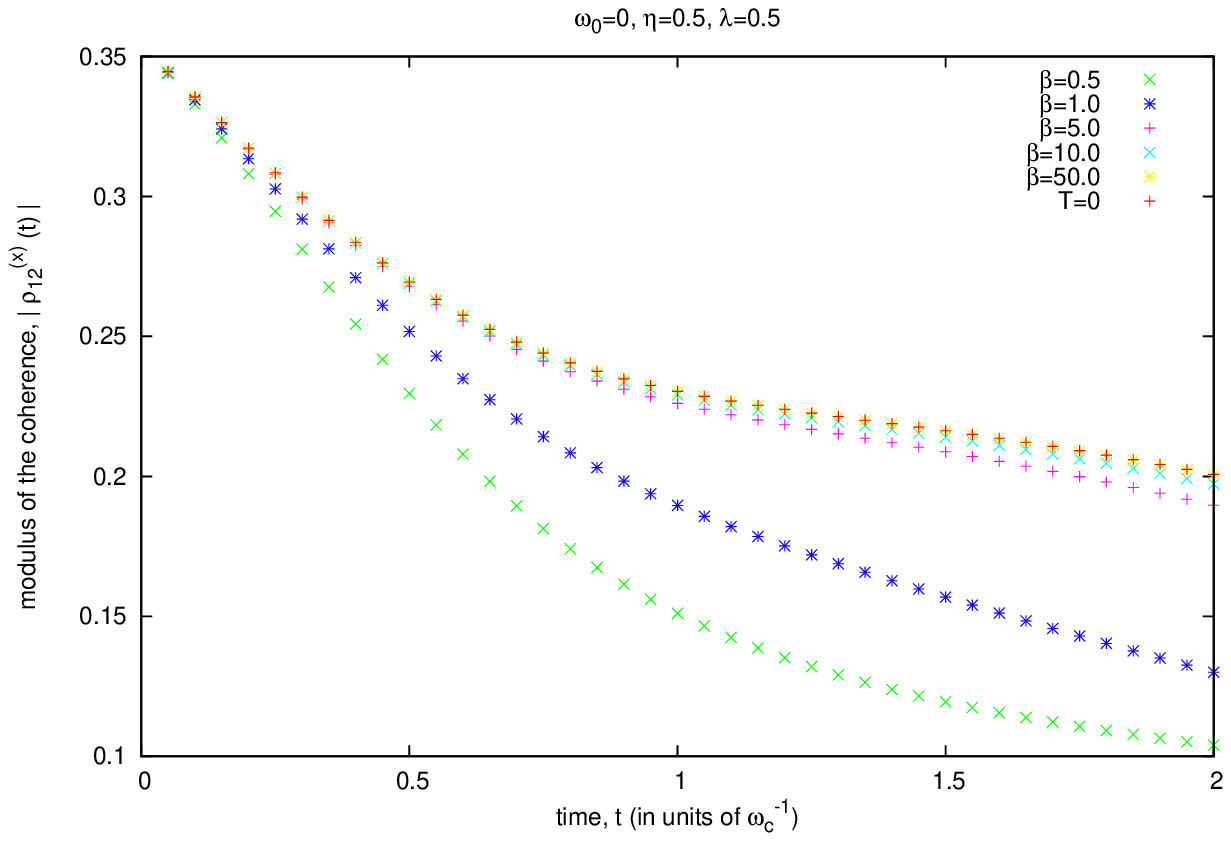}

\includegraphics[width=0.5\textwidth]{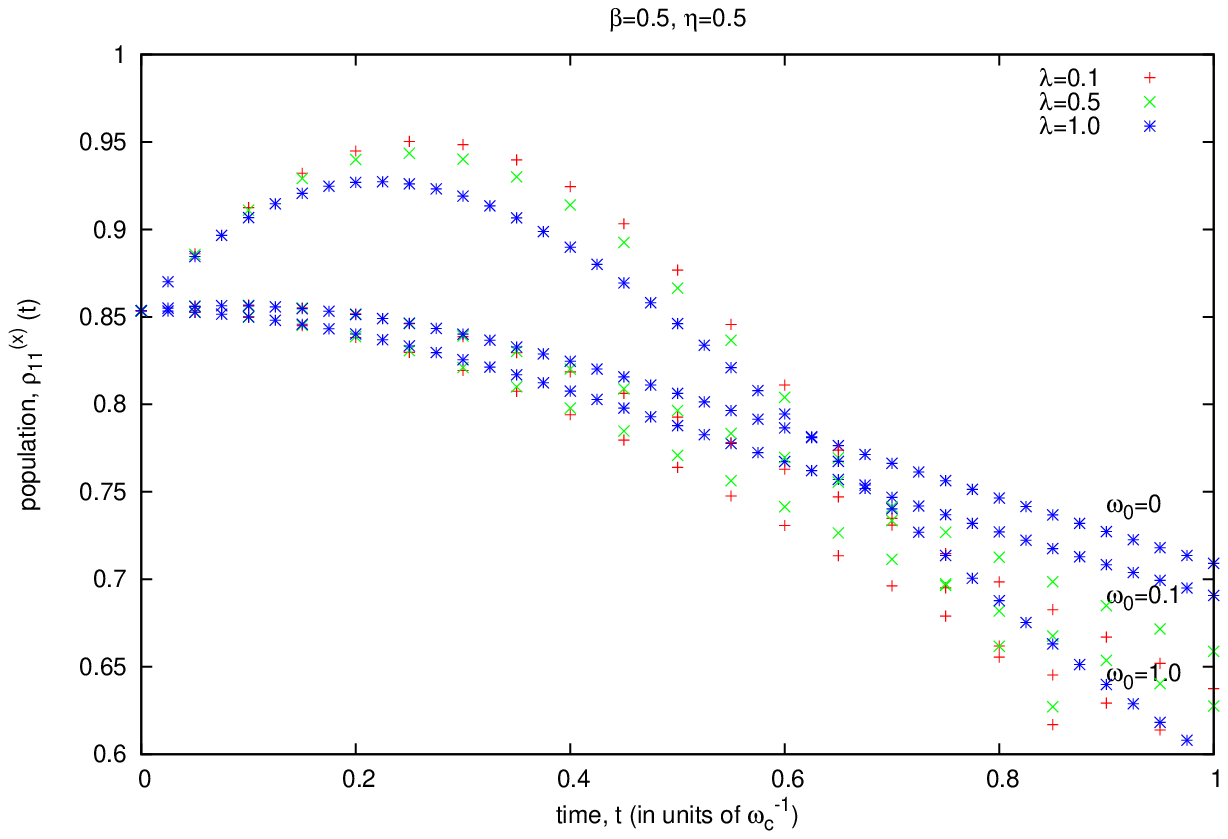}\includegraphics[width=0.5\textwidth]{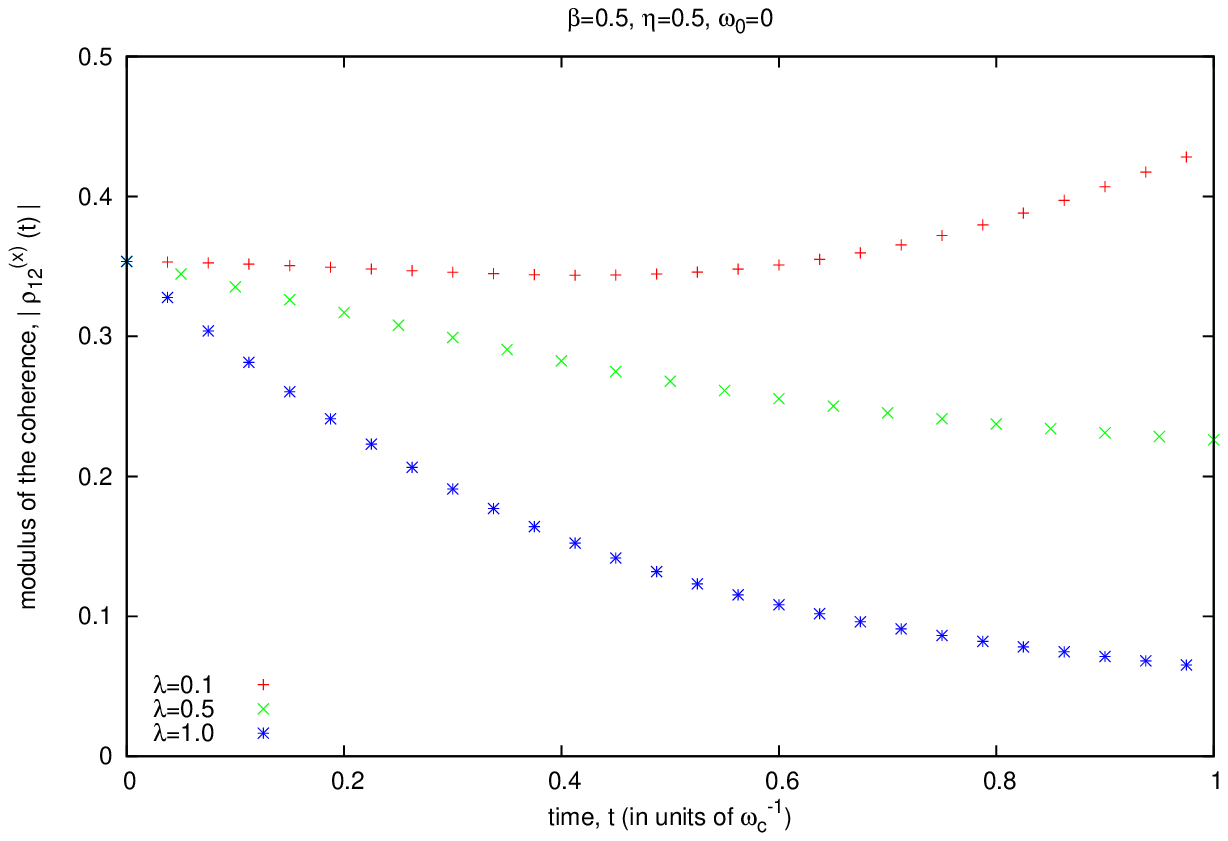}

\includegraphics[width=0.5\textwidth]{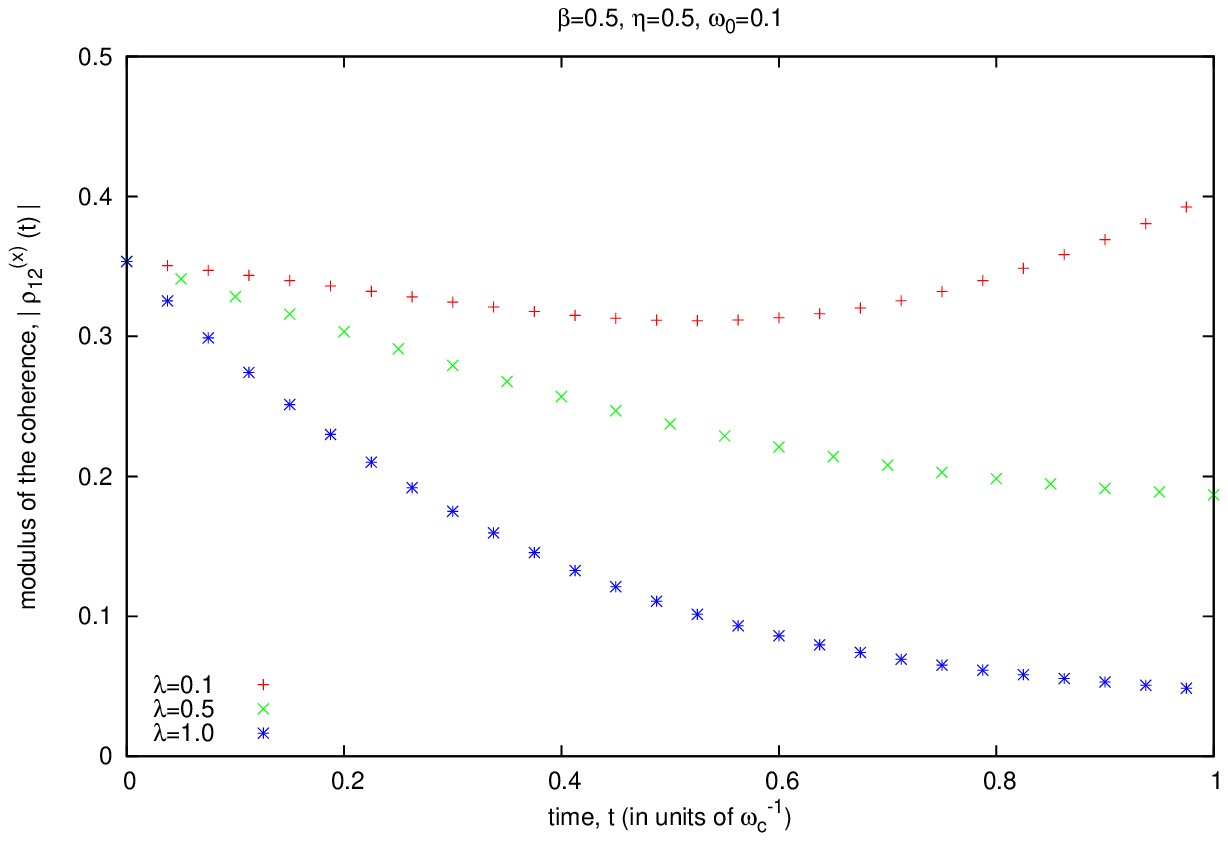}\includegraphics[width=0.5\textwidth]{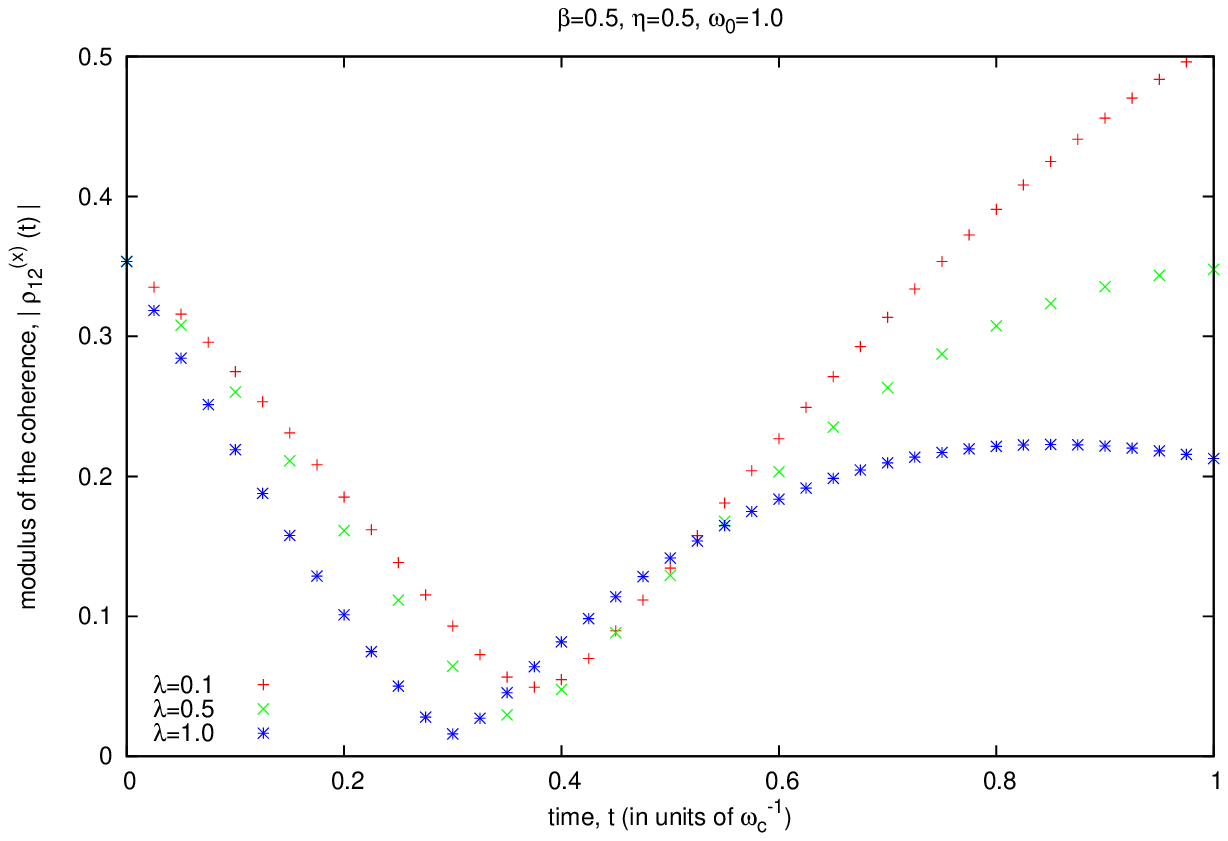}

\caption{Effect of the introduction of $T$, followed by the introduction of
$T$ and $\omega_{0}$. The last three graphs contain the coherences
corresponding to the populations plotted in the third one. To avoid
confusion caused by the superposition of different curves, they were
broken in three.}
\end{figure}

\section{Conclusions and Perspectives}

In this work, we employed the formalism that we developed to treat
the hybrid formulation of the master equation \cite{key-20}, which
helped us to simplify the analytic solution, although the system of
equations encountered still had to be solved numerically. In both
cases, phase and amplitude damping, the interaction between the system
and the measurement apparatus diminishes the effects of the noise
on the population of the main system, allowing us to generalize our
previous conclusions in this sense \cite{key-23}. This means that
in both cases we are capable of performing a measurement more efficiently
if instead of measuring it instantly at $t>0$, we perform a finite-time
measurement from $0$ to $t$, even at finite temperature. Likewise,
for finite-time measurements, an increase in the system-apparatus
coupling constant $\lambda$ is capable of making their results more
reliable.

The possibility of application of this kind of measurement to improve
the quality of the results depends, however, on the practical existence
of measurements that can be performed for a finite period of time,
in accordance to the model. In the following appendix, we propose
a verification of our hypothesis by a simple experimental setting
that, if yielding positive results, would not only affect fundamental
aspects of quantum theory but, also, other areas such as quantum information
theory \cite{key-26}.

Besides the assumption that measurements are finite in time, we took
the additional step of approximating the system-apparatus interaction
as a Markovian process. Even though having the advantage of simplifying
our calculations, this is a hypothesis far less straightforward to
justify than the finite-time measurement. Future prospects of our
research include eliminating this approximation by letting the $\lambda^{2}$
vary in time\cite{QZE_Turku} and seeing the effects of this in the
noisy measurement process.

\section*{Acknowledgments}

C. A. Brasil acknowledges support from Coordenação de Aperfeiçoamento
de Pessoal de Nível Superior (CAPES) and Fundação de Amparo à Pesquisa
do Estado de São Paulo (FAPESP), project number 2011/19848-4, Brazil,
and to A. O. Caldeira for hospitality and useful discussions.

L. A. de Castro acknowledges support from Coordenação de Aperfeiçoamento
de Pessoal de Nível Superior (CAPES).

R. d. J. Napolitano acknowledges support from Conselho Nacional de
Desenvolvimento Científico e Tecnológico (CNPq), Brazil.

\section*{Appendix: Proposed Experimental Test}

Beyond what has been shown in \cite{key-28}, here we propose an alternative
experimental application of our formalism.\textcolor{red}{{} }Throughout
this and previous articles, we made the bold assumption that the measurement
can ultimately be described as a finite-time interaction between the
system and a large number of (traced-out) degrees of freedom of the
apparatus. Moreover, we assumed that this interaction could be described
as Markovian. The accuracy of these assumptions is far from uncontroversial,
but could be verified in the experimental test described below, which
is based on the theory of weak measurements. \cite{WeakMeasurements}
We should note, before proceeding, that in this test we are mostly
interested in the validity of the first assumption, that is, the possibility
of making finite-time measurements in practice. In case the Markovian
approximation is not valid the experiment can still be used to test
the extension of our model to non-Markovian measurement apparatus,
currently a work in progress.

In this experimental setting, we analyse what happens when we measure
the magnetic moment of a spin-1/2 particle with two Stern-Gerlach
apparatus, which corresponds to the kind of two-state measurement
described in the body of this article. The first apparatus performs
a {}``weak'' measurement, i. e., one that does not completely destroy
the coherences; and the second promoting the full collapse of the
wave function. As the second equipment's task is simply to measure
the state of the spins of the particle after they have been through
the first equipment, we shall not concern ourselves with its inner
workings, and it will not be necessary for this second apparatus to
actually be a Stern-Gerlach.

We choose the direction of the $z$ axis so the final measurement
is done along it. As this experiment will yield trivial results if
we choose the observable measured by the weak measurement as the same
the second one, we will tilt the first apparatus with an angle $\beta$
in respect with the $z$ axis. In this case, the observable being
measured by it will be:

\[
\hat{\sigma}_{\beta}=\cos\beta\hat{\sigma}_{z}+\sin\beta\hat{\sigma}_{x},
\]
 where we are calling the direction along which the particles move
the $y$ axis.

We shall not make any assumptions about the initial state of the spins
of the particles, allowing it to be represented by any $2\times2$
density matrix $\hat{\rho}\left(0\right)$. When these particles cross
the first equipment, which we will consider noiseless for simplicity,
it will reduce their coherences (in the basis of eigenstates of $\hat{\sigma}_{\beta}$)
to a fraction $0<b<1$ of their original value, while leaving the
populations intact. For {}``strong'', or {}``complete'' measurements,
the wave function collapse would result in the coherences completely
vanishing ($b=0$), but for a weak measurement $b$ is actually closer
to one.

Explicitly, we can write that, if a particle enters the apparatus
at $t=0$ and leaves it at $\tau$, its final state will be:

\[
\hat{\rho}\left(\tau\right)=\frac{1}{2}\left[1+b\left(\tau\right)\right]\hat{\rho}\left(0\right)+\frac{1}{2}\left[1-b\left(\tau\right)\right]\hat{\sigma}_{\beta}\hat{\rho}\left(0\right)\hat{\sigma}_{\beta}.
\]
 In the equation above, we have given the fraction $b$ a time dependence,
which is precisely the hypothesis we wish to test. If the measurement
is instantaneous, this function will be constant, while it will display
other kinds of behavior if it is actually a dynamical process. Therefore,
our experiment should be capable of providing us information about
the form of $b\left(t\right)$.

The final measurement, if performed correctly, will give us simply
the expectation value of the variable $\hat{\sigma}_{z}$ at $t=\tau$,
regardless of how much time it takes to extract this information from
the quantum system, or of how this process is performed. Therefore,
from the equation above, we can conclude that the expectancy value
of the experiment will be:

\[
\left\langle \hat{\sigma}_{z}\right\rangle \left(\tau\right)=\left[\cos^{2}\beta+b\left(\tau\right)\sin^{2}\beta\right]\left\langle \hat{\sigma}_{z}\right\rangle \left(0\right)+\frac{1}{2}\left[1-b\left(\tau\right)\right]\sin\left(2\beta\right)\left\langle \hat{\sigma}_{x}\right\rangle \left(0\right),
\]
 if we consider that the system does not evolve between the two measurements,
i. e. $\omega_{0}=0$.

Comparing this theoretical formula with the results of the experiments
described below, one would be able to deduce the form of the function
$b\left(t\right)$. Indeed, the results of $\left\langle \hat{\sigma}_{z}\right\rangle \left(\tau\right)$
and $\left\langle \hat{\sigma}_{z}\right\rangle \left(0\right)$can
be measured directly by turning on or off the first Stern-Gerlach
apparatus. If $\left\langle \hat{\sigma}_{x}\right\rangle \left(0\right)$
cannot be known from the preparation of the spins, we can eliminate
it from the equation by taking $\beta=\pi/2$:

\[
\left\langle \hat{\sigma}_{z}\right\rangle \left(\tau\right)=b\left(\tau\right)\left\langle \hat{\sigma}_{z}\right\rangle \left(0\right),\quad\beta=\frac{\pi}{2}.
\]
 This is probably the preferred setting of the experiment, but we
will proceed nevertheless treating $\beta$ as any angle, to keep
our results as general as possible. The only thing we have to keep
in mind is that we cannot choose $\beta=n\pi$, $n\in\mathbb{Z},$as
this would reduce the expression above to something trivial.

\subsection{Checking if the measurement is {}``weak''}

The experiment proposed requires the assumption that the first measurement
is weak. If it turns out that the first Stern-Gerlach is actually
completely collapsing the wave function, there will be no meaningful
results.

In this undesirable situation, we would have $b=0$, yielding a measurement
of:

\[
\left\langle \hat{\sigma}_{z}\right\rangle _{b=0}\left(\tau\right)=\cos^{2}\beta\left\langle \hat{\sigma}_{z}\right\rangle \left(0\right)+\frac{1}{2}\sin\left(2\beta\right)\left\langle \hat{\sigma}_{x}\right\rangle \left(0\right).
\]

We should, therefore, make sure our test results are sufficiently
distant from those expected when the measurement is not weak. The
difference between and the preferred and failed cases is of:

\[
\Delta_{z}\equiv\left|\left\langle \hat{\sigma}_{z}\right\rangle _{b\approx1}\left(\tau\right)-\left\langle \hat{\sigma}_{z}\right\rangle _{b=0}\left(\tau\right)\right|\approx\left|\cos^{2}\beta\left\langle \hat{\sigma}_{z}\right\rangle \left(0\right)-\frac{1}{2}\sin\left(2\beta\right)\left\langle \hat{\sigma}_{x}\right\rangle \left(0\right)\right|.
\]
 Therefore, it is desirable to choose the initial conditions and the
angle $\beta$ so that $\Delta_{z}$ is bigger than our experimental
errors, in order to distinguish the case when the measurement is weak
from the case when it is too strong.

In the best setting, where $\beta=\pi/2$, the undesirable situation
has $\left\langle \hat{\sigma}_{z}\right\rangle _{b=0}\left(\tau\right)=0$.
This means that, as long as we choose an initial condition satisfying
$\left|\left\langle \hat{\sigma}_{z}\right\rangle \left(0\right)\right|>0$,
we just have to measure an average value distinguishable from zero
to be sure we are indeed performing {}``weak'' measurements. The
error of the experiment, therefore, must be kept below $\Delta_{z}=\left|\left\langle \hat{\sigma}_{z}\right\rangle \left(0\right)\right|>0$.
On the other hand, it may also be important to verify if $\left\langle \hat{\sigma}_{z}\right\rangle \left(\tau\right)$
is distinguishable from $\left\langle \hat{\sigma}_{z}\right\rangle \left(0\right)$,
i. e., that \textbf{some} intermediary measurement is being performed.

\subsection{Varying times of measurement}

The previous experiment can be only used to make sure that there is
a $b$ function in action, but gives no information about the time
dependency of that function. To acquire information about it, we must
vary the measurement periods $\tau$ and see what effect these have
in the final result. There are two possible ways of varying the time
of exposure in a Stern-Gerlach equipment: you can change the velocity
with which the particles cross the apparatus, or you can change the
length of their path. It is just important not to change the strength
of measurement (by altering the gradient of the magnetic field, for
example), instead of the time of exposure of the spin.

In its most general form, the average value measured will be:

\[
\left\langle \hat{\sigma}_{z}\right\rangle \left(\tau\right)=A+b\left(\tau\right)C,
\]
 where $A$, $C$ are constants. Once we calculate them using the
known values of the initial conditions and the tilting of the intermediary
measurement $\beta$, we can plot the values of $b\left(\tau\right)$,
thereby verifying whether it is constant or if measurement indeed
has a time dependency. For the situation where $\beta=\pi/2,$ the
function becomes simply a renormalization of the measured average
values:

\[
b\left(\tau\right)=\frac{\left\langle \hat{\sigma}_{z}\right\rangle \left(\tau\right)}{\left\langle \hat{\sigma}_{z}\right\rangle \left(0\right)}.
\]
 If the measurement is an instantaneous phenomenon that does not depend
on the time of interaction of the apparatus, we should encounter a
constant, while a smooth dynamical measurement would produce a positive
function with the limits $b\left(\tau\to0\right)=1$, $b\left(\tau\to\infty\right)=0$.

\subsection{Connection with our model of measurement}

In our model of measurement as a Markovian interaction between the
system and the measurement apparatus, the $b\left(\tau\right)$ function
has the form of a negative exponential:

\[
b\left(\tau\right)=e^{-2\lambda^{2}\tau},
\]
 where $\lambda$ is the constant that represents the strength of
the coupling between the system and the measurement apparatus. This
is a function that satisfies the limits given in the previous section,
thus being a plausible candidate for the results that can be found
experimentally.

If controlling the time the particles remain under measurement is
not a feasible task in some laboratory, one could vary instead the
coupling strength with the measurement apparatus (represented by the
rate of change of the magnetic field, in the case of the Stern-Gerlach
experiment). If the dependence found between $b$ and $\lambda^{2}$
in this case is exponential, it will not be a confirmation that measurement
is a finite-time phenomenon, but it could be a good first step towards
confirming our theory. Conversely, if we are capable of controlling
$\tau$ and measuring $b$ with good precision, we can use the equation
above to calculate $\lambda^{2}$, transforming this theoretical parameter
into something concrete.

We should note, finally, that the heart of the Markovian approximation
is that $\lambda^{2}$ is a positive constant. In case it is not,
we would still find some change in the value of $b\left(t\right)$,
even though not as simple as a negative exponential. Therefore, this
experiment could still be used to determine the finite-time length
of the measurement, while our model would have to be altered to account
for varying of $\lambda^{2}$. As mentioned in the final section of
article, these adaptations are currently work in progress.

\end{document}